%% file: sigmars.tex
\documentstyle[12pt,psfig]{article}

\begin{document}

\begin{flushright}
hep-th/0401085
\end{flushright}

\vspace{5mm}

\begin{center}
{\Large \bf Black Brane World from Gravitating Half $\sigma$-lump}\\[12mm]
Donghyun Kim~~and~~Yoonbai Kim\\[2mm]
{\it BK21 Physics Research Division and Institute of Basic Science,
Sungkyunkwan University,\\
Suwon 440-746, Korea}\\
{\tt dhkim$@$newton.skku.ac.kr,~~ yoonbai$@$skku.edu}\\[6mm]
JungJai Lee\\[2mm]
{\it Department of Physics, Daejin University,
GyeongGi, Pocheon 487-711, Korea}\\
{\it Department of Physics, North Carolina
State University, Raleigh, NC 27695-8202, USA}\\
{\tt jjlee$@$daejin.ac.kr}

\end{center}

\vspace{5mm}

\begin{abstract}
We study O($N+1$) nonlinear $\sigma$-model in $(p+1+N)$-dimensional curved
spacetime with a negative cosmological constant, and find a new $\sigma$-lump
solution with half-integer winding and divergent energy. 
When the spatial structure of $N$ extra-dimensions
is determined by this global defect, a black $\sigma p$-brane surrounded
by the degenerated horizon is formed and its near-horizon geometry is
identified as a warp geometry of a cigar type.
\end{abstract}


\newpage

\setcounter{equation}{0}
\section{Introduction}

From the early stage of brane world 
scenario~\cite{Arkani-Hamed:1998rs} with warp 
geometry~\cite{Randall:1999ee,Randall:1999vf}, one of the intriguing issues
has been a construction of the brane world with a thick brane.
The popular method is to assume the bulk fields and to find a gravitating 
static defect solution which forms a thick brane in the extra-dimensions.

When the extra-dimension is one, a natural solitonic configuration is 
domain wall~\cite{DeWolfe:1999cp}. For two or more spatial extra-dimensions,
rotational symmetry is assumed and then the corresponding internal 
symmetry was also continuous. Without a negative bulk cosmological constant,
the global vortex in two extra-dimensions encounters a mild naked 
singularity\cite{Cohen:1999ia} as has been done in 
(2+1)-dimensions~\cite{Cohen:1988sg}.
Inclusion of a negative cosmological constant with the global vortex
reproduces a singularity-free warp metric with a thick 
brane~\cite{Gregory:1999gv} as the extension from zero-dimensional 
brane~\cite{Kim:1997ba}. The obtained intriguing thick brane world 
models  include global and local 
vortices~\cite{Gregory:1999gv} and 
\cite{Olasagasti:2000gx}-\cite{Gregory:2002tp},
global and local monopoles~\cite{Gherghetta:2000jf}-\cite{Roessl:2002rv}, 
topological lumps in nonlinear
$\sigma$-model~\cite{Olasagasti:2001hm}, and their 
higher dimensional analogues~\cite{Olasagasti:2000gx,Gherghetta:2000jf,
Gherghetta:2000rt,Moon:2000hn,Moon:2001ck}. 
Their zero thickness limits are also 
studied~\cite{Gherghetta:2000qi}.

The familiar brane world metric obtained in the above is summarized by
\begin{equation}\label{acmetric}
ds^{2}=e^{2A(R)}(dt^{2}-dx^{i 2})-dR^{2}-{C^{2}(R)}d\Omega^{2}_{N-1}.
\end{equation}
Here the warp factor $A(R)$ of interest is usually proportional to $-R$,
$C(R)$ is either a constant for cigar geometry or an exponential form,
and the angular part $d\Omega^{2}_{N-1}$ has solid deficit angle due to
the global defects with a long range energy tail. 
If we look into the first two terms in the warp metric (\ref{acmetric}),
an interesting structure of a degenerated horizon at $r_{{\rm H}}
\ne 0$ arises through 
a transformation to Poincar\'{e} coordinates
\begin{equation}\label{bbmet}
ds^{2}=(r-r_{{\rm H}})^{2}(dt^{2}-dx^{i 2})-\frac{dr^{2}}{(r-r_{{\rm H}})^{2}}
-{C^{2}(R(r))}d\Omega^{2}_{N-1}.
\end{equation}
This black brane structure is materialized by the global 
defects\cite{Moon:2000hn} 
irrespective of the dimensions of the black
brane~\cite{Kim:1997ba}.
About a possible source of this horizon structure, it seems likely to be 
the divergent energy of global defects, which is not so harmful in the 
brane world model because of the negative vacuum energy proportional to
the spatial volume of extra-dimensions at each point on the brane.
An appropriate model for settling this question is nonlinear $\sigma$-model.
In (2+1)-dimensional anti-de Sitter spacetime, 
O(3) nonlinear $\sigma$-model is known to support a half $\sigma$-lump with
divergent energy in addition to a topological $\sigma$-lump with finite
energy~\cite{Kim:1998gx}. The half $\sigma$-lump can form an extremal
charged BTZ black hole but the topological $\sigma$-lump cannot in 
(2+1)-dimensions.

In this paper, we will examine O($N+1$) nonlinear $\sigma$-model 
in ($p+1+N$)-dimensional bulk with a flat $p$-brane and $N$ spatial
extra-dimensions.
In ref.~\cite{Olasagasti:2001hm}, the topological $\sigma$-lump was studied
and the obtained warp geometry does not show the structure of
black brane world as expected. We will show that there exists the 
half $\sigma$-lump and find the corresponding degenerated horizon where
a cigar type geometry with an exponentially-decaying warp factor is realized.

The rest of the paper is organized as follows. In sections 2 and 3,
we consider O($N+1$) nonlinear $\sigma$-model in ($D=p+1+N$)-dimensional bulk
and find the nontopological half $\sigma$-lump solution of which 
warp geometry is obtained near-degenerated-horizon.
In section 4, we study in detail the properties of the warp geometry and 
its horizon. We conclude with summary and discussion in section 5.

\setcounter{equation}{0}
\section{O($N+1$) Nonlinear $\sigma$ Model in $(p+1+N)$-Dimensions}
We begin this section with introducing
O$(N+1)$ nonlinear $\sigma$-model in $D$-dimensional curved spacetime 
described by the action
\begin{equation}
S=\int d^D x\sqrt{(-1)^{D+1}g_{D}}\left[
-\frac{M^{D-2}_*}{16\pi}(R+2\tilde{\Lambda})+
\frac12 g^{AB}\nabla_A\phi^I\nabla_B\phi^I -
\frac{\lambda(x)}{2} (\phi^J\phi^J -v^2) \right],
\end{equation}
where $A, B, ...$, denote $D$-dimensional bulk indices and internal
indices, $I,J, ...$, run from 1 to $N+1$.
Note that both scalar field $\phi^{I}(x)$ and vacuum expectation value $v$
have the same mass dimension $(D-2)/2$, and an auxiliary field $\lambda(x)$
mass dimension two. $M_*$ is the mass scale of $D$-dimensional gravity and
bulk cosmological constant $\tilde{\Lambda}$ has mass dimension two,
which is assumed to be negative.

Euler-Lagrange equation for the scalar field after eliminating the 
auxiliary field $\lambda(x)$ is
\begin{equation}\label{seq}
\Box \phi^{I} - \left( \phi^{J}~{\frac{\Box}{v^2}}~\phi^{J} \right) 
\phi^{I} = 0.
\end{equation}
Einstein equations are given by
\begin{equation}\label{Rab}
{R^{A}}_{B} = \frac{8\pi}{M_{*}^{p+1}} {{\cal{T}}^{A}}_{B}
- \frac{2 \tilde{\Lambda}}{p+N-1}{\delta^{A}}_{B}, 
\end{equation}
where energy-momentum tensor $T^{AB}$ reads
\begin{eqnarray}\label{Tab}
T^{AB} = \nabla^A \phi^I\nabla^B \phi^I
-g^{AB} \left[\frac{1}{2} g^{CD}\nabla_{C} \phi^I \nabla_{D} \phi^I
- \frac{\lambda(x)}{2}(\phi^I \phi^I - v^2) \right],
\end{eqnarray}
and
\begin{equation}\label{modiE-M}
T \equiv {T^{M}}_{M} ,\qquad {{\cal{T}}^{A}}_{B} \equiv {T^{A}}_{B}
-\frac{T}{p+N-1} {\delta^{A}}_{B}.
\end{equation}

We are interested in the defect configurations of which spatial transverse
dimension is $p$. Nontrivial winding exists between $N$-dimensional
extra-dimensions and internal $N$-dimensional sphere of 
$\phi^{I}\phi^{I}=v^{2}$.
Thus, the dimension of bulk spacetime is given by
$D = p+N+1$ and we will call such defect as a $\sigma$-{\it lump} 
in the context of
soliton solution or a $\sigma p$-{\it brane} for the description of
brane world
from here on since the defect does not depend on the $p$-transverse spatial
coordinates
$\{ x^{i} | i=1,2,\cdots , p\}$.
The rotationally-symmetric defects are supposed to 
live in the extra-dimensions so that
an appropriate choice of metric is
Poincar\'{e} type coordinates:
\begin{equation}\label{onmetric}
ds^{2}= B(r)e^{2\Phi(r)} ( dt^2 -  d{x^{i2}} ) -
\frac{dr^2}{B(r)} -r^2 d{\Omega}^{2}_{N-1} ,
\end{equation}
where $d{\Omega}^{2}_{N-1}$ denotes solid angle in the $N$ extra-dimensions :
$d\Omega^{2}_{N-1}=d\theta^{2}_{1}+\sin^{2}\theta_{1}d\theta^{2}_{2}
+\dots +\sin^{2}\theta_{1}\cdots\sin^{2}\theta_{N-2}d\theta^{2}_{N-1}$
in the spherical coordinates $(r,\theta_{1},\theta_{2}, \cdots , \theta_{N-1})$.
We will show in the next section 
that the brane world of our interest is obtained either when 
it is bounded by a degenerated Cauchy horizon of $p+N-1$ dimensions or when the asymptotic region of sufficiently large distance is considered. Therefore, 
in what follows, we shall also call the horizon as {\it black brane horizon}
and the $(p+N+1)$-dimensional bulk spacetime as {\it black brane world}.

For static defect solutions, we take a hedgehog ansatz
\begin{equation}\label{onansatz}
\phi^{I} = v \hat{\phi^{I}},\\
\end{equation}
where
$\hat{\phi^{I}}=(\sin{F(r)}\sin{n\theta_1}\cdots
\sin{n\theta_{N-2}}\sin{n\theta_{N-1}},
\sin{F(r)}\sin{n\theta_1}\cdots\sin{n\theta_{N-2}}
\cos{n\theta_{N-1}},\\
\sin{F(r)} \sin{n\theta_1}\cdots\sin{n\theta_{N-3}}\cos{n\theta_{N-2}},
\cdot \cdot \cdot \cdot \cdot,
\sin{F(r)} \sin{n\theta_1} \cos{n\theta_2},
\sin{F(r)} \cos{n\theta_1},
\cos{F(r)}\hspace{1mm})$.
In order to keep the rotational symmetry in the extra-dimensions,
$n$ is an arbitrary natural number for $N=2$ but should be unity
for $N \geq 3$.

For the metric (\ref{onmetric}) and the ansatz (\ref{onansatz}), the scalar 
field equation (\ref{seq}) becomes
\begin{equation}\label{onscalareq}
BF'' +\left[ \frac{p+2}{2}B' + (p+1)B\Phi' +
\frac{N-1}{r}B\right]F'
- (N-1)\frac{n^{2}}{r^2} \sin{F}\cos{F}= 0,
\end{equation}
where prime $'$ denotes derivative of $r$.
The Einstein equations (\ref{Rab}) have the following nonvanishing components
\begin{eqnarray}
{R^{t}}_{t}
&=&\frac{B''}{2} +B \Phi''
+ \frac{p}{4}\frac{{B'}^2}{B} + \frac{2p+3}{2} B'\Phi' + (p+1) B
{\Phi'}^{2} +
\frac{N-1}{2} \frac{B'}{r} + (N-1)\frac{B \Phi'}{r}\nonumber\\
&=& 8\pi G_{D}{{\cal{T}}^{t}}_{t} -
\frac{2 \tilde{\Lambda}}{p+N-1} \label{onRtt},\\
{R^{r}}_{r}
&=&(p+1)\left(\frac{B''}{2}+B\Phi''+\frac{3}{2}B'\Phi'+B{\Phi'}^{2}\right)
+\frac{N-1}{2}\frac{B'}{r}\nonumber\\
&=& 8\pi G_{D} {{\cal{T}}^{r}}_{r} -
\frac{2 \tilde{\Lambda}}{p+N-1} \label{onRrr},\\
{R^{\theta_{a}}}_{\theta_{a}}
&=&\frac{p+2}{2} \frac{B'}{r} +
(p+1)\frac{B \Phi'}{r} +(N-2)\frac{B-1}{r^2}\nonumber\\
&=& 8\pi G_{D} {{\cal{T}}^{\theta_a}}_{\theta_a} - 
\frac{2 \tilde{\Lambda}}{p+N-1} \label{onRthth1},
\end{eqnarray}
where nonvanishing components of the energy-momentum tensor
(\ref{modiE-M}) are
\begin{eqnarray}
{{\cal{T}}^{t}}_{t} &=& {{\cal{T}}^{i}}_{i} = -
\frac{N-2}{2(p+1)}\left[B{F'}^2
+ (N-1)\frac{n^2}{r^2} \sin^2 F \right],\\
{{\cal{T}}^{r}}_{r} &=&
-\frac{N-2}{2(p+1)}\left[\frac{2p+N}{N-2} B F'^{2}
+(N-1)\frac{n^2}{r^2} \sin^{2} F \right],\\
{{\cal{T}}^{\theta_{a}}}_{\theta_{a}} &=&
-\frac{N-2}{2(p+1)}
\left[B{F'}^2 + \frac{(N-1)(N-2)+2(p+1)}{N-2} \frac{n^2}{r^2}
\sin^{2} F\right],
\end{eqnarray}
and their trace in Eq.~(\ref{modiE-M}) is
\begin{equation}
T = \frac{p+N-1}{2} \left( B F'^{2} 
+ \frac{N-1}{2} \frac{n^2}{r^2} \sin^2 F\right).
\end{equation}
In the above we have used rescaled variables for both the coordinates and
the cosmological constant as $x^{A} (\equiv v^{2/(D-2)} x^{A})$
and $\Lambda (\equiv \tilde{\Lambda}/v^{4/(D-2)})$. Dimensionless Newton's
constant is also defined by $G_{D} \equiv v^2 / {M_{*}}^{D-2}$.
The right-hand sides of the Einstein equations (\ref{onRtt})-(\ref{onRthth1}) 
involve three types of energy-momentum contributions, i.e., 
they are short-range derivative of the scalar amplitude $F$,
solitonic long-range term proportional to $n^2$, and 
negative vacuum energy $\Lambda$ proportional
to the volume of the bulk.

Independent components of the Einstein equations are summarized 
by two nonlinear
equations including first-order differential terms. Specifically, they are
$B, B'$, and $\Phi'$ in addition to $F$ and $F'$:
One is the $\theta_a \theta_a$-component (\ref{onRthth1})
and the other is obtained by subtraction of the $rr$-component multiplied by
$p+1$ from the $tt$-component (\ref{onRtt})
\begin{eqnarray}
&&\frac{1}{4}\frac{{B'}^2}{B} +  B' \Phi' +B {\Phi'}^2 
+ \frac{N-1}{r}\left[ \frac{1}{2(p+1)} B'
+ \frac{1}{p} B \Phi'\right]
= 8 \pi G_{D} \frac{N-2}{2(p+1)^2}\nonumber\\
&&\times\left[\frac{2(p+1) - p(N-2)}{p(N-2)}B{F'}^2
-(N-1)\frac{n^2}{r^2}\sin^2 F \right]
- \frac{2 \Lambda}{(p+1)(p+N-1)}.
\label{onRtt-subeq}
\end{eqnarray}

In the case of two extra-dimensions $(N=2)$, one may suspect possibility of 
self-duality of O(3) nonlinear $\sigma$-model, described by the first-order
equation. Once the cosmological constant is turned on, it has been proved 
that any static soliton configuration of the
first-order self-dual equation does not satisfy the second-order
Euler-Lagrange equation (\ref{seq}) under any static metric even for the
$p=0$ brane case~\cite{Kim:1998gx}. 
In fact, for the $p=0$ brane of O(3) nonlinear $\sigma$-model, self-duality can be saturated only under the stationary metric and the
obtained self-dual topological lump solution constitutes a spacetime with
closed time-like curves~\cite{Kim:1998cb}. 
It means that there does not exist any
Bogomolnyi type bound in this brane world irrespective of $p$ and $N$.

\section{Half $\sigma$-lump as a $\sigma p$-brane}

In this section, by examining the equations of motion, we will find
gravitating static half $\sigma$-lump solution of which nature 
is nontopological.
Spacetime structure formed by the half lump configuration involves an extremal
charged black $\sigma p$-brane with a single degenerated horizon.

Boundary conditions at the origin, $r=0$, are determined by nonsingularity of
the fields and the metric functions:
\begin{equation}
F(0)=0,~B(0)=1,~\Phi(0)=\Phi_{0},
\end{equation}
where $\Phi_{0}$ is always set to be zero by reparametrization of the spacetime variables
of the $p$-brane. Near the origin, the series solutions up to leading term are
\begin{eqnarray}
F(r) &\approx& F_0 r^{n} + \cdots , \label{f0r}\\
B(r) &\approx&1- \frac{2}{N-1} \left[ \frac{p-N+1}{N(p+N-1)} 
|\Lambda|\right.\nonumber\\
&&\left.+ \frac{2(p+1)-(N-2)(p-N+1)}{(p+1)}
2\pi G_D n^2 {F_0}^2 \delta_{n1}\right] r^2 + \cdots,\label{b0r}\\
\Phi(r) &\approx& \Phi_0 + \frac{1}{N-1} \left[ \frac{p}{N(p+N-1)} 
|\Lambda|\right.\nonumber\\
&&\left. + \frac{2(p+1)-p(N-2)}{p+1} 2\pi G_D n^2 {F_0}^2 \delta_{n1}
\right] r^2
+\cdots,\label{Phi0r}\\
B(r)e^{2[\Phi(r)-\Phi_{0}]} &\approx& 1+ {2}\left[ \frac{1}{N(p+N-1)}|\Lambda|
+ \frac{N-2}{p+1} 2\pi G_{D} n^2 {F_0}^2 \delta_{n1}\right] r^2 
+ \cdots ,\label{PB0r}
\end{eqnarray}
where $F_{0}$ is an undetermined constant decided by proper 
behavior of the scalar
field at opposite boundary, e.g., spatial infinity at $r=\infty$ or a
horizon at $r=r_{\rm H}$ if it exists. 

Since the metric function $\Phi$ itself does not appear in the
Einstein equations (\ref{onRthth1})-(\ref{onRtt-subeq}), those equations 
are understood as
a first-order differential equation for $B(r)$ and an algebraic equation for
$\Phi'(r)$ with initial values, i.e., the boundary conditions are $B(0)=1$
or $\Phi'(0)=0$.
It means that there is only one free parameter $F_{0}$ in the
series solutions (\ref{f0r})-(\ref{Phi0r}) to adjust for the existence
and location of the horizon, or equivalently the minimum value of $B(r)$,
so that expected structure of the horizon is determined by a fine-tuning of
the parameter $F_{0}$ for given $M_{\ast}$, $|\Lambda|$, and
$\sqrt{\lambda}v$.  Since $B(r) e^{2 [\Phi(r)-\Phi_{0}]}$ in Eq.~(\ref{PB0r}) 
is always increasing
near the origin, one cannot get some useful information on the 
existence of the horizon
from the series expansion near the origin. As confirmed by 
the numerical work later,
the metric $B(r)$ is likely to be decreasing near the origin.
A plausible condition to get a horizon $r_{\rm H}$
specified by $B(r_{\rm H})=0$ is to make $B(r)$ a decreasing function near the
origin, i.e., the conditions are $(dB/dr)|_{r=0}=0$ and 
$(d^{2}B/dr^{2})|_{r=0}<0$. 
Since the metric function $B(r)$ is convex up around its minimum point,
the position $r_{-}$ of vanishing $B(r)$ in Eq.~(\ref{b0r}) is smaller than
the horizon
\begin{equation}\label{rminus}
r_{-}=\left\{{{\frac{2}{N-1}\left[\frac{p-N+1}{N(p+N-1)}|\Lambda|
+\frac{2(p+1)-(N-2)(p-N+1)}{p+1}
2\pi n^2 G_{D} F^{2}_{0}\delta_{1n}\right]}}\right\}^{-\frac{1}{2}}
< r_{\rm H}.
\end{equation}

When the metric function $B(r)$ has the minimum at $r=r_{\rm min}$, one can 
classify the cases into three by signature of the value of 
$B(r_{\rm min})$, i.e., $B(r_{\rm min})>0$, =0, or $<0$.
Above all let us take into account $B(r_{\rm min})=0$ case
which implies coincidence of the horizon $r_{\rm H}$ and the minimum 
$r_{\rm min}$. The reason why is the formation of a brane world of an 
exponentially-decaying warp factor
is constituted in the case of $B(r_{\rm H})=0$ as we will show. 
Our basic assumption is finiteness of the field $F(r)$ and its derivative
$F^{'}(r)$ at the horizon $r_{\rm H}$, which reflects nonsingular
nature of the
configuration of our interest. Then it is
natural to assume vanishing of both the metric $B$ and its derivative $B^{'}$
at the horizon as $B(r_{\rm H})=B'(r_{\rm H})=0$. The resultant leading
behavior of the metric function $B(r)$ is
\begin{equation}\label{bhor}
B(r)\approx B_{\rm H}(r_{\rm H}-r)^{2}+B_{s}(r_{\rm H}-r)^{2+s}+ \cdots ,
\end{equation}
where $s\ge 1$.
Since the metric function $\Phi (r)$ appears only as its exponential in the 
metric or derivative
of it, a logarithmic singularity of $\Phi$ can be harmless for an appropriate
value of $\alpha$ including $\alpha=0$ :
\begin{equation}\label{phor}
e^{2[\Phi(r) - \Phi_{\rm H}]} \approx (r_{\rm H} - r)^{-2\alpha},
\end{equation}
where $\Phi_{\rm H}$ is another undetermined constant. 
Later, for arbitrary $\alpha$, we will find an appropriate coordinate
transformation.
In the scalar equation (\ref{onscalareq}), all the terms should vanish at the
horizon except for the last trigonometric function term due to the behavior 
of the metric functions
(\ref{bhor})-(\ref{phor}). It asks vanishing of the last term such that
the scalar amplitude should reach one of the following boundary value
at the horizon $r_{\rm H}$ :
\begin{eqnarray}\label{fbdhor}
F(r_{\rm H})=\left\{
          \begin{array}{ll}
          m \pi~~ &\mbox{from the sine term,}  \\
          (m + \frac{1}{2} )\pi ~~&\mbox{from the cosine term}.
          \end{array}\right.
\end{eqnarray}
Inserting the expansions (\ref{bhor})--(\ref{phor}) 
with the boundary condition of the scalar field (\ref{fbdhor})
into the $\theta_{a}\theta_{a}$-component (\ref{onRthth1}) of the Einstein 
equations, we express boundary value of the scalar amplitude $F$ at the
horizon as 
\begin{equation}\label{sinrh}
\sin F(r_{\rm H})=\pm \sqrt{\frac{1}{8\pi G_{D}n^{2}}
\left[\frac{2(p+1)}{2(p+1) + (N-1)(N-2)}\right]
\left[(N-2)+\frac{2|\Lambda|}{p+N-1}r_{\rm H}^{2}\right]} \, .
\end{equation}
For the topological $\sigma$-lump with $F(r_{\rm H})=\pi$, a mismatch in
Eq.~(\ref{sinrh}) prohibits
any nontrivial solution as far as the cosmological constant $\Lambda$ is 
negative.  On the other hand, for the nontopological $\sigma$-lump of
half-winding, we can
express the location of the horizon $r_{\rm H}$
in terms of the parameters of the theory
\begin{eqnarray}\label{rhori}
r_{\rm H}&=&\sqrt{\frac{p+N-1}{2|\Lambda|}
\left\{{8\pi G_{D}n^{2}}\left[ 1 + \frac{(N-1)(N-2)}{2(p+1)}\right]
-(N-2)\right\}} \nonumber\\
&\stackrel{N=2}{\Longrightarrow}& \sqrt{\frac{4(p+1)\pi G_{D}}{|\Lambda|}}
\, n\, .
\end{eqnarray}
As shown in the case of two extra-dimensions ($N = 2$), an intriguing but 
expected property is the fact that  radius of the horizon $r_{\rm H}$ 
is proportional to the vorticity $n$ which is a dimensionless parameter
representing popularity of overlapped $\sigma$-lumps. 

Positivity of the horizon $r_{\rm H}$ in Eq.~(\ref{rhori}) requires a condition
\begin{equation}\label{supe}
8\pi G_{D} n^2 \left[ 1 + \frac{(N-1)(N-2)}{2(p+1)}\right]> N-2,
\end{equation}
and it means that the natural scale of bulk nonlinear $\sigma$-field is
supermassive or Planck scale since the vorticity $n$ is unity as far as
$N$ is larger than two. When $n=1$, comparison between Eq.~(\ref{rminus}) and
Eq.~(\ref{sinrh}) provides a restriction on the shooting parameter $F_{0}$ of
Eq.~(\ref{f0r}):
\begin{eqnarray}\label{f0v}
F_0 &>&\left\{\frac{|\Lambda|}{4 \pi G_{D}}\frac{2(p+1)}{(p+N-1)[2(p+1)
-(N-2)(p-N+1)]}\right.
\nonumber\\ 
&&\left.\times\left[ \frac{(p+1)(N-1)}{4 \pi G_{D}[2(p+1)+(N-1)(N-2)]
-(p+1)(N-2)}
-\frac{p-N+1}{N}\right]\right\}^{\frac{1}{2}} \!.
\end{eqnarray}
When $p-N+1 \leq 0$, positivity of the square root in Eq.~(\ref{f0v}) asks 
supermassive symmetry breaking scale for any reasonable, finite $N$: 
\begin{equation}\label{pn1}
4 \pi G_{D} > \frac{(p+1)(N-1)}{2(p+1) + (N-1)(N-2)} \, .
\end{equation}
When $p-N+1 > 0$, it gives another range with an
upper bound for the scale of the bulk nonlinear $\sigma$-field
\begin{equation}\label{pn2}
\frac{(p+1)(N-2)}{2(p+1)+(N-1)(N-2)}< 4 \pi G_{D}
< \frac{(p+1)[N(N-1)+(N-2)(p-N+1)]}{(p-N+1)[2(p+1)+(N-1)(N-2)]}.
\end{equation}
For multi-defects $(n>1)$ in two extra-dimensions, we also have a
similar condition by comparing Eq.~(\ref{rminus}) to
Eq.~(\ref{rhori}) such as $4 \pi G_{D} n^{2} (p-1) > 1$.
If sufficiently-large number of the half $\sigma$-lumps are superimposed, 
an intriguing observation
is made under the crude approximation: the scale of spontaneous 
symmetry breaking $v^{2/(D-2)}$ needs not to be supermassive,
and a possible hierarchy between the Planck scale $M_{*}$ and 
the symmetry breaking scale $v^{2/(D-2)}$
is probably controlled by huge topological winding $n$.

Finally, leading term of the $tt$-component of the Einstein equations
(\ref{onRtt}) and second-order expansion of the 
$\theta_{a}\theta_{a}$-component
fix the coefficients of the metric functions (\ref{bhor})-(\ref{phor}) as
\begin{eqnarray}
B_{\rm H}&=& 
\frac{8 |\Lambda|}{(p+N-1)\left\{2 + (p+1)^2 \kappa  \mp (p+1) 
\sqrt{[4+\kappa(p+2)^2]\kappa}\right\}}\, ,\label{bh}\\
\alpha &=& 1 - \frac{p+1}{2}\kappa \pm \sqrt{\left[ 1 
+ \left(\frac{p+1}{2}\right)^2 
\kappa \right]\kappa} \, , \label{alpha}
\end{eqnarray}
where
\begin{eqnarray}\label{kapp}
\kappa &=& \frac{1}{2(p+1)} - \frac{(p+1)(N-1)(N-2)}{2(p+1) + (N-1)(N-2) 
- \frac{(p+1)(N-2)}{4 \pi G_{D} n^2}}\, . 
\end{eqnarray}
Furthermore, the power of the second term of $B$ in Eq.~(\ref{bhor}) is
also fixed by one, $s=1$.
Therefore, the lapse function in front of time variable $t$ results in
\begin{eqnarray}\label{metho}
\lefteqn{B(r) e^{2 [\Phi(r) - \Phi_{\rm H}]} \approx} \nonumber\\ 
&&\frac{8|\Lambda|}{(p+N-1)\left\{2 + (p+1)^2 \kappa \mp (p+1) 
\sqrt{[4+\kappa(p+2)^2]\kappa}\right\}}\nonumber\\
&&\times [\sigma (r_{\rm H} - r)]^{\{(p+1)
\kappa \mp \sqrt{\left[ 4 + {(p+1)^2} \kappa \right]\kappa}\}} ,
\end{eqnarray}
where $\sigma = +1$
for the interior region and $\sigma = -1$ for the exterior region. 
Between two possible solutions in Eqs.~(\ref{bh})--(\ref{metho}), a necessary 
condition for vanishing $B e^{2(\Phi-\Phi_{\rm H})}$ at $r_{\rm H}$ selects 
the lower sign in Eqs.~(\ref{bh})--(\ref{metho}) when $\kappa$ is positive. 
To be specific, the condition of positive $\kappa$ is rephrased by 
one of the following two possible conditions for $4\pi G_{D}n^{2}$:
Either
\begin{eqnarray}
4\pi G_{D} n^2 < \frac{(p+1)(N-2)}{2(p+1)+(N-1)(N-2)} \label{pn}
\end{eqnarray}
or
$$
4\pi G_{D} n^2 >\frac{(p+1)(N-2)}{2(p+1)+(N-1)(N-2)-2(p+1)^2 (N-1)(N-2)}
$$
together with
\begin{equation}\label{4p}
\frac{2(p+1)}{2p^2 + 4p + 1} > (N-1)(N-2) .
\end{equation}
Comparison of Eq.~(\ref{pn}) with Eqs.~(\ref{pn1})--(\ref{pn2}) 
tells us that the
condition given by Eq.~(\ref{pn}) seems unlikely.
Similarly, a probable condition is provided by the intersection of 
Eq.~(\ref{4p}) and Eq.~(\ref{pn1}) or 
that of Eq.~(\ref{4p}) and Eq.~(\ref{pn2}).

For a better understanding of leading behavior of the scalar field $F(r)$ near
the horizon $r_{\rm H}$, we consider a linearized equation for 
$\delta F(r)$ defined by
$F(r)=\frac{\pi}{2} + \delta F(r)$. As an approximation of $B(r)$ and $\Phi(r)$
we bring up the series solutions (\ref{bhor})-(\ref{phor}).
Substitution of these into the scalar equation (\ref{onscalareq}) leads to
a linear equation of $\delta F$
\begin{equation}\label{lineq}
x^{2} \frac{d^2 \delta F}{dx^2} + [1+(p+1)(1-\alpha)] x \frac{d \delta F}{dx}
- \delta (p+1)(1-\alpha)^2 \delta F = 0,
\end{equation}
where 
$1/x \equiv r_{\rm H}-r$ and
$\delta = {(N-1)n^{2}}/{\kappa \{ 4 \pi G_{D} n^2 [2(p+1) +(N-1)(N-2)] 
-(p+1)(N-2)\}}$.
Note that the previous condition for Eq.~(\ref{4p}) asks $\delta$ to be 
positive.
Then an exact solution of Eq.~(\ref{lineq}) is
\begin{equation}\label{parteq}
\delta F(x) = C_{+} x^{k_+} + C_{-} x^{k_-},
\end{equation}
where 
$k_{\pm}=\frac{1}{4}(p+1)^2 \kappa \left[ 1 +  
\sqrt{[1 + (\frac{2}{p+1})^2 \frac{1}{\kappa}]}\right]
\left[-1 \pm \sqrt{1+ \frac{4 \delta}{p+1}} \right] > 0$ , for both upper
and lower signs,
and $C_{\pm}$ are determined by proper
behaviors of the scalar field at both boundaries, $x = r_{\rm H}$
and $x = 0$.
\begin{figure}[ht]
\begin{center}
\input{srsfig1}
\end{center}
\caption{A representative shape of $\delta F(r)$ for 
a $\sigma$3-brane with three
extra-dimensions when $|\Lambda|=1$, $4 \pi G_{D} =1$, and $n=1$.
Coefficients, $C_{+}$ and $C_{-}$, are chosen as $-19.6016$ and $3.395$,
respectively. Aforementioned singularity appears at the horizon $r_{\rm H}$.}
\label{srsfig1}
\end{figure}
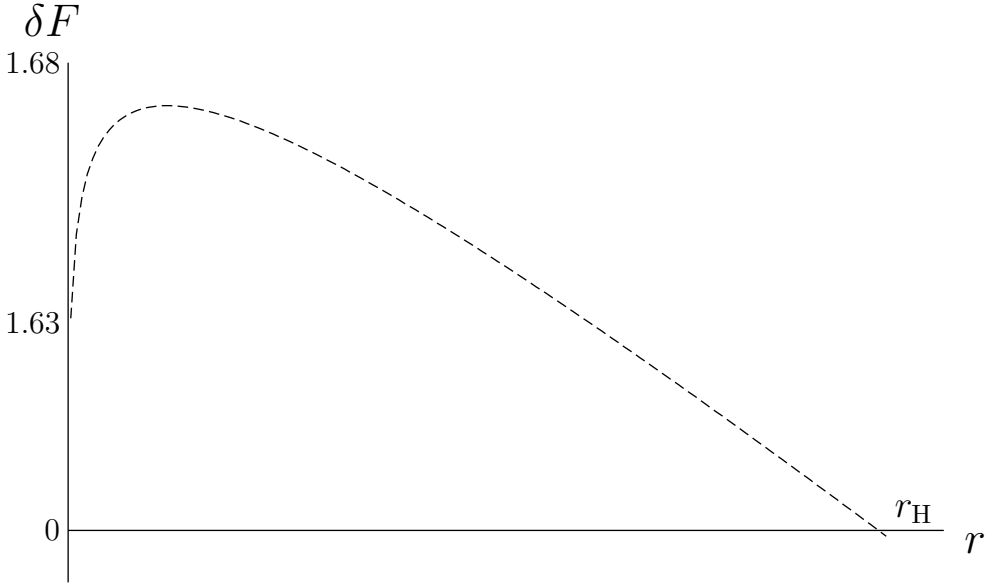
A specific form of $\delta F$ is shown in Fig.~{\ref{srsfig1}}. Trying a power
series solution $\delta F\approx F_{1}(r_{\rm H}-r)^{t}$ in the scalar
equation (\ref{lineq}), we cannot fix the coefficient $F_{1}$ but $t$
becomes
\begin{eqnarray}
&&t = \frac{1}{4} (p+1)^2 \kappa \left[ 1 + \sqrt{ 1 + 
(\frac{2}{p+1})^{2} \frac{1}{\kappa}}\right] 
\\
&&\times \left\{ -1 \pm \sqrt{1 + \frac{2 n^2 (N-1)}
{\kappa(p+1)
\left\{8 \pi G_{D} n^2 [2(p+1) + (N-1)(N-2)]-2(p+1)(N-2)\right\}}} \right\} .
\nonumber
\end{eqnarray}
Since the terms in the square root is positive, $t$ is also positive for
both upper and lower signs, which lets the perturbation allowable.
In order to describe our world of $p=3$, we have $t=0.161$ for two
extra-dimensions $(N=2)$ and $t=0.019$ for three extra-dimensions $(N=3)$
in the Planck scale such as $4 \pi G_{D} n^2 =1$.
The obtained values of $t$ are not natural numbers so that
$F(r)$ is mostly nonanalytic near the horizon but both $F(r_{\rm H})$ and
$F^{'}(r_{\rm H})$ are finite as expected.

Since second-order expansion of the Einstein equations
(\ref{onRtt})-(\ref{onRthth1}) does not involve contribution 
of the scalar field,
the coefficients of second-order terms in Eqs.~(\ref{bhor})--(\ref{phor}) are
given as
\begin{eqnarray}
&&B_{s}= \frac{2}{(p+1) \alpha + (p+4)}
\left\{\frac{p+N-1}{2|\Lambda|}
\left\{{8\pi G_{D}n^{2}}\left[ 1 + \frac{(N-1)(N-2)}{2(p+1)}\right]
-(N-2)\right\}\right\}^{-\frac{1}{2}}
\nonumber\\ 
&&\hspace{7mm}\times \left\{ \frac{ 8 |\Lambda|}{(p+N-1)[2 
+ (p+1)^2 \kappa \mp (p+1) 
\sqrt{[4+\kappa(p+2)^2]\kappa}]}
\right\}\nonumber\\
&&\hspace{7mm}\times \left\{ p[(p+1) \alpha - (p+2)] 
\sqrt{\frac{p+N-1}{2|\Lambda|}
\left\{{8\pi G_{D}n^{2}}\left[ 1 + \frac{(N-1)(N-2)}{2(p+1)}\right]
-(N-2)\right\}}\right.\nonumber\\
&&\hspace{7mm}\left.+ p(N-2) -  \alpha (p+1)(N-1)\right\} ,\label{bs}\\
&&\Phi_{\rm H}= \frac{2 - \alpha}{(p+1) \alpha + (p+4)}
\left\{(\alpha - 1) (p+1) -1 
+ \left[\frac{p(N-2) - \alpha (p+1)(N-1)}{p} 
\right] \right.\nonumber\\
&&\hspace{7mm}\left. \times \left\{\frac{p+N-1}{2|\Lambda|}
\left\{{8\pi G_{D}n^{2}}\left[ 1 + \frac{(N-1)(N-2)}{2(p+1)}\right]
-(N-2)\right\}\right\}^{-\frac{1}{2}}  
\right\}\nonumber\\
&&\hspace{7mm}\times \left\{ \frac{8 |\Lambda|}{(p+N-1)[2 
+ (p+1)^2 \kappa \mp (p+1) 
\sqrt{[4+\kappa(p+2)^2]\kappa}]}
\right\}\nonumber\\
&&\hspace{7mm}-\frac{\alpha(N-1)}{p}\left\{\frac{p+N-1}{2|\Lambda|}
\left\{{8\pi G_{D}n^{2}}\left[ 1 + \frac{(N-1)(N-2)}{2(p+1)}\right]
-(N-2)\right\}\right\}^{-\frac{1}{2}} ,\label{phih}
\end{eqnarray}
where $\alpha$ is given in Eq.~(\ref{alpha}).
When $r$ is larger than $r_{\rm H}$, the scalar amplitude $F(r)$
has its boundary value, $F(r)=\pi/2$, as mentioned
in Eq.~(\ref{fbdhor}). 

Though the scalar equation (\ref{onscalareq})
is trivially satisfied outside the horizon, the right-hand side of the 
Einstein equation
(\ref{onRthth1}) tells us that at the exterior region the negative vacuum
energy from the cosmological constant dominates and the long-range topological
term is a subleading term proportional to ${\cal O}(r^{-2})$.
It does not violate no hair conjecture because there is no distinction 
between the source of the topological term and the charge of an 
antisymmetric tensor field of rank $N-1$ for an observer
in the exterior region~\cite{Kim:1998cb,Kim:1998gx}.
Systematic expansion at asymptotic region gives the case that
the field and metric functions behave approximately as those of pure
anti-de Sitter spacetime:
\begin{eqnarray}
B(r) &\sim& \frac{2|\Lambda|}{(p+N)(p+N-1)} r^2
- \frac{2p+N-1}{p(p+1) + (N-2)(2p+N-1)} \nonumber\\
&&\times \left\{8 \pi G_{D} n^2 \left[1+\frac{(N-1)(N-2)}{2(p+1)} 
\right]-(N-2)\right\},
\label{binfr}\\
\Phi(r)- \Phi_{\infty}
&\sim&
\frac{(p+N)(p+N-1)}{4[(p+1)+(N-2)(2p+N-1)]}\nonumber\\
&&\times \left\{8 \pi G_{D} n^2 \left[1+\frac{(N-1)(N-2)}{2(p+1)} \right]
-(N-2)\right\}\frac{1}{|\Lambda|}
\frac{1}{r^2},
\label{pinfr}\nonumber\\
B(r) e^{2[\Phi(r) - \Phi_{\infty}]} &\sim& 
 \frac{2 |\Lambda|}{(p+N)(p+N-1)} r^{2} \nonumber\\
&&\hspace{-15mm}-\left\{ \frac{2p+N-1}{p(p+1) + (N-2)(2p+N-1)}
-\frac{1}{2[(p+1) + (N-2)(2p+N-1)]} 
\right\}\nonumber\\
&&\hspace{-15mm}\times  \left\{8 \pi G_{D} n^{2} \left[ 1 
+ \frac{(N-1)(N-2)}{2(p+1)}\right] - (N-2)\right\}\, ,
\label{bpinfr}
\end{eqnarray}
where $\Phi_{\infty}$ is determined by the proper boundary values
of the metric functions $\Phi$ and $B$ at the origin or at the horizon
$r_{\rm H}$ if it exists. Since the leading term of the metric function
$\Phi$ is a constant and the subleading term decays rapidly for sufficiently
large $r$, geometry of the exterior region is mostly governed by the metric
function $B$. The existence of a horizon requires necessarily negativity of
the constant term of the metric $B$ (\ref{binfr}) so that we reproduce
the condition of a supermassive scale
for $N>2$ in Eq.~(\ref{supe})
obtained at the interior region of the horizon. Another rough estimation of the
radius of the horizon can be performed from $e^{2\Phi(r_{+})} B(r_{+})=0$
in Eq.~(\ref{pinfr}) or $B({\tilde{r}}_{+})=0$ in Eq.~(\ref{binfr}):
\begin{eqnarray}\label{rpl}
&&{\tilde{r}}_{+}=\sqrt{\frac{(p+N)(2p+N-1)}{p(p+1)+(N-2)(2p+N-1)}} 
\, r_{\rm H}
\nonumber\\
&&>{r}_{+}=\sqrt{\frac{(p+N)(2p+N-1)}{p(p+1) + (N-2)(2p+N-1)}
-\frac{p+N}{2[(p+1) + (N-2)(2p+N-1)]}} \, r_{\rm H}\nonumber\\
&&>r_{\rm H}\,.
\end{eqnarray}
Since both prefactors in front of $r_{\rm H}$ in Eq.~(\ref{rpl}) are a few for
every $N$ $(N\ge 2)$, we read validity of the above expansion and signature
of the next order term in the expansion should be negative. Note that
$\mathop{\lim}\limits_{N \to \infty}{\tilde{r}}_{+}
= \mathop{\lim}\limits_{N \to \infty} r_{+} = r_{\rm H}$.
\begin{figure}[ht]
\begin{center}
\input{srsfig2a}
\end{center}
\vspace{-11mm}
\begin{center}{(a)}
\end{center}
\vspace{-6mm}
\begin{center}
\input{srsfig2b}
\end{center}
\vspace{-11mm}
\begin{center}{(b)}
\end{center}
\vspace{-6mm}
\caption{(a) Scalar amplitude $F(r)$ for various half-winding
solitons when $p=3$, $N=3$, $8\pi Gv^{2}=1.58$,
$|\Lambda|/v^{2}=0.65$, and $n=1$. The number of nodes increases
as initial shooting parameter $F_{0}$ increases. (b) Plots of
$B(r)$ show development of a horizon at $vr_{\rm H}=1.94$ as the
initial shooting parameter $F_{0}$ increases. For both figures,
$F_{0}= 0.23$ for (i), $F_{0}=0.43$ for (ii), $F_{0}=0.53$ for
(iii), $F_{0}=0.63$ for (iv), $F_{0}=0.73$ for (v).}
\label{srsfig2}
\end{figure}

Though the scalar field $F(r)$ and the metric function $\Phi$ are nonanalytic 
at the horizon $r_{\rm H}$, it does not mean existence of a physical 
singularity at the horizon. It can be proven by examining Kretschmann scalar
$R^{ABCD}R_{ABCD}$.
Computation of the Kretschmann scalar for the metric (\ref{onmetric}) is 
\begin{eqnarray}\label{Kre}
\lefteqn{R^{ABCD}R_{ABCD} =} \nonumber\\
&=&+(p+1)(B''^{2}+6B'' B' \Phi' +4B'' B\Phi''
+4B'' B \Phi'^{2} +12B'B\Phi'\Phi''+4B^{2} {\Phi''}^{2} 
+8B^{2} \Phi'')\nonumber\\
&&+ p(p+1) \left( \frac{{B'}^4}{8 B^{2}} + \frac{{B'}^3 \Phi'}{B}\right)
+(p+1)(p+3)(3 {B'}^2 {\Phi'}^2 + 4 B B' {\Phi'}^3 ) 
+ 3 (p^2 + p + 1) B^2 {\Phi'}^4 \nonumber\\
&& + \frac{N-1}{r^2} [(p+2) B'^2 + 4(p+1)(B B' \Phi' + B^2 \Phi'^2)].
\end{eqnarray}
At the position of the $\sigma p$-brane it has a finite value as expected :
\begin{eqnarray}\label{Kretori}
\lefteqn{\left. R^{ABCD}R_{ABCD} \right.|_{r=0}=} \nonumber\\
&&\frac{16(p+1)}{(N-1)^2} \left\{
\left[ \frac{p-N+1}{N(p+N-1)} |\Lambda|
+ \frac{2(p+1)-(N-2)(p-N+1)}{(p+1)}
2\pi G_D n^2 {F_0}^2 \delta_{n1}\right]^2 \right.\nonumber\\
&&\left.\times\left\{ 1+ 8 \left[ \frac{p}{N(p+N-1)} |\Lambda|
+ \frac{2(p+1)-p(N-2)}{p+1} 2\pi G_D n^2 {F_0}^2 \delta_{n1}\right]
\right\} \right.\nonumber\\
&&\left.-\left[ \frac{p}{N(p+N-1)} |\Lambda|
+ \frac{2(p+1)-p(N-2)}{p+1} 2\pi G_D n^2 {F_0}^2 \delta_{n1}\right]
\right.\nonumber\\
&&\left.\times \left\{ 2\left[ \frac{p-N+1}{N(p+N-1)} |\Lambda|
+ \frac{2(p+1)-(N-2)(p-N+1)}{(p+1)}
2\pi G_D n^2 {F_0}^2 \delta_{n1}\right] -(N-1)\right\} \right.\nonumber\\
&&\left.+ \left[ \frac{p}{N(p+N-1)} |\Lambda|
+ \frac{2(p+1)-p(N-2)}{p+1} 2\pi G_D n^2 {F_0}^2 \delta_{n1}\right]^2
\right\},
\end{eqnarray}
and substitution of Eqs.~(\ref{bhor})--(\ref{phor}) into Eq.~(\ref{Kre})
gives
\begin{eqnarray}
\lefteqn{\left. R^{ABCD}R_{ABCD}\right|_{r=r_{\rm H}}=} \nonumber\\
&& \frac{4 (p+1) \kappa |\Lambda|}
{(p+N-1)[-2(p+1) + \kappa (p^2 + 4p +5) \mp (p+1) 
\sqrt{[4+\kappa(p+1)^2]\kappa}]}\nonumber\\
&&\times \left\{ \left[2(-p+3) + (16 p^{2} - 6p - 25 ) \kappa 
+ (p+1)(32 p^2 + 13 p - 42 ) \kappa^{2}
- \frac{1}{2} (8p+3) (p+1)^{3} \kappa^{3} \right]\right.\nonumber\\
&&\left.\pm 2 [- 2 (4 p - 1 ) + p (24 p + 31 ) \kappa 
- (8p+3)(p+1)^{2} \kappa^{2}]
\sqrt{[ 4 + {(p+1)^{2}}\kappa]\kappa}\right\}\,, 
\end{eqnarray}
where $\kappa$ is given in Eq.~(\ref{kapp}). One can easily see 
that value of the Kretschmann scalar is finite at $r = r_{\rm H}$ and then 
the singularity at $r_{\rm H}$ is nothing but 
a coordinate singularity which also supports the claim 
that the position $r_{\rm H}$ is the location of the horizon 
of an extremal black hole. 
Furthermore, the terms in the expressions (\ref{Kre}) and (\ref{Kretori})
tell us that the whole spacetime is free from physical singularity.

{}From now on let us consider the cases of nonvanishing $B(r_{\rm min})$. When
$B(r_{\rm min})>0$, $\Phi(r)$ and $F(r)$ are smooth around $r_{{\rm min}}$
so that the bulk is smoothly connected to the pure anti-de Sitter
space (See the lines (i)--(iv) in Fig.~\ref{srsfig2}-(b)). 
Therefore the leading behavior of the metric near $r\approx r_{{\rm min}}$
dictates flat spacetime.

When $B(r_{\rm min})<0$, 
there should exist at least two positions represented by $r_{\rm H}$ to make 
the metric function vanish such as
$B(r_{\rm H}) = 0$. Then, the leading behavior of $B(r)$ is approximated by
\begin{equation}\label{mbh}
B(r) \approx B_{\rm H} (r_{\rm H} - r) + \cdots .
\end{equation}
Our basic assumption is the same as those of the extremal case such as
finiteness of the scalar field $F(r)$ and its derivative, and mild allowable 
singular form of the metric function $\Phi$ in Eq.~(\ref{phor}).
At any point $r_{\rm H}$ with vanishing $B(r)$, the Einstein equations 
(\ref{onRthth1}) and (\ref{onRtt-subeq}) are summarized as
\begin{eqnarray}
&&\left. \left[B B' \Phi' + \frac{1}{4} B'^{2} + (B \Phi')^{2}\right] 
\right|_{r=r_{\rm H}} = 0, \label{heq1}\\
&&\left. \left[B^2 \Phi'' + \frac{1}{2} B B' \Phi' - \frac{1}{4} B'^{2}
\right] \right|_{r=r_{\rm H}} = 0. \label{heq2}
\end{eqnarray}
Since $B(r_{{\rm H}})=0$, both Eqs.~(\ref{heq1})--(\ref{heq2}) require
$B'(r_{{\rm H}})=0$ which forces impossibility of geometry with two
horizons.

Throughout numerical computation, a prototype of $\sigma 3$-brane solution 
with unit topological winding ($n=1$) is displayed in Figure~\ref{srsfig2}.
The extra-dimensions are three, the symmetry breaking scale is almost Planck
scale ($8\pi Gv^{2}=1.58$), and the absolute value of negative cosmological
constant is also the square of the Planck scale ($|\Lambda|/v^{2}=0.65$). 
As shown in the solid line (v) of Fig.~\ref{srsfig2}-(a), the scalar amplitude
starts from zero at the origin, $F(0)=0$, and arrives at the boundary value of 
half-winding at the horizon, $F(r_{\rm H})=\pi/2$ after several oscillations 
around the boundary value. In the exterior region, $r>r_{\rm H}$, the
scalar amplitude remains at the boundary value. A horizon is developed at the
position $r_{\rm H}$ as shown in the solid line of Fig.~\ref{srsfig2}-(b)
by the half $\sigma$-lump configuration of $F(r_{\rm H})=\pi/2$.
Impossibility of the geometry with two horizons is also confirmed by
numerical works.

In this section, we look for static defect of 
O($N+1$) nonlinear $\sigma$-model, which live in the
$N$ extra-dimensions with a negative cosmological constant.
We find, in addition to usual topological $\sigma$-lump,
a new half $\sigma$-lump solution of which energy is not finite.
This divergent long-tail of the energy density is not harmful in 
anti-de Sitter spacetime but, intriguingly enough, it can form 
a degenerated horizon. In the next section, we will show that it is
the horizon of an extremal black $\sigma p$-brane and a coordinate 
transformation uncovers a near-horizon warp geometry of a black
brane world.

\section{Warp Geometry near the Horizon}
When we completely neglected the effect of matter fields, the geometry
is solely governed by the negative cosmological constant and, from 
Eqs.~(\ref{binfr})--(\ref{bpinfr}), its metric is
\begin{eqnarray}
d\tilde{s}^{2} \approx \tilde{r}^2 d\tilde{x}^{\mu} d\tilde{x}_{\mu} 
-\frac{d\tilde{r}^{2}}{\tilde{r}^2}
-\tilde{r}^{2}d\tilde{\Omega}^{2}_{N-1},
\end{eqnarray}
where we rescale the variables as $\tilde{s} = \sqrt{\Lambda_{{\rm eff}}}
\,\, s$,
$\tilde{r} = \Lambda_{{\rm eff}}\,\, {r}$, $\tilde{x}^{\mu} 
= e^{-\Phi_\infty}x^{\mu}$,
$d\tilde{\Omega}^{2}_{N-1} = d{\Omega^{2}_{N-1}}/\Lambda_{{\rm eff}}$ and 
$\Lambda_{{\rm eff}} 
= 2|\Lambda|/(p+N)(p+N-1)$. Note that there is a solid deficit
angle in extra-dimensions. A coordinate transformation to the familiar
coordinate system (\ref{acmetric}), 
$\tilde{r} = e^{\pm \rho} (0 \leq \rho \leq \infty)$, 
leads to a warp geometry with an exponentially-varying circumference
\begin{eqnarray}\label{wmet1}
d\tilde{s}^2 = e^{\pm 2 \rho} d\tilde{x}^{\mu} d\tilde{x}_{\mu}
- d \rho^2 - e^{\pm 2 \rho} d\tilde{\Omega}^2_{N-1}.
\end{eqnarray}
Since we started without effect of the matter fields, the warp factor 
in Eq.~(\ref{wmet1}) should
be universal independent of matter contents inside.

In what follows we interpret physical meaning of 
the near-horizon geometry obtained
in the previous section. At the interior region of the degenerated horizon at 
$r_{\rm H}$, it is depicted by
\begin{eqnarray}
d s^2 \approx \Lambda_{\rm H} ({r}_{\rm H} - r )^{\alpha_{\rm H}} 
dx^{\mu} d x_{\mu}
- \frac{d r^2}{B_{\rm H}({r}_{\rm H} - r)^2}
- r^2 d{\Omega}^{2}_{N-1}, 
\end{eqnarray}
where $r_{\rm H}, B_{\rm H}, \alpha_{\rm H},$ and $\Lambda_{\rm H}$ are given
in Eq.~(\ref{rhori}) and Eq.~(\ref{bh})--Eq.~(\ref{metho}).
Let us introduce a new radial variable
$\bar{r}$ as $e^{\pm 2 \bar{r}} = (r_{\rm H} - r)^{\alpha_{\rm H}}$ for small 
positive $e^{\pm 2 \bar{r}}$ together with rescaling 
$\bar{s} = ({\alpha_{\rm H}} \sqrt{B_{\rm H}} /2)s$,
$\bar{x}^{\mu} =( \alpha_{\rm H} \sqrt{{\Lambda_{\rm H}}{B_{\rm H}}} / 2) 
x^{\mu}$,
and $d \bar{\Omega}^{2}_{N-1} = ({\alpha_{\rm H}}^2 {B_{\rm H}}/4 )
d{\Omega^{2}_{N-1}}$. Here, the solid deficit angle at the horizon 
is due to the long tail of energy density of the half $\sigma {\it p}$-lump.
Then, the resultant metric describes a cigar type geometry with 
an exponentially decaying warp factor:
\begin{eqnarray}
d \bar{s}^2 \approx e^{\pm 2 \bar{r}} d\bar{x}^{\mu} d\bar{x}_{\mu}
- d{\bar{r}}^2 - r^{2}_{\rm H} d\bar{\Omega}^{2}_{N-1}.
\end{eqnarray}
Simultaneously, another coordinate transformation 
$(R-{r_{\rm H}})^2 = e^{\pm 2 \bar{r}}$ gives
\begin{eqnarray}
d \bar{s}^2 \approx (R - r_{\rm H})^2 d\bar{x}^{\mu} d\bar{x}_{\mu} 
- \frac{dR^2}{(R-r_{\rm H})^2} - r^{2}_{\rm H} d\bar{\Omega}^{2}_{N-1},
\end{eqnarray} 
which shows clearly the degenerated horizon at $R = r_{\rm H}$ 
of an extremal black $\sigma p$-brane.
 
We also study the detailed property of the black $\sigma p$-brane 
world and demonstrate clearly that the
obtained bulk spacetime near the horizon coincides exactly with the
Randall-Sundrum type brane world with one $\sigma p$-brane and $N$ 
extra-dimensions.
Since the spacetime structure of the $\sigma p$-brane is flat and we assumed
rotational symmetry in the $N$ extra-dimensions, nontrivial part of the bulk
geometry appears in the warp factor in front of the spacetime metric of the
$\sigma p$-brane and the radial component of the extra-dimensions as shown in
Eq.~(\ref{bbmet}). For a given 
time, the spatial geometry of the extra-dimensions may mostly 
be depicted by the
following quantities because of rotational symmetry 
of the extra-dimensions. They
are the radial distance from the origin $r=0$, $\displaystyle{
{\cal R}(r)=\int_{0}^{r}\frac{dr'}{\sqrt{B(r')}} }$,
the circumferential volume, $\displaystyle{l(r)=\int d{\Omega}_{N-1} 
r^{N-1} }$, and the spatial volume of the interior region of the bulk per unit 
volume of the $\sigma p$-brane, $\displaystyle{
{\cal V}(r)=\int d{\Omega}_{N-1}
\int_{0}^{r}dr' {r'}^{N-1}e^{(p+1)\Phi(r')}B(r')^{p/2}.
}$
If we use
the volume of an $(N-1)$ sphere of unit radius, $\int d \Omega_{N-1}
= 2 \pi^{N/2} / \Gamma (N/2)$, then we have 
$ \int d \bar{\Omega}_{N-1} = [2\pi^{N/2}/{\Gamma(N/2)}]
\times[\alpha_{\rm H}\sqrt{B_{\rm H}}/2]^{N-1}$, 
where the second factor stands for 
existence of a solid deficit angle due to the global defect sitting at 
the origin.
For spatial hypersurface of $p+N$ dimensions, the radial distance
from the origin $r=0$ to $r_{\rm H}$ is logarithmically divergent:
\begin{equation}
{\cal R}(r_{\rm H})=\int_{0}^{r_{\rm H}}\frac{dr'}{\sqrt{B(r')}}\sim
-\lim_{r\rightarrow r_{\rm H}}\ln|r_{\rm H}-r|\rightarrow\infty ,
\end{equation}
but the circumference is finite :
\begin{equation}
l(r_{\rm H})=\frac{2\pi^{N/2}}{\Gamma(N/2)} 
\left(\frac{\alpha_{\rm H}}{2} \sqrt{B_{\rm H}} r_{\rm H}\right)^{N-1}.
\end{equation}
Despite of the infinite radial distance and the finite circumferential volume,
the spatial volume per unit $\sigma p$-brane
volume bounded by the horizon is finite due to the warp factor in front of
the $\sigma p$-brane metric (\ref{bbmet}):
\begin{equation}\label{volh}
{\cal V}(r_{\rm H})=\frac{2\pi^{N/2}}{\Gamma(N/2)}\int d^{p}x
\int_{0}^{r_{\rm H}}dr' {r'}^{N-1}e^{(p+1)\Phi(r')}B(r')^{p/2}\bigg/
\int d^{p}x\sim \mbox{finite}.
\end{equation}

For the case of two extra-dimensions ($N=2$), the value of $r_{\rm H}$ is
proportional to the winding number $n$ in Eq.~(\ref{rhori}) so that the 
volume (\ref{volh}) can be large as $n$ becomes large. 
It suggests that our  black $\sigma p$-brane world interpolates
Randall-Sundrum type with two branes for small $n$ to Arkani
Hamed-Dimopoulos-Dvali type for sufficiently large $n$. 
Therefore, a geometric fine-tuning by a large extra-dimensions 
can be replaced by
that of large population of half $\sigma p$-lumps.

\begin{figure}[h]
\centerline{
\psfig{figure=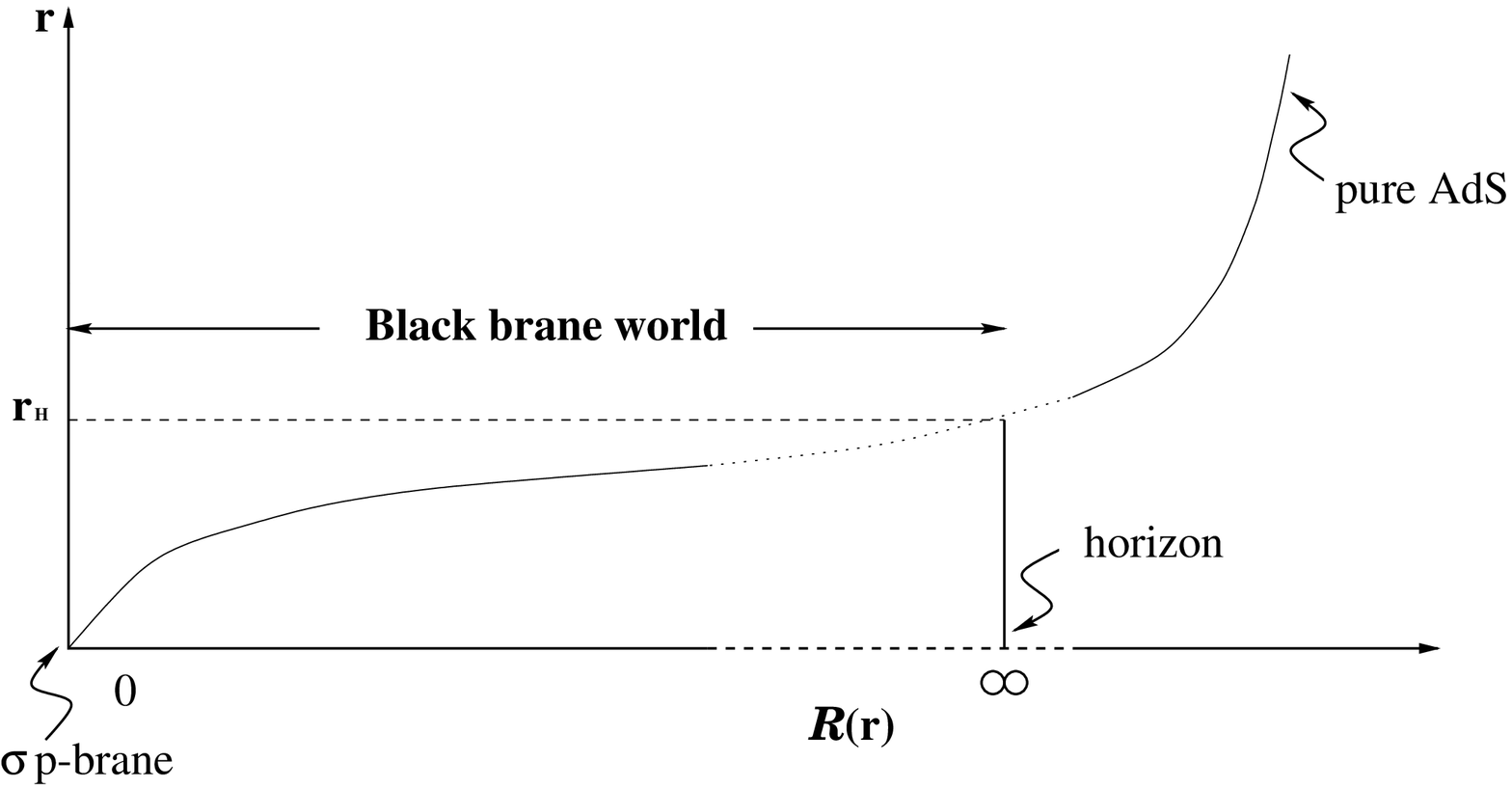,height=7.5cm}}
\caption{Schematic shape of a black $\sigma p$-brane world.}
\label{srsfig3}
\end{figure}

\setcounter{equation}{0}
\section{Summary}

We studied O($N+1$) nonlinear $\sigma$-model in $(p+1+N)$-dimensional 
curved spacetime with a negative cosmological constant. In addition to
the well-known topological $\sigma$-lump solution, we found half
$\sigma$-lump, another
class of solution similar to the global defect in linear $\sigma$-model.
When spatial $N$ extra-dimensions are formed
by this half $\sigma$-lump, our flat $p$-brane is located at the center of
the global defect surrounded by the degenerated horizon
of the black $\sigma$p-brane and its near-horizon geometry is 
a warp geometry of cigar type
(See Fig.~\ref{srsfig3}). Existence of this
half $\sigma$-lump can be detected by a solid deficit angle 
at the horizon due to
its divergent energy.
At the asymptotic region of anti-de Sitter space, another warp geometry 
is generated as usual (See Fig.~\ref{srsfig3}). 

Final comments are in order. Since the obtained horizon structure is 
extremal, we are temped to claim that the warp geometry near 
the degenerated horizon
is a sort of energetically-favored configurations according to the
experience in four-dimensional Reissner-Nordstrom black hole. 
However, justification of this statement should be postponed until we have
complicated stability analysis as in 
ref.~\cite{Kobayashi:2001jd,Freedman:2003ax}.
Though we found both static topological and half $\sigma$-lump
solutions, they are not BPS objects in anit-de Sitter spacetime and
a lesson from (2+1)-dimensions is that the BPS lump is given by
a stationary solution with closed timelike curves~\cite{Kim:1998cb}.
Recently, new global defect solutions have been 
found~\cite{Bazeia:2003qt}. It may be intriguing to test the above warp
structure by employing those objects.

\section*{Acknowledgements}
We would like to thank J. Troost for valuable discussions.
This work is supported by Korea Research Foundation Grants
KRF-2001-015-DP0082 and 
is the result of research activities (Astrophysical Research
Center for the Structure and Evolution of the Cosmos (ARCSEC))
supported by Korea Science $\&$ Engineering Foundation.

\end{document}

%% file: srsfig1.tex
\begingroup%
  \makeatletter%
  \newcommand{\GNUPLOTspecial}{%
    \@sanitize\catcode`\%=14\relax\special}%
  \setlength{\unitlength}{0.1bp}%
{\GNUPLOTspecial{!
/gnudict 256 dict def
gnudict begin
/Color false def
/Solid false def
/gnulinewidth 5.000 def
/userlinewidth gnulinewidth def
/vshift -33 def
/dl {10 mul} def
/hpt_ 31.5 def
/vpt_ 31.5 def
/hpt hpt_ def
/vpt vpt_ def
/M {moveto} bind def
/L {lineto} bind def
/R {rmoveto} bind def
/V {rlineto} bind def
/vpt2 vpt 2 mul def
/hpt2 hpt 2 mul def
/Lshow { currentpoint stroke M
  0 vshift R show } def
/Rshow { currentpoint stroke M
  dup stringwidth pop neg vshift R show } def
/Cshow { currentpoint stroke M
  dup stringwidth pop -2 div vshift R show } def
/UP { dup vpt_ mul /vpt exch def hpt_ mul /hpt exch def
  /hpt2 hpt 2 mul def /vpt2 vpt 2 mul def } def
/DL { Color {setrgbcolor Solid {pop []} if 0 setdash }
 {pop pop pop Solid {pop []} if 0 setdash} ifelse } def
/BL { stroke userlinewidth 2 mul setlinewidth } def
/AL { stroke userlinewidth 2 div setlinewidth } def
/UL { dup gnulinewidth mul /userlinewidth exch def
      10 mul /udl exch def } def
/PL { stroke userlinewidth setlinewidth } def
/LTb { BL [] 0 0 0 DL } def
/LTa { AL [1 udl mul 2 udl mul] 0 setdash 0 0 0 setrgbcolor } def
/LT0 { PL [] 1 0 0 DL } def
/LT1 { PL [4 dl 2 dl] 0 1 0 DL } def
/LT2 { PL [2 dl 3 dl] 0 0 1 DL } def
/LT3 { PL [1 dl 1.5 dl] 1 0 1 DL } def
/LT4 { PL [5 dl 2 dl 1 dl 2 dl] 0 1 1 DL } def
/LT5 { PL [4 dl 3 dl 1 dl 3 dl] 1 1 0 DL } def
/LT6 { PL [2 dl 2 dl 2 dl 4 dl] 0 0 0 DL } def
/LT7 { PL [2 dl 2 dl 2 dl 2 dl 2 dl 4 dl] 1 0.3 0 DL } def
/LT8 { PL [2 dl 2 dl 2 dl 2 dl 2 dl 2 dl 2 dl 4 dl] 0.5 0.5 0.5 DL } def
/Pnt { stroke [] 0 setdash
   gsave 1 setlinecap M 0 0 V stroke grestore } def
/Dia { stroke [] 0 setdash 2 copy vpt add M
  hpt neg vpt neg V hpt vpt neg V
  hpt vpt V hpt neg vpt V closepath stroke
  Pnt } def
/Pls { stroke [] 0 setdash vpt sub M 0 vpt2 V
  currentpoint stroke M
  hpt neg vpt neg R hpt2 0 V stroke
  } def
/Box { stroke [] 0 setdash 2 copy exch hpt sub exch vpt add M
  0 vpt2 neg V hpt2 0 V 0 vpt2 V
  hpt2 neg 0 V closepath stroke
  Pnt } def
/Crs { stroke [] 0 setdash exch hpt sub exch vpt add M
  hpt2 vpt2 neg V currentpoint stroke M
  hpt2 neg 0 R hpt2 vpt2 V stroke } def
/TriU { stroke [] 0 setdash 2 copy vpt 1.12 mul add M
  hpt neg vpt -1.62 mul V
  hpt 2 mul 0 V
  hpt neg vpt 1.62 mul V closepath stroke
  Pnt  } def
/Star { 2 copy Pls Crs } def
/BoxF { stroke [] 0 setdash exch hpt sub exch vpt add M
  0 vpt2 neg V  hpt2 0 V  0 vpt2 V
  hpt2 neg 0 V  closepath fill } def
/TriUF { stroke [] 0 setdash vpt 1.12 mul add M
  hpt neg vpt -1.62 mul V
  hpt 2 mul 0 V
  hpt neg vpt 1.62 mul V closepath fill } def
/TriD { stroke [] 0 setdash 2 copy vpt 1.12 mul sub M
  hpt neg vpt 1.62 mul V
  hpt 2 mul 0 V
  hpt neg vpt -1.62 mul V closepath stroke
  Pnt  } def
/TriDF { stroke [] 0 setdash vpt 1.12 mul sub M
  hpt neg vpt 1.62 mul V
  hpt 2 mul 0 V
  hpt neg vpt -1.62 mul V closepath fill} def
/DiaF { stroke [] 0 setdash vpt add M
  hpt neg vpt neg V hpt vpt neg V
  hpt vpt V hpt neg vpt V closepath fill } def
/Pent { stroke [] 0 setdash 2 copy gsave
  translate 0 hpt M 4 {72 rotate 0 hpt L} repeat
  closepath stroke grestore Pnt } def
/PentF { stroke [] 0 setdash gsave
  translate 0 hpt M 4 {72 rotate 0 hpt L} repeat
  closepath fill grestore } def
/Circle { stroke [] 0 setdash 2 copy
  hpt 0 360 arc stroke Pnt } def
/CircleF { stroke [] 0 setdash hpt 0 360 arc fill } def
/C0 { BL [] 0 setdash 2 copy moveto vpt 90 450  arc } bind def
/C1 { BL [] 0 setdash 2 copy        moveto
       2 copy  vpt 0 90 arc closepath fill
               vpt 0 360 arc closepath } bind def
/C2 { BL [] 0 setdash 2 copy moveto
       2 copy  vpt 90 180 arc closepath fill
               vpt 0 360 arc closepath } bind def
/C3 { BL [] 0 setdash 2 copy moveto
       2 copy  vpt 0 180 arc closepath fill
               vpt 0 360 arc closepath } bind def
/C4 { BL [] 0 setdash 2 copy moveto
       2 copy  vpt 180 270 arc closepath fill
               vpt 0 360 arc closepath } bind def
/C5 { BL [] 0 setdash 2 copy moveto
       2 copy  vpt 0 90 arc
       2 copy moveto
       2 copy  vpt 180 270 arc closepath fill
               vpt 0 360 arc } bind def
/C6 { BL [] 0 setdash 2 copy moveto
      2 copy  vpt 90 270 arc closepath fill
              vpt 0 360 arc closepath } bind def
/C7 { BL [] 0 setdash 2 copy moveto
      2 copy  vpt 0 270 arc closepath fill
              vpt 0 360 arc closepath } bind def
/C8 { BL [] 0 setdash 2 copy moveto
      2 copy vpt 270 360 arc closepath fill
              vpt 0 360 arc closepath } bind def
/C9 { BL [] 0 setdash 2 copy moveto
      2 copy  vpt 270 450 arc closepath fill
              vpt 0 360 arc closepath } bind def
/C10 { BL [] 0 setdash 2 copy 2 copy moveto vpt 270 360 arc closepath fill
       2 copy moveto
       2 copy vpt 90 180 arc closepath fill
               vpt 0 360 arc closepath } bind def
/C11 { BL [] 0 setdash 2 copy moveto
       2 copy  vpt 0 180 arc closepath fill
       2 copy moveto
       2 copy  vpt 270 360 arc closepath fill
               vpt 0 360 arc closepath } bind def
/C12 { BL [] 0 setdash 2 copy moveto
       2 copy  vpt 180 360 arc closepath fill
               vpt 0 360 arc closepath } bind def
/C13 { BL [] 0 setdash  2 copy moveto
       2 copy  vpt 0 90 arc closepath fill
       2 copy moveto
       2 copy  vpt 180 360 arc closepath fill
               vpt 0 360 arc closepath } bind def
/C14 { BL [] 0 setdash 2 copy moveto
       2 copy  vpt 90 360 arc closepath fill
               vpt 0 360 arc } bind def
/C15 { BL [] 0 setdash 2 copy vpt 0 360 arc closepath fill
               vpt 0 360 arc closepath } bind def
/Rec   { newpath 4 2 roll moveto 1 index 0 rlineto 0 exch rlineto
       neg 0 rlineto closepath } bind def
/Square { dup Rec } bind def
/Bsquare { vpt sub exch vpt sub exch vpt2 Square } bind def
/S0 { BL [] 0 setdash 2 copy moveto 0 vpt rlineto BL Bsquare } bind def
/S1 { BL [] 0 setdash 2 copy vpt Square fill Bsquare } bind def
/S2 { BL [] 0 setdash 2 copy exch vpt sub exch vpt Square fill Bsquare } bind def
/S3 { BL [] 0 setdash 2 copy exch vpt sub exch vpt2 vpt Rec fill Bsquare } bind def
/S4 { BL [] 0 setdash 2 copy exch vpt sub exch vpt sub vpt Square fill Bsquare } bind def
/S5 { BL [] 0 setdash 2 copy 2 copy vpt Square fill
       exch vpt sub exch vpt sub vpt Square fill Bsquare } bind def
/S6 { BL [] 0 setdash 2 copy exch vpt sub exch vpt sub vpt vpt2 Rec fill Bsquare } bind def
/S7 { BL [] 0 setdash 2 copy exch vpt sub exch vpt sub vpt vpt2 Rec fill
       2 copy vpt Square fill
       Bsquare } bind def
/S8 { BL [] 0 setdash 2 copy vpt sub vpt Square fill Bsquare } bind def
/S9 { BL [] 0 setdash 2 copy vpt sub vpt vpt2 Rec fill Bsquare } bind def
/S10 { BL [] 0 setdash 2 copy vpt sub vpt Square fill 2 copy exch vpt sub exch vpt Square fill
       Bsquare } bind def
/S11 { BL [] 0 setdash 2 copy vpt sub vpt Square fill 2 copy exch vpt sub exch vpt2 vpt Rec fill
       Bsquare } bind def
/S12 { BL [] 0 setdash 2 copy exch vpt sub exch vpt sub vpt2 vpt Rec fill Bsquare } bind def
/S13 { BL [] 0 setdash 2 copy exch vpt sub exch vpt sub vpt2 vpt Rec fill
       2 copy vpt Square fill Bsquare } bind def
/S14 { BL [] 0 setdash 2 copy exch vpt sub exch vpt sub vpt2 vpt Rec fill
       2 copy exch vpt sub exch vpt Square fill Bsquare } bind def
/S15 { BL [] 0 setdash 2 copy Bsquare fill Bsquare } bind def
/D0 { gsave translate 45 rotate 0 0 S0 stroke grestore } bind def
/D1 { gsave translate 45 rotate 0 0 S1 stroke grestore } bind def
/D2 { gsave translate 45 rotate 0 0 S2 stroke grestore } bind def
/D3 { gsave translate 45 rotate 0 0 S3 stroke grestore } bind def
/D4 { gsave translate 45 rotate 0 0 S4 stroke grestore } bind def
/D5 { gsave translate 45 rotate 0 0 S5 stroke grestore } bind def
/D6 { gsave translate 45 rotate 0 0 S6 stroke grestore } bind def
/D7 { gsave translate 45 rotate 0 0 S7 stroke grestore } bind def
/D8 { gsave translate 45 rotate 0 0 S8 stroke grestore } bind def
/D9 { gsave translate 45 rotate 0 0 S9 stroke grestore } bind def
/D10 { gsave translate 45 rotate 0 0 S10 stroke grestore } bind def
/D11 { gsave translate 45 rotate 0 0 S11 stroke grestore } bind def
/D12 { gsave translate 45 rotate 0 0 S12 stroke grestore } bind def
/D13 { gsave translate 45 rotate 0 0 S13 stroke grestore } bind def
/D14 { gsave translate 45 rotate 0 0 S14 stroke grestore } bind def
/D15 { gsave translate 45 rotate 0 0 S15 stroke grestore } bind def
/DiaE { stroke [] 0 setdash vpt add M
  hpt neg vpt neg V hpt vpt neg V
  hpt vpt V hpt neg vpt V closepath stroke } def
/BoxE { stroke [] 0 setdash exch hpt sub exch vpt add M
  0 vpt2 neg V hpt2 0 V 0 vpt2 V
  hpt2 neg 0 V closepath stroke } def
/TriUE { stroke [] 0 setdash vpt 1.12 mul add M
  hpt neg vpt -1.62 mul V
  hpt 2 mul 0 V
  hpt neg vpt 1.62 mul V closepath stroke } def
/TriDE { stroke [] 0 setdash vpt 1.12 mul sub M
  hpt neg vpt 1.62 mul V
  hpt 2 mul 0 V
  hpt neg vpt -1.62 mul V closepath stroke } def
/PentE { stroke [] 0 setdash gsave
  translate 0 hpt M 4 {72 rotate 0 hpt L} repeat
  closepath stroke grestore } def
/CircE { stroke [] 0 setdash 
  hpt 0 360 arc stroke } def
/Opaque { gsave closepath 1 setgray fill grestore 0 setgray closepath } def
/DiaW { stroke [] 0 setdash vpt add M
  hpt neg vpt neg V hpt vpt neg V
  hpt vpt V hpt neg vpt V Opaque stroke } def
/BoxW { stroke [] 0 setdash exch hpt sub exch vpt add M
  0 vpt2 neg V hpt2 0 V 0 vpt2 V
  hpt2 neg 0 V Opaque stroke } def
/TriUW { stroke [] 0 setdash vpt 1.12 mul add M
  hpt neg vpt -1.62 mul V
  hpt 2 mul 0 V
  hpt neg vpt 1.62 mul V Opaque stroke } def
/TriDW { stroke [] 0 setdash vpt 1.12 mul sub M
  hpt neg vpt 1.62 mul V
  hpt 2 mul 0 V
  hpt neg vpt -1.62 mul V Opaque stroke } def
/PentW { stroke [] 0 setdash gsave
  translate 0 hpt M 4 {72 rotate 0 hpt L} repeat
  Opaque stroke grestore } def
/CircW { stroke [] 0 setdash 
  hpt 0 360 arc Opaque stroke } def
/BoxFill { gsave Rec 1 setgray fill grestore } def
end
}}%
\begin{picture}(3600,2160)(0,0)%
{\GNUPLOTspecial{"
gnudict begin
gsave
0 0 translate
0.100 0.100 scale
0 setgray
newpath
1.000 UL
LTb
1.000 UL
LT0
150 296 M
3300 0 V
1.000 UL
LT0
150 100 M
0 1960 V
1.000 UL
LT1
160 1096 M
21 317 V
21 139 V
20 87 V
21 60 V
20 45 V
21 35 V
21 27 V
20 22 V
21 17 V
21 14 V
20 11 V
21 9 V
20 7 V
21 5 V
21 3 V
20 3 V
21 1 V
21 0 V
20 -1 V
21 -2 V
20 -2 V
21 -3 V
21 -4 V
20 -4 V
21 -5 V
21 -5 V
20 -6 V
21 -6 V
20 -6 V
21 -7 V
21 -7 V
20 -8 V
21 -8 V
21 -8 V
20 -8 V
21 -9 V
20 -8 V
21 -10 V
21 -9 V
20 -9 V
21 -10 V
21 -10 V
20 -10 V
21 -10 V
20 -10 V
21 -11 V
21 -11 V
20 -10 V
21 -11 V
21 -11 V
20 -12 V
21 -11 V
20 -11 V
21 -12 V
21 -12 V
20 -11 V
21 -12 V
21 -12 V
20 -12 V
21 -12 V
20 -13 V
21 -12 V
21 -12 V
20 -13 V
21 -12 V
21 -13 V
20 -13 V
21 -12 V
20 -13 V
21 -13 V
21 -13 V
20 -13 V
21 -13 V
21 -13 V
20 -14 V
21 -13 V
20 -13 V
21 -14 V
21 -13 V
20 -13 V
21 -14 V
21 -14 V
20 -13 V
21 -14 V
20 -14 V
21 -13 V
21 -14 V
20 -14 V
21 -14 V
21 -14 V
20 -14 V
21 -14 V
20 -14 V
21 -14 V
21 -14 V
20 -14 V
21 -14 V
21 -15 V
20 -14 V
21 -14 V
20 -15 V
21 -14 V
21 -14 V
20 -15 V
21 -14 V
21 -15 V
20 -14 V
21 -15 V
20 -14 V
21 -15 V
21 -14 V
20 -15 V
21 -15 V
21 -14 V
20 -15 V
21 -15 V
20 -15 V
21 -15 V
21 -14 V
20 -15 V
21 -15 V
21 -15 V
20 -15 V
21 -15 V
20 -15 V
21 -15 V
21 -15 V
20 -15 V
21 -15 V
21 -15 V
20 -15 V
21 -15 V
20 -15 V
21 -15 V
21 -15 V
20 -16 V
21 -15 V
21 -15 V
20 -15 V
21 -15 V
20 -16 V
21 -15 V
21 -15 V
20 -15 V
21 -16 V
21 -15 V
20 -15 V
21 -16 V
20 -15 V
stroke
grestore
end
showpage
}}%
\put(90,2160){\makebox(0,0)[b]{\shortstack{\Large{$\delta F$}}}}%
\put(3409,366){\makebox(0,0)[r]{\large{$r_{\rm H}$}}}%
\put(3609,246){\makebox(0,0)[r]{\Large{${r}$}}}%
\put(120,2060){\makebox(0,0)[r]{1.68}}%
\put(120,1080){\makebox(0,0)[r]{1.63}}%
\put(120,296){\makebox(0,0)[r]{0}}%
\end{picture}%
\endgroup
 

%% file: srsfig2a.tex
\begingroup%
  \makeatletter%
  \newcommand{\GNUPLOTspecial}{%
    \@sanitize\catcode`\%=14\relax\special}%
  \setlength{\unitlength}{0.1bp}%
{\GNUPLOTspecial{!
/gnudict 256 dict def
gnudict begin
/Color false def
/Solid false def
/gnulinewidth 5.000 def
/userlinewidth gnulinewidth def
/vshift -33 def
/dl {10 mul} def
/hpt_ 31.5 def
/vpt_ 31.5 def
/hpt hpt_ def
/vpt vpt_ def
/M {moveto} bind def
/L {lineto} bind def
/R {rmoveto} bind def
/V {rlineto} bind def
/vpt2 vpt 2 mul def
/hpt2 hpt 2 mul def
/Lshow { currentpoint stroke M
  0 vshift R show } def
/Rshow { currentpoint stroke M
  dup stringwidth pop neg vshift R show } def
/Cshow { currentpoint stroke M
  dup stringwidth pop -2 div vshift R show } def
/UP { dup vpt_ mul /vpt exch def hpt_ mul /hpt exch def
  /hpt2 hpt 2 mul def /vpt2 vpt 2 mul def } def
/DL { Color {setrgbcolor Solid {pop []} if 0 setdash }
 {pop pop pop Solid {pop []} if 0 setdash} ifelse } def
/BL { stroke userlinewidth 2 mul setlinewidth } def
/AL { stroke userlinewidth 2 div setlinewidth } def
/UL { dup gnulinewidth mul /userlinewidth exch def
      10 mul /udl exch def } def
/PL { stroke userlinewidth setlinewidth } def
/LTb { BL [] 0 0 0 DL } def
/LTa { AL [1 udl mul 2 udl mul] 0 setdash 0 0 0 setrgbcolor } def
/LT0 { PL [] 1 0 0 DL } def
/LT1 { PL [4 dl 2 dl] 0 1 0 DL } def
/LT2 { PL [2 dl 3 dl] 0 0 1 DL } def
/LT3 { PL [1 dl 1.5 dl] 1 0 1 DL } def
/LT4 { PL [5 dl 2 dl 1 dl 2 dl] 0 1 1 DL } def
/LT5 { PL [4 dl 3 dl 1 dl 3 dl] 1 1 0 DL } def
/LT6 { PL [2 dl 2 dl 2 dl 4 dl] 0 0 0 DL } def
/LT7 { PL [2 dl 2 dl 2 dl 2 dl 2 dl 4 dl] 1 0.3 0 DL } def
/LT8 { PL [2 dl 2 dl 2 dl 2 dl 2 dl 2 dl 2 dl 4 dl] 0.5 0.5 0.5 DL } def
/Pnt { stroke [] 0 setdash
   gsave 1 setlinecap M 0 0 V stroke grestore } def
/Dia { stroke [] 0 setdash 2 copy vpt add M
  hpt neg vpt neg V hpt vpt neg V
  hpt vpt V hpt neg vpt V closepath stroke
  Pnt } def
/Pls { stroke [] 0 setdash vpt sub M 0 vpt2 V
  currentpoint stroke M
  hpt neg vpt neg R hpt2 0 V stroke
  } def
/Box { stroke [] 0 setdash 2 copy exch hpt sub exch vpt add M
  0 vpt2 neg V hpt2 0 V 0 vpt2 V
  hpt2 neg 0 V closepath stroke
  Pnt } def
/Crs { stroke [] 0 setdash exch hpt sub exch vpt add M
  hpt2 vpt2 neg V currentpoint stroke M
  hpt2 neg 0 R hpt2 vpt2 V stroke } def
/TriU { stroke [] 0 setdash 2 copy vpt 1.12 mul add M
  hpt neg vpt -1.62 mul V
  hpt 2 mul 0 V
  hpt neg vpt 1.62 mul V closepath stroke
  Pnt  } def
/Star { 2 copy Pls Crs } def
/BoxF { stroke [] 0 setdash exch hpt sub exch vpt add M
  0 vpt2 neg V  hpt2 0 V  0 vpt2 V
  hpt2 neg 0 V  closepath fill } def
/TriUF { stroke [] 0 setdash vpt 1.12 mul add M
  hpt neg vpt -1.62 mul V
  hpt 2 mul 0 V
  hpt neg vpt 1.62 mul V closepath fill } def
/TriD { stroke [] 0 setdash 2 copy vpt 1.12 mul sub M
  hpt neg vpt 1.62 mul V
  hpt 2 mul 0 V
  hpt neg vpt -1.62 mul V closepath stroke
  Pnt  } def
/TriDF { stroke [] 0 setdash vpt 1.12 mul sub M
  hpt neg vpt 1.62 mul V
  hpt 2 mul 0 V
  hpt neg vpt -1.62 mul V closepath fill} def
/DiaF { stroke [] 0 setdash vpt add M
  hpt neg vpt neg V hpt vpt neg V
  hpt vpt V hpt neg vpt V closepath fill } def
/Pent { stroke [] 0 setdash 2 copy gsave
  translate 0 hpt M 4 {72 rotate 0 hpt L} repeat
  closepath stroke grestore Pnt } def
/PentF { stroke [] 0 setdash gsave
  translate 0 hpt M 4 {72 rotate 0 hpt L} repeat
  closepath fill grestore } def
/Circle { stroke [] 0 setdash 2 copy
  hpt 0 360 arc stroke Pnt } def
/CircleF { stroke [] 0 setdash hpt 0 360 arc fill } def
/C0 { BL [] 0 setdash 2 copy moveto vpt 90 450  arc } bind def
/C1 { BL [] 0 setdash 2 copy        moveto
       2 copy  vpt 0 90 arc closepath fill
               vpt 0 360 arc closepath } bind def
/C2 { BL [] 0 setdash 2 copy moveto
       2 copy  vpt 90 180 arc closepath fill
               vpt 0 360 arc closepath } bind def
/C3 { BL [] 0 setdash 2 copy moveto
       2 copy  vpt 0 180 arc closepath fill
               vpt 0 360 arc closepath } bind def
/C4 { BL [] 0 setdash 2 copy moveto
       2 copy  vpt 180 270 arc closepath fill
               vpt 0 360 arc closepath } bind def
/C5 { BL [] 0 setdash 2 copy moveto
       2 copy  vpt 0 90 arc
       2 copy moveto
       2 copy  vpt 180 270 arc closepath fill
               vpt 0 360 arc } bind def
/C6 { BL [] 0 setdash 2 copy moveto
      2 copy  vpt 90 270 arc closepath fill
              vpt 0 360 arc closepath } bind def
/C7 { BL [] 0 setdash 2 copy moveto
      2 copy  vpt 0 270 arc closepath fill
              vpt 0 360 arc closepath } bind def
/C8 { BL [] 0 setdash 2 copy moveto
      2 copy vpt 270 360 arc closepath fill
              vpt 0 360 arc closepath } bind def
/C9 { BL [] 0 setdash 2 copy moveto
      2 copy  vpt 270 450 arc closepath fill
              vpt 0 360 arc closepath } bind def
/C10 { BL [] 0 setdash 2 copy 2 copy moveto vpt 270 360 arc closepath fill
       2 copy moveto
       2 copy vpt 90 180 arc closepath fill
               vpt 0 360 arc closepath } bind def
/C11 { BL [] 0 setdash 2 copy moveto
       2 copy  vpt 0 180 arc closepath fill
       2 copy moveto
       2 copy  vpt 270 360 arc closepath fill
               vpt 0 360 arc closepath } bind def
/C12 { BL [] 0 setdash 2 copy moveto
       2 copy  vpt 180 360 arc closepath fill
               vpt 0 360 arc closepath } bind def
/C13 { BL [] 0 setdash  2 copy moveto
       2 copy  vpt 0 90 arc closepath fill
       2 copy moveto
       2 copy  vpt 180 360 arc closepath fill
               vpt 0 360 arc closepath } bind def
/C14 { BL [] 0 setdash 2 copy moveto
       2 copy  vpt 90 360 arc closepath fill
               vpt 0 360 arc } bind def
/C15 { BL [] 0 setdash 2 copy vpt 0 360 arc closepath fill
               vpt 0 360 arc closepath } bind def
/Rec   { newpath 4 2 roll moveto 1 index 0 rlineto 0 exch rlineto
       neg 0 rlineto closepath } bind def
/Square { dup Rec } bind def
/Bsquare { vpt sub exch vpt sub exch vpt2 Square } bind def
/S0 { BL [] 0 setdash 2 copy moveto 0 vpt rlineto BL Bsquare } bind def
/S1 { BL [] 0 setdash 2 copy vpt Square fill Bsquare } bind def
/S2 { BL [] 0 setdash 2 copy exch vpt sub exch vpt Square fill Bsquare } bind def
/S3 { BL [] 0 setdash 2 copy exch vpt sub exch vpt2 vpt Rec fill Bsquare } bind def
/S4 { BL [] 0 setdash 2 copy exch vpt sub exch vpt sub vpt Square fill Bsquare } bind def
/S5 { BL [] 0 setdash 2 copy 2 copy vpt Square fill
       exch vpt sub exch vpt sub vpt Square fill Bsquare } bind def
/S6 { BL [] 0 setdash 2 copy exch vpt sub exch vpt sub vpt vpt2 Rec fill Bsquare } bind def
/S7 { BL [] 0 setdash 2 copy exch vpt sub exch vpt sub vpt vpt2 Rec fill
       2 copy vpt Square fill
       Bsquare } bind def
/S8 { BL [] 0 setdash 2 copy vpt sub vpt Square fill Bsquare } bind def
/S9 { BL [] 0 setdash 2 copy vpt sub vpt vpt2 Rec fill Bsquare } bind def
/S10 { BL [] 0 setdash 2 copy vpt sub vpt Square fill 2 copy exch vpt sub exch vpt Square fill
       Bsquare } bind def
/S11 { BL [] 0 setdash 2 copy vpt sub vpt Square fill 2 copy exch vpt sub exch vpt2 vpt Rec fill
       Bsquare } bind def
/S12 { BL [] 0 setdash 2 copy exch vpt sub exch vpt sub vpt2 vpt Rec fill Bsquare } bind def
/S13 { BL [] 0 setdash 2 copy exch vpt sub exch vpt sub vpt2 vpt Rec fill
       2 copy vpt Square fill Bsquare } bind def
/S14 { BL [] 0 setdash 2 copy exch vpt sub exch vpt sub vpt2 vpt Rec fill
       2 copy exch vpt sub exch vpt Square fill Bsquare } bind def
/S15 { BL [] 0 setdash 2 copy Bsquare fill Bsquare } bind def
/D0 { gsave translate 45 rotate 0 0 S0 stroke grestore } bind def
/D1 { gsave translate 45 rotate 0 0 S1 stroke grestore } bind def
/D2 { gsave translate 45 rotate 0 0 S2 stroke grestore } bind def
/D3 { gsave translate 45 rotate 0 0 S3 stroke grestore } bind def
/D4 { gsave translate 45 rotate 0 0 S4 stroke grestore } bind def
/D5 { gsave translate 45 rotate 0 0 S5 stroke grestore } bind def
/D6 { gsave translate 45 rotate 0 0 S6 stroke grestore } bind def
/D7 { gsave translate 45 rotate 0 0 S7 stroke grestore } bind def
/D8 { gsave translate 45 rotate 0 0 S8 stroke grestore } bind def
/D9 { gsave translate 45 rotate 0 0 S9 stroke grestore } bind def
/D10 { gsave translate 45 rotate 0 0 S10 stroke grestore } bind def
/D11 { gsave translate 45 rotate 0 0 S11 stroke grestore } bind def
/D12 { gsave translate 45 rotate 0 0 S12 stroke grestore } bind def
/D13 { gsave translate 45 rotate 0 0 S13 stroke grestore } bind def
/D14 { gsave translate 45 rotate 0 0 S14 stroke grestore } bind def
/D15 { gsave translate 45 rotate 0 0 S15 stroke grestore } bind def
/DiaE { stroke [] 0 setdash vpt add M
  hpt neg vpt neg V hpt vpt neg V
  hpt vpt V hpt neg vpt V closepath stroke } def
/BoxE { stroke [] 0 setdash exch hpt sub exch vpt add M
  0 vpt2 neg V hpt2 0 V 0 vpt2 V
  hpt2 neg 0 V closepath stroke } def
/TriUE { stroke [] 0 setdash vpt 1.12 mul add M
  hpt neg vpt -1.62 mul V
  hpt 2 mul 0 V
  hpt neg vpt 1.62 mul V closepath stroke } def
/TriDE { stroke [] 0 setdash vpt 1.12 mul sub M
  hpt neg vpt 1.62 mul V
  hpt 2 mul 0 V
  hpt neg vpt -1.62 mul V closepath stroke } def
/PentE { stroke [] 0 setdash gsave
  translate 0 hpt M 4 {72 rotate 0 hpt L} repeat
  closepath stroke grestore } def
/CircE { stroke [] 0 setdash 
  hpt 0 360 arc stroke } def
/Opaque { gsave closepath 1 setgray fill grestore 0 setgray closepath } def
/DiaW { stroke [] 0 setdash vpt add M
  hpt neg vpt neg V hpt vpt neg V
  hpt vpt V hpt neg vpt V Opaque stroke } def
/BoxW { stroke [] 0 setdash exch hpt sub exch vpt add M
  0 vpt2 neg V hpt2 0 V 0 vpt2 V
  hpt2 neg 0 V Opaque stroke } def
/TriUW { stroke [] 0 setdash vpt 1.12 mul add M
  hpt neg vpt -1.62 mul V
  hpt 2 mul 0 V
  hpt neg vpt 1.62 mul V Opaque stroke } def
/TriDW { stroke [] 0 setdash vpt 1.12 mul sub M
  hpt neg vpt 1.62 mul V
  hpt 2 mul 0 V
  hpt neg vpt -1.62 mul V Opaque stroke } def
/PentW { stroke [] 0 setdash gsave
  translate 0 hpt M 4 {72 rotate 0 hpt L} repeat
  Opaque stroke grestore } def
/CircW { stroke [] 0 setdash 
  hpt 0 360 arc Opaque stroke } def
/BoxFill { gsave Rec 1 setgray fill grestore } def
end
}}%
\begin{picture}(3600,2160)(0,0)%
{\GNUPLOTspecial{"
gnudict begin
gsave
0 0 translate
0.100 0.100 scale
0 setgray
newpath
1.000 UL
LTb
400 300 M
63 0 V
2987 0 R
-63 0 V
400 700 M
63 0 V
2987 0 R
-63 0 V
400 1100 M
63 0 V
2987 0 R
-63 0 V
400 1500 M
63 0 V
2987 0 R
-63 0 V
400 1900 M
63 0 V
2987 0 R
-63 0 V
400 300 M
0 63 V
0 1697 R
0 -63 V
908 300 M
0 63 V
0 1697 R
0 -63 V
1417 300 M
0 63 V
0 1697 R
0 -63 V
1925 300 M
0 63 V
0 1697 R
0 -63 V
2433 300 M
0 63 V
0 1697 R
0 -63 V
2942 300 M
0 63 V
0 1697 R
0 -63 V
3450 300 M
0 63 V
0 1697 R
0 -63 V
1.000 UL
LTb
400 300 M
3050 0 V
0 1760 V
-3050 0 V
400 300 L
1.000 UL
LT1
1385 300 M
0 1760 V
1.000 UL
LT0
400 1557 M
3050 0 V
1.000 UL
LT0
400 300 M
5 6 V
5 6 V
5 6 V
5 5 V
5 6 V
6 6 V
5 6 V
5 6 V
5 6 V
5 5 V
5 6 V
5 6 V
5 6 V
5 6 V
5 6 V
5 6 V
5 6 V
6 5 V
5 6 V
5 6 V
5 6 V
5 6 V
5 6 V
5 6 V
5 6 V
5 6 V
5 5 V
5 6 V
5 6 V
6 6 V
5 6 V
5 6 V
5 6 V
5 6 V
5 6 V
5 6 V
5 6 V
5 6 V
5 6 V
5 6 V
5 6 V
6 6 V
5 6 V
5 6 V
5 6 V
5 6 V
5 6 V
5 6 V
5 7 V
5 6 V
5 6 V
5 6 V
5 6 V
6 6 V
5 6 V
5 7 V
5 6 V
5 6 V
5 6 V
5 7 V
5 6 V
5 6 V
5 6 V
5 7 V
5 6 V
6 6 V
5 7 V
5 6 V
5 7 V
5 6 V
5 6 V
5 7 V
5 6 V
5 7 V
5 6 V
5 7 V
5 6 V
6 7 V
5 7 V
5 6 V
5 7 V
5 7 V
5 6 V
5 7 V
5 7 V
5 7 V
5 6 V
5 7 V
5 7 V
6 7 V
5 7 V
5 7 V
5 7 V
5 7 V
5 7 V
5 7 V
5 7 V
5 7 V
5 8 V
5 7 V
5 7 V
6 7 V
5 8 V
5 7 V
5 8 V
5 7 V
5 8 V
5 7 V
5 8 V
5 8 V
5 7 V
5 8 V
5 8 V
6 8 V
5 8 V
5 8 V
5 8 V
5 8 V
5 8 V
5 8 V
5 9 V
5 8 V
5 9 V
5 8 V
5 9 V
6 8 V
5 9 V
5 9 V
5 9 V
5 9 V
5 9 V
5 9 V
5 10 V
5 9 V
5 10 V
5 9 V
5 10 V
6 10 V
5 10 V
5 10 V
5 10 V
5 11 V
5 10 V
5 11 V
5 11 V
5 11 V
5 11 V
5 12 V
5 11 V
6 12 V
5 13 V
5 12 V
5 13 V
5 12 V
5 14 V
5 13 V
5 14 V
5 15 V
5 14 V
5 15 V
5 16 V
6 16 V
5 17 V
5 18 V
5 18 V
5 19 V
5 20 V
5 21 V
5 23 V
5 24 V
5 25 V
5 27 V
5 28 V
6 28 V
5 25 V
5 17 V
5 10 V
5 3 V
5 -3 V
5 -6 V
5 -9 V
5 -10 V
5 -13 V
5 -14 V
5 -16 V
6 -17 V
5 -18 V
5 -21 V
5 -22 V
5 -26 V
5 -29 V
5 -35 V
5 -47 V
5 -61 V
5 15 V
5 18 V
5 8 V
6 3 V
5 2 V
5 0 V
5 0 V
5 0 V
5 -1 V
5 0 V
5 -1 V
5 0 V
5 0 V
5 -1 V
5 0 V
6 0 V
5 0 V
5 -1 V
5 0 V
5 0 V
5 0 V
5 0 V
5 0 V
5 -1 V
5 0 V
5 0 V
5 0 V
6 0 V
5 0 V
5 0 V
5 0 V
5 0 V
5 0 V
5 0 V
5 0 V
5 0 V
5 0 V
5 0 V
5 0 V
6 0 V
5 0 V
5 -1 V
5 0 V
5 0 V
5 0 V
5 0 V
5 0 V
5 0 V
5 0 V
5 0 V
5 0 V
6 0 V
5 0 V
5 0 V
5 0 V
5 0 V
5 0 V
5 0 V
5 0 V
5 0 V
5 0 V
5 0 V
5 0 V
6 0 V
5 0 V
5 0 V
5 0 V
5 0 V
5 0 V
5 0 V
5 0 V
5 0 V
5 0 V
5 0 V
5 0 V
6 0 V
5 0 V
5 0 V
5 0 V
5 0 V
5 0 V
5 0 V
5 0 V
5 0 V
5 0 V
5 0 V
5 0 V
6 0 V
5 0 V
5 0 V
5 0 V
5 0 V
5 0 V
5 0 V
5 0 V
5 0 V
5 0 V
5 0 V
5 0 V
6 0 V
5 0 V
5 0 V
5 0 V
5 0 V
5 0 V
5 0 V
5 0 V
5 0 V
5 0 V
5 0 V
5 0 V
6 0 V
5 0 V
5 0 V
5 0 V
5 0 V
5 0 V
5 0 V
5 0 V
5 0 V
5 0 V
5 0 V
5 0 V
6 0 V
5 0 V
5 0 V
5 0 V
5 0 V
5 0 V
5 0 V
5 0 V
5 0 V
5 0 V
5 0 V
5 0 V
6 0 V
5 0 V
5 0 V
5 0 V
5 0 V
5 0 V
5 0 V
5 0 V
5 0 V
5 0 V
5 0 V
5 0 V
6 0 V
5 0 V
5 0 V
5 0 V
5 0 V
5 0 V
5 0 V
5 0 V
5 0 V
5 0 V
5 0 V
5 0 V
6 0 V
5 0 V
5 0 V
5 0 V
5 0 V
5 0 V
5 0 V
5 0 V
5 0 V
5 0 V
5 0 V
5 0 V
6 0 V
5 0 V
5 0 V
5 0 V
5 0 V
5 0 V
5 0 V
5 0 V
5 0 V
5 0 V
5 0 V
5 0 V
6 0 V
5 0 V
5 0 V
5 0 V
5 0 V
5 0 V
5 0 V
5 0 V
5 0 V
5 0 V
5 0 V
5 0 V
6 0 V
5 0 V
5 0 V
5 0 V
5 0 V
5 0 V
5 0 V
5 0 V
5 0 V
5 0 V
currentpoint stroke M
5 0 V
5 0 V
6 0 V
5 0 V
5 0 V
5 0 V
5 0 V
5 0 V
5 0 V
5 0 V
5 0 V
5 0 V
5 0 V
5 0 V
6 0 V
5 0 V
5 0 V
5 0 V
5 0 V
5 0 V
5 0 V
5 0 V
5 0 V
5 0 V
5 0 V
5 0 V
6 0 V
5 0 V
5 0 V
5 0 V
5 0 V
5 0 V
5 0 V
5 0 V
5 0 V
5 0 V
5 0 V
5 0 V
6 0 V
5 0 V
5 0 V
5 0 V
5 0 V
5 0 V
5 0 V
5 0 V
5 0 V
5 0 V
5 0 V
5 0 V
6 0 V
5 0 V
5 0 V
5 0 V
5 0 V
5 0 V
5 0 V
5 0 V
5 0 V
5 0 V
5 0 V
5 0 V
6 0 V
5 0 V
5 0 V
5 0 V
5 0 V
5 0 V
5 0 V
5 0 V
5 0 V
5 0 V
5 0 V
5 0 V
6 0 V
5 0 V
5 0 V
5 0 V
5 0 V
5 0 V
5 0 V
5 0 V
5 0 V
5 0 V
5 0 V
5 0 V
6 0 V
5 0 V
5 0 V
5 0 V
5 0 V
5 0 V
5 0 V
5 0 V
5 0 V
5 0 V
5 0 V
5 0 V
6 0 V
5 0 V
5 0 V
5 0 V
5 0 V
5 0 V
5 0 V
5 0 V
5 0 V
5 0 V
5 0 V
5 0 V
6 0 V
5 0 V
5 0 V
5 0 V
5 0 V
5 0 V
5 0 V
5 0 V
5 0 V
5 0 V
5 0 V
5 0 V
6 0 V
5 0 V
5 0 V
5 0 V
5 0 V
5 0 V
5 0 V
5 0 V
5 0 V
5 0 V
5 0 V
5 0 V
6 0 V
5 0 V
5 0 V
5 0 V
5 0 V
5 0 V
5 0 V
5 0 V
5 0 V
5 0 V
5 0 V
5 0 V
6 0 V
5 0 V
5 0 V
5 0 V
5 0 V
5 0 V
5 0 V
5 0 V
5 0 V
5 0 V
5 0 V
5 0 V
6 0 V
5 0 V
5 0 V
5 0 V
5 0 V
5 0 V
5 0 V
5 0 V
5 0 V
5 0 V
5 0 V
5 0 V
6 0 V
5 0 V
5 0 V
5 0 V
5 0 V
5 0 V
5 0 V
5 0 V
5 0 V
5 0 V
5 0 V
5 0 V
6 0 V
5 0 V
5 0 V
5 0 V
5 0 V
5 0 V
5 0 V
5 0 V
5 0 V
5 0 V
5 0 V
5 0 V
6 0 V
5 0 V
5 0 V
5 0 V
5 0 V
5 0 V
5 0 V
1.000 UL
LT2
400 300 M
5 2 V
5 2 V
5 2 V
5 1 V
5 2 V
6 2 V
5 2 V
5 2 V
5 2 V
5 1 V
5 2 V
5 2 V
5 2 V
5 2 V
5 2 V
5 1 V
5 2 V
6 2 V
5 2 V
5 2 V
5 2 V
5 1 V
5 2 V
5 2 V
5 2 V
5 2 V
5 2 V
5 1 V
5 2 V
6 2 V
5 2 V
5 2 V
5 2 V
5 1 V
5 2 V
5 2 V
5 2 V
5 2 V
5 2 V
5 1 V
5 2 V
6 2 V
5 2 V
5 2 V
5 2 V
5 1 V
5 2 V
5 2 V
5 2 V
5 2 V
5 2 V
5 1 V
5 2 V
6 2 V
5 2 V
5 2 V
5 1 V
5 2 V
5 2 V
5 2 V
5 2 V
5 1 V
5 2 V
5 2 V
5 2 V
6 2 V
5 2 V
5 1 V
5 2 V
5 2 V
5 2 V
5 2 V
5 1 V
5 2 V
5 2 V
5 2 V
5 2 V
6 1 V
5 2 V
5 2 V
5 2 V
5 1 V
5 2 V
5 2 V
5 2 V
5 2 V
5 1 V
5 2 V
5 2 V
6 2 V
5 1 V
5 2 V
5 2 V
5 2 V
5 2 V
5 1 V
5 2 V
5 2 V
5 2 V
5 1 V
5 2 V
6 2 V
5 2 V
5 1 V
5 2 V
5 2 V
5 2 V
5 1 V
5 2 V
5 2 V
5 2 V
5 1 V
5 2 V
6 2 V
5 2 V
5 1 V
5 2 V
5 2 V
5 2 V
5 1 V
5 2 V
5 2 V
5 1 V
5 2 V
5 2 V
6 2 V
5 1 V
5 2 V
5 2 V
5 1 V
5 2 V
5 2 V
5 1 V
5 2 V
5 2 V
5 2 V
5 1 V
6 2 V
5 2 V
5 1 V
5 2 V
5 2 V
5 1 V
5 2 V
5 2 V
5 1 V
5 2 V
5 2 V
5 1 V
6 2 V
5 2 V
5 1 V
5 2 V
5 1 V
5 2 V
5 2 V
5 1 V
5 2 V
5 2 V
5 1 V
5 2 V
6 1 V
5 2 V
5 2 V
5 1 V
5 2 V
5 1 V
5 2 V
5 2 V
5 1 V
5 2 V
5 1 V
5 2 V
6 2 V
5 1 V
5 2 V
5 1 V
5 2 V
5 1 V
5 2 V
5 2 V
5 1 V
5 2 V
5 1 V
5 2 V
6 1 V
5 2 V
5 1 V
5 2 V
5 1 V
5 2 V
5 1 V
5 2 V
5 1 V
5 2 V
5 1 V
5 2 V
6 1 V
5 2 V
5 1 V
5 2 V
5 1 V
5 2 V
5 1 V
5 2 V
5 1 V
5 2 V
5 1 V
5 2 V
6 1 V
5 1 V
5 2 V
5 1 V
5 2 V
5 1 V
5 2 V
5 1 V
5 1 V
5 2 V
5 1 V
5 2 V
6 1 V
5 1 V
5 2 V
5 1 V
5 1 V
5 2 V
5 1 V
5 2 V
5 1 V
5 1 V
5 2 V
5 1 V
6 1 V
5 2 V
5 1 V
5 1 V
5 2 V
5 1 V
5 1 V
5 2 V
5 1 V
5 1 V
5 1 V
5 2 V
6 1 V
5 1 V
5 2 V
5 1 V
5 1 V
5 1 V
5 2 V
5 1 V
5 1 V
5 1 V
5 2 V
5 1 V
6 1 V
5 1 V
5 2 V
5 1 V
5 1 V
5 1 V
5 1 V
5 2 V
5 1 V
5 1 V
5 1 V
5 1 V
6 1 V
5 2 V
5 1 V
5 1 V
5 1 V
5 1 V
5 1 V
5 1 V
5 2 V
5 1 V
5 1 V
5 1 V
6 1 V
5 1 V
5 1 V
5 1 V
5 2 V
5 1 V
5 1 V
5 1 V
5 1 V
5 1 V
5 1 V
5 1 V
6 1 V
5 1 V
5 1 V
5 1 V
5 1 V
5 1 V
5 1 V
5 1 V
5 1 V
5 1 V
5 1 V
5 1 V
6 1 V
5 1 V
5 1 V
5 1 V
5 1 V
5 1 V
5 1 V
5 1 V
5 1 V
5 1 V
5 1 V
5 1 V
6 1 V
5 1 V
5 1 V
5 1 V
5 1 V
5 1 V
5 1 V
5 0 V
5 1 V
5 1 V
5 1 V
5 1 V
6 1 V
5 1 V
5 1 V
5 1 V
5 0 V
5 1 V
5 1 V
5 1 V
5 1 V
5 1 V
5 1 V
5 0 V
6 1 V
5 1 V
5 1 V
5 1 V
5 1 V
5 0 V
5 1 V
5 1 V
5 1 V
5 1 V
5 0 V
5 1 V
6 1 V
5 1 V
5 0 V
5 1 V
5 1 V
5 1 V
5 1 V
5 0 V
5 1 V
5 1 V
5 1 V
5 0 V
6 1 V
5 1 V
5 0 V
5 1 V
5 1 V
5 1 V
5 0 V
5 1 V
5 1 V
5 0 V
5 1 V
5 1 V
6 0 V
5 1 V
5 1 V
5 1 V
5 0 V
5 1 V
5 1 V
5 0 V
5 1 V
5 1 V
5 0 V
5 1 V
6 0 V
5 1 V
5 1 V
5 0 V
5 1 V
5 1 V
5 0 V
5 1 V
5 0 V
5 1 V
currentpoint stroke M
5 1 V
5 0 V
6 1 V
5 1 V
5 0 V
5 1 V
5 0 V
5 1 V
5 0 V
5 1 V
5 1 V
5 0 V
5 1 V
5 0 V
6 1 V
5 0 V
5 1 V
5 1 V
5 0 V
5 1 V
5 0 V
5 1 V
5 0 V
5 1 V
5 0 V
5 1 V
6 0 V
5 1 V
5 0 V
5 1 V
5 1 V
5 0 V
5 1 V
5 0 V
5 1 V
5 0 V
5 1 V
5 0 V
6 1 V
5 0 V
5 1 V
5 0 V
5 0 V
5 1 V
5 0 V
5 1 V
5 0 V
5 1 V
5 0 V
5 1 V
6 0 V
5 1 V
5 0 V
5 1 V
5 0 V
5 1 V
5 0 V
5 0 V
5 1 V
5 0 V
5 1 V
5 0 V
6 1 V
5 0 V
5 0 V
5 1 V
5 0 V
5 1 V
5 0 V
5 1 V
5 0 V
5 0 V
5 1 V
5 0 V
6 1 V
5 0 V
5 0 V
5 1 V
5 0 V
5 1 V
5 0 V
5 0 V
5 1 V
5 0 V
5 1 V
5 0 V
6 0 V
5 1 V
5 0 V
5 0 V
5 1 V
5 0 V
5 1 V
5 0 V
5 0 V
5 1 V
5 0 V
5 0 V
6 1 V
5 0 V
5 0 V
5 1 V
5 0 V
5 0 V
5 1 V
5 0 V
5 0 V
5 1 V
5 0 V
5 0 V
6 1 V
5 0 V
5 0 V
5 1 V
5 0 V
5 0 V
5 1 V
5 0 V
5 0 V
5 1 V
5 0 V
5 0 V
6 1 V
5 0 V
5 0 V
5 1 V
5 0 V
5 0 V
5 1 V
5 0 V
5 0 V
5 0 V
5 1 V
5 0 V
6 0 V
5 1 V
5 0 V
5 0 V
5 0 V
5 1 V
5 0 V
5 0 V
5 1 V
5 0 V
5 0 V
5 0 V
6 1 V
5 0 V
5 0 V
5 1 V
5 0 V
5 0 V
5 0 V
5 1 V
5 0 V
5 0 V
5 0 V
5 1 V
6 0 V
5 0 V
5 0 V
5 1 V
5 0 V
5 0 V
5 0 V
5 1 V
5 0 V
5 0 V
5 0 V
5 1 V
6 0 V
5 0 V
5 0 V
5 1 V
5 0 V
5 0 V
5 0 V
5 1 V
5 0 V
5 0 V
5 0 V
5 1 V
6 0 V
5 0 V
5 0 V
5 1 V
5 0 V
5 0 V
5 0 V
5 0 V
5 1 V
5 0 V
5 0 V
5 0 V
6 1 V
5 0 V
5 0 V
5 0 V
5 0 V
5 1 V
5 0 V
1.000 UL
LT4
400 300 M
5 3 V
5 4 V
5 3 V
5 4 V
5 3 V
6 4 V
5 3 V
5 4 V
5 3 V
5 3 V
5 4 V
5 3 V
5 4 V
5 3 V
5 4 V
5 3 V
5 4 V
6 3 V
5 3 V
5 4 V
5 3 V
5 4 V
5 3 V
5 4 V
5 3 V
5 4 V
5 3 V
5 3 V
5 4 V
6 3 V
5 4 V
5 3 V
5 4 V
5 3 V
5 4 V
5 3 V
5 3 V
5 4 V
5 3 V
5 4 V
5 3 V
6 4 V
5 3 V
5 4 V
5 3 V
5 3 V
5 4 V
5 3 V
5 4 V
5 3 V
5 4 V
5 3 V
5 4 V
6 3 V
5 4 V
5 3 V
5 4 V
5 3 V
5 3 V
5 4 V
5 3 V
5 4 V
5 3 V
5 4 V
5 3 V
6 4 V
5 3 V
5 4 V
5 3 V
5 4 V
5 3 V
5 4 V
5 3 V
5 3 V
5 4 V
5 3 V
5 4 V
6 3 V
5 4 V
5 3 V
5 4 V
5 3 V
5 4 V
5 3 V
5 4 V
5 3 V
5 4 V
5 3 V
5 4 V
6 3 V
5 4 V
5 3 V
5 4 V
5 3 V
5 4 V
5 3 V
5 4 V
5 3 V
5 4 V
5 3 V
5 4 V
6 3 V
5 4 V
5 3 V
5 4 V
5 3 V
5 4 V
5 3 V
5 4 V
5 3 V
5 4 V
5 3 V
5 4 V
6 3 V
5 4 V
5 3 V
5 4 V
5 3 V
5 4 V
5 3 V
5 4 V
5 3 V
5 4 V
5 3 V
5 4 V
6 3 V
5 4 V
5 3 V
5 4 V
5 3 V
5 4 V
5 4 V
5 3 V
5 4 V
5 3 V
5 4 V
5 3 V
6 4 V
5 3 V
5 4 V
5 3 V
5 4 V
5 3 V
5 4 V
5 3 V
5 4 V
5 3 V
5 4 V
5 3 V
6 4 V
5 4 V
5 3 V
5 4 V
5 3 V
5 4 V
5 3 V
5 4 V
5 3 V
5 4 V
5 3 V
5 4 V
6 3 V
5 4 V
5 3 V
5 4 V
5 3 V
5 4 V
5 3 V
5 4 V
5 3 V
5 4 V
5 3 V
5 4 V
6 3 V
5 4 V
5 3 V
5 4 V
5 3 V
5 4 V
5 3 V
5 4 V
5 3 V
5 4 V
5 3 V
5 4 V
6 3 V
5 4 V
5 3 V
5 4 V
5 3 V
5 4 V
5 3 V
5 4 V
5 3 V
5 4 V
5 3 V
5 3 V
6 4 V
5 3 V
5 4 V
5 3 V
5 4 V
5 3 V
5 3 V
5 4 V
5 3 V
5 3 V
5 4 V
5 3 V
6 4 V
5 3 V
5 3 V
5 4 V
5 3 V
5 3 V
5 4 V
5 3 V
5 3 V
5 4 V
5 3 V
5 3 V
6 3 V
5 4 V
5 3 V
5 3 V
5 3 V
5 4 V
5 3 V
5 3 V
5 3 V
5 3 V
5 4 V
5 3 V
6 3 V
5 3 V
5 3 V
5 3 V
5 3 V
5 3 V
5 3 V
5 3 V
5 4 V
5 3 V
5 3 V
5 3 V
6 3 V
5 2 V
5 3 V
5 3 V
5 3 V
5 3 V
5 3 V
5 3 V
5 3 V
5 3 V
5 2 V
5 3 V
6 3 V
5 3 V
5 3 V
5 2 V
5 3 V
5 3 V
5 2 V
5 3 V
5 3 V
5 2 V
5 3 V
5 3 V
6 2 V
5 3 V
5 2 V
5 3 V
5 2 V
5 3 V
5 2 V
5 2 V
5 3 V
5 2 V
5 3 V
5 2 V
6 2 V
5 3 V
5 2 V
5 2 V
5 2 V
5 3 V
5 2 V
5 2 V
5 2 V
5 2 V
5 2 V
5 2 V
6 2 V
5 2 V
5 2 V
5 2 V
5 2 V
5 2 V
5 2 V
5 2 V
5 2 V
5 2 V
5 2 V
5 2 V
6 2 V
5 1 V
5 2 V
5 2 V
5 2 V
5 1 V
5 2 V
5 2 V
5 1 V
5 2 V
5 2 V
5 1 V
6 2 V
5 1 V
5 2 V
5 1 V
5 2 V
5 1 V
5 2 V
5 1 V
5 2 V
5 1 V
5 1 V
5 2 V
6 1 V
5 2 V
5 1 V
5 1 V
5 1 V
5 2 V
5 1 V
5 1 V
5 1 V
5 2 V
5 1 V
5 1 V
6 1 V
5 1 V
5 2 V
5 1 V
5 1 V
5 1 V
5 1 V
5 1 V
5 1 V
5 1 V
5 1 V
5 1 V
6 1 V
5 1 V
5 1 V
5 1 V
5 1 V
5 1 V
5 1 V
5 1 V
5 1 V
5 1 V
5 0 V
5 1 V
6 1 V
5 1 V
5 1 V
5 1 V
5 0 V
5 1 V
5 1 V
5 1 V
5 1 V
5 0 V
5 1 V
5 1 V
6 1 V
5 0 V
5 1 V
5 1 V
5 0 V
5 1 V
5 1 V
5 0 V
5 1 V
5 1 V
5 0 V
5 1 V
6 1 V
5 0 V
5 1 V
5 1 V
5 0 V
5 1 V
5 0 V
5 1 V
5 0 V
5 1 V
currentpoint stroke M
5 1 V
5 0 V
6 1 V
5 0 V
5 1 V
5 0 V
5 1 V
5 0 V
5 1 V
5 0 V
5 1 V
5 0 V
5 1 V
5 0 V
6 1 V
5 0 V
5 1 V
5 0 V
5 1 V
5 0 V
5 0 V
5 1 V
5 0 V
5 1 V
5 0 V
5 1 V
6 0 V
5 0 V
5 1 V
5 0 V
5 1 V
5 0 V
5 0 V
5 1 V
5 0 V
5 0 V
5 1 V
5 0 V
6 1 V
5 0 V
5 0 V
5 1 V
5 0 V
5 0 V
5 1 V
5 0 V
5 0 V
5 1 V
5 0 V
5 0 V
6 1 V
5 0 V
5 0 V
5 0 V
5 1 V
5 0 V
5 0 V
5 1 V
5 0 V
5 0 V
5 0 V
5 1 V
6 0 V
5 0 V
5 1 V
5 0 V
5 0 V
5 0 V
5 1 V
5 0 V
5 0 V
5 0 V
5 1 V
5 0 V
6 0 V
5 0 V
5 1 V
5 0 V
5 0 V
5 0 V
5 1 V
5 0 V
5 0 V
5 0 V
5 1 V
5 0 V
6 0 V
5 0 V
5 0 V
5 1 V
5 0 V
5 0 V
5 0 V
5 0 V
5 1 V
5 0 V
5 0 V
5 0 V
6 0 V
5 1 V
5 0 V
5 0 V
5 0 V
5 0 V
5 1 V
5 0 V
5 0 V
5 0 V
5 0 V
5 1 V
6 0 V
5 0 V
5 0 V
5 0 V
5 0 V
5 1 V
5 0 V
5 0 V
5 0 V
5 0 V
5 0 V
5 1 V
6 0 V
5 0 V
5 0 V
5 0 V
5 0 V
5 0 V
5 1 V
5 0 V
5 0 V
5 0 V
5 0 V
5 0 V
6 1 V
5 0 V
5 0 V
5 0 V
5 0 V
5 0 V
5 0 V
5 0 V
5 1 V
5 0 V
5 0 V
5 0 V
6 0 V
5 0 V
5 0 V
5 1 V
5 0 V
5 0 V
5 0 V
5 0 V
5 0 V
5 0 V
5 0 V
5 1 V
6 0 V
5 0 V
5 0 V
5 0 V
5 0 V
5 0 V
5 0 V
5 0 V
5 1 V
5 0 V
5 0 V
5 0 V
6 0 V
5 0 V
5 0 V
5 0 V
5 0 V
5 1 V
5 0 V
5 0 V
5 0 V
5 0 V
5 0 V
5 0 V
6 0 V
5 0 V
5 0 V
5 1 V
5 0 V
5 0 V
5 0 V
5 0 V
5 0 V
5 0 V
5 0 V
5 0 V
6 0 V
5 0 V
5 1 V
5 0 V
5 0 V
5 0 V
5 0 V
1.000 UL
LT3
400 300 M
5 5 V
5 5 V
5 5 V
5 5 V
5 5 V
6 5 V
5 5 V
5 5 V
5 5 V
5 5 V
5 6 V
5 5 V
5 5 V
5 5 V
5 5 V
5 5 V
5 5 V
6 5 V
5 5 V
5 5 V
5 5 V
5 5 V
5 5 V
5 5 V
5 5 V
5 5 V
5 6 V
5 5 V
5 5 V
6 5 V
5 5 V
5 5 V
5 5 V
5 5 V
5 5 V
5 5 V
5 6 V
5 5 V
5 5 V
5 5 V
5 5 V
6 5 V
5 5 V
5 6 V
5 5 V
5 5 V
5 5 V
5 5 V
5 5 V
5 6 V
5 5 V
5 5 V
5 5 V
6 5 V
5 6 V
5 5 V
5 5 V
5 5 V
5 6 V
5 5 V
5 5 V
5 5 V
5 6 V
5 5 V
5 5 V
6 6 V
5 5 V
5 5 V
5 6 V
5 5 V
5 5 V
5 6 V
5 5 V
5 5 V
5 6 V
5 5 V
5 6 V
6 5 V
5 6 V
5 5 V
5 6 V
5 5 V
5 5 V
5 6 V
5 6 V
5 5 V
5 6 V
5 5 V
5 6 V
6 5 V
5 6 V
5 6 V
5 5 V
5 6 V
5 5 V
5 6 V
5 6 V
5 6 V
5 5 V
5 6 V
5 6 V
6 5 V
5 6 V
5 6 V
5 6 V
5 6 V
5 6 V
5 6 V
5 5 V
5 6 V
5 6 V
5 6 V
5 6 V
6 6 V
5 6 V
5 6 V
5 6 V
5 6 V
5 7 V
5 6 V
5 6 V
5 6 V
5 6 V
5 6 V
5 7 V
6 6 V
5 6 V
5 7 V
5 6 V
5 6 V
5 7 V
5 6 V
5 7 V
5 6 V
5 7 V
5 6 V
5 7 V
6 7 V
5 6 V
5 7 V
5 7 V
5 7 V
5 6 V
5 7 V
5 7 V
5 7 V
5 7 V
5 7 V
5 7 V
6 7 V
5 7 V
5 8 V
5 7 V
5 7 V
5 8 V
5 7 V
5 7 V
5 8 V
5 7 V
5 8 V
5 8 V
6 7 V
5 8 V
5 8 V
5 8 V
5 8 V
5 8 V
5 8 V
5 8 V
5 8 V
5 8 V
5 8 V
5 9 V
6 8 V
5 9 V
5 8 V
5 9 V
5 9 V
5 8 V
5 9 V
5 9 V
5 9 V
5 9 V
5 10 V
5 9 V
6 9 V
5 10 V
5 9 V
5 10 V
5 9 V
5 10 V
5 10 V
5 10 V
5 10 V
5 9 V
5 10 V
5 10 V
6 10 V
5 10 V
5 10 V
5 10 V
5 9 V
5 10 V
5 9 V
5 9 V
5 9 V
5 9 V
5 8 V
5 8 V
6 8 V
5 7 V
5 6 V
5 7 V
5 5 V
5 5 V
5 5 V
5 4 V
5 4 V
5 3 V
5 3 V
5 3 V
6 2 V
5 2 V
5 1 V
5 1 V
5 1 V
5 1 V
5 0 V
5 0 V
5 0 V
5 0 V
5 -1 V
5 0 V
6 -1 V
5 0 V
5 -1 V
5 -1 V
5 -1 V
5 -1 V
5 -1 V
5 -2 V
5 -1 V
5 -1 V
5 -1 V
5 -2 V
6 -1 V
5 -1 V
5 -2 V
5 -1 V
5 -1 V
5 -2 V
5 -1 V
5 -2 V
5 -1 V
5 -1 V
5 -2 V
5 -1 V
6 -1 V
5 -2 V
5 -1 V
5 -1 V
5 -2 V
5 -1 V
5 -1 V
5 -2 V
5 -1 V
5 -1 V
5 -1 V
5 -1 V
6 -2 V
5 -1 V
5 -1 V
5 -1 V
5 -1 V
5 -1 V
5 -1 V
5 -2 V
5 -1 V
5 -1 V
5 -1 V
5 -1 V
6 -1 V
5 -1 V
5 -1 V
5 -1 V
5 -1 V
5 0 V
5 -1 V
5 -1 V
5 -1 V
5 -1 V
5 -1 V
5 -1 V
6 0 V
5 -1 V
5 -1 V
5 -1 V
5 -1 V
5 0 V
5 -1 V
5 -1 V
5 0 V
5 -1 V
5 -1 V
5 -1 V
6 0 V
5 -1 V
5 -1 V
5 0 V
5 -1 V
5 0 V
5 -1 V
5 -1 V
5 0 V
5 -1 V
5 0 V
5 -1 V
6 0 V
5 -1 V
5 0 V
5 -1 V
5 0 V
5 -1 V
5 0 V
5 -1 V
5 0 V
5 -1 V
5 0 V
5 -1 V
6 0 V
5 -1 V
5 0 V
5 -1 V
5 0 V
5 0 V
5 -1 V
5 0 V
5 -1 V
5 0 V
5 0 V
5 -1 V
6 0 V
5 0 V
5 -1 V
5 0 V
5 -1 V
5 0 V
5 0 V
5 -1 V
5 0 V
5 0 V
5 -1 V
5 0 V
6 0 V
5 0 V
5 -1 V
5 0 V
5 0 V
5 -1 V
5 0 V
5 0 V
5 0 V
5 -1 V
5 0 V
5 0 V
6 -1 V
5 0 V
5 0 V
5 0 V
5 -1 V
5 0 V
5 0 V
5 0 V
5 -1 V
5 0 V
5 0 V
5 0 V
6 0 V
5 -1 V
5 0 V
5 0 V
5 0 V
5 0 V
5 -1 V
5 0 V
5 0 V
5 0 V
5 0 V
5 -1 V
6 0 V
5 0 V
5 0 V
5 0 V
5 -1 V
5 0 V
5 0 V
5 0 V
5 0 V
5 0 V
currentpoint stroke M
5 -1 V
5 0 V
6 0 V
5 0 V
5 0 V
5 0 V
5 -1 V
5 0 V
5 0 V
5 0 V
5 0 V
5 0 V
5 0 V
5 -1 V
6 0 V
5 0 V
5 0 V
5 0 V
5 0 V
5 0 V
5 -1 V
5 0 V
5 0 V
5 0 V
5 0 V
5 0 V
6 0 V
5 0 V
5 -1 V
5 0 V
5 0 V
5 0 V
5 0 V
5 0 V
5 0 V
5 0 V
5 0 V
5 -1 V
6 0 V
5 0 V
5 0 V
5 0 V
5 0 V
5 0 V
5 0 V
5 0 V
5 -1 V
5 0 V
5 0 V
5 0 V
6 0 V
5 0 V
5 0 V
5 0 V
5 0 V
5 0 V
5 0 V
5 -1 V
5 0 V
5 0 V
5 0 V
5 0 V
6 0 V
5 0 V
5 0 V
5 0 V
5 0 V
5 0 V
5 0 V
5 -1 V
5 0 V
5 0 V
5 0 V
5 0 V
6 0 V
5 0 V
5 0 V
5 0 V
5 0 V
5 0 V
5 0 V
5 0 V
5 -1 V
5 0 V
5 0 V
5 0 V
6 0 V
5 0 V
5 0 V
5 0 V
5 0 V
5 0 V
5 0 V
5 0 V
5 0 V
5 0 V
5 0 V
5 0 V
6 -1 V
5 0 V
5 0 V
5 0 V
5 0 V
5 0 V
5 0 V
5 0 V
5 0 V
5 0 V
5 0 V
5 0 V
6 0 V
5 0 V
5 0 V
5 0 V
5 0 V
5 0 V
5 -1 V
5 0 V
5 0 V
5 0 V
5 0 V
5 0 V
6 0 V
5 0 V
5 0 V
5 0 V
5 0 V
5 0 V
5 0 V
5 0 V
5 0 V
5 0 V
5 0 V
5 0 V
6 0 V
5 0 V
5 0 V
5 -1 V
5 0 V
5 0 V
5 0 V
5 0 V
5 0 V
5 0 V
5 0 V
5 0 V
6 0 V
5 0 V
5 0 V
5 0 V
5 0 V
5 0 V
5 0 V
5 0 V
5 0 V
5 0 V
5 0 V
5 0 V
6 0 V
5 0 V
5 0 V
5 0 V
5 -1 V
5 0 V
5 0 V
5 0 V
5 0 V
5 0 V
5 0 V
5 0 V
6 0 V
5 0 V
5 0 V
5 0 V
5 0 V
5 0 V
5 0 V
5 0 V
5 0 V
5 0 V
5 0 V
5 0 V
6 0 V
5 0 V
5 0 V
5 0 V
5 0 V
5 0 V
5 0 V
5 0 V
5 0 V
5 0 V
5 0 V
5 0 V
6 -1 V
5 0 V
5 0 V
5 0 V
5 0 V
5 0 V
5 0 V
1.000 UL
LT5
400 300 M
5 4 V
5 5 V
5 4 V
5 4 V
5 4 V
6 4 V
5 5 V
5 4 V
5 4 V
5 4 V
5 5 V
5 4 V
5 4 V
5 4 V
5 5 V
5 4 V
5 4 V
6 4 V
5 5 V
5 4 V
5 4 V
5 4 V
5 5 V
5 4 V
5 4 V
5 4 V
5 5 V
5 4 V
5 4 V
6 4 V
5 5 V
5 4 V
5 4 V
5 5 V
5 4 V
5 4 V
5 4 V
5 5 V
5 4 V
5 4 V
5 5 V
6 4 V
5 4 V
5 4 V
5 5 V
5 4 V
5 4 V
5 5 V
5 4 V
5 4 V
5 5 V
5 4 V
5 4 V
6 4 V
5 5 V
5 4 V
5 4 V
5 5 V
5 4 V
5 4 V
5 5 V
5 4 V
5 4 V
5 5 V
5 4 V
6 5 V
5 4 V
5 4 V
5 5 V
5 4 V
5 4 V
5 5 V
5 4 V
5 5 V
5 4 V
5 4 V
5 5 V
6 4 V
5 5 V
5 4 V
5 4 V
5 5 V
5 4 V
5 5 V
5 4 V
5 5 V
5 4 V
5 5 V
5 4 V
6 4 V
5 5 V
5 4 V
5 5 V
5 4 V
5 5 V
5 4 V
5 5 V
5 4 V
5 5 V
5 4 V
5 5 V
6 5 V
5 4 V
5 5 V
5 4 V
5 5 V
5 4 V
5 5 V
5 4 V
5 5 V
5 5 V
5 4 V
5 5 V
6 4 V
5 5 V
5 5 V
5 4 V
5 5 V
5 5 V
5 4 V
5 5 V
5 5 V
5 4 V
5 5 V
5 5 V
6 4 V
5 5 V
5 5 V
5 5 V
5 4 V
5 5 V
5 5 V
5 5 V
5 4 V
5 5 V
5 5 V
5 5 V
6 5 V
5 4 V
5 5 V
5 5 V
5 5 V
5 5 V
5 5 V
5 4 V
5 5 V
5 5 V
5 5 V
5 5 V
6 5 V
5 5 V
5 5 V
5 5 V
5 5 V
5 5 V
5 5 V
5 5 V
5 5 V
5 5 V
5 5 V
5 5 V
6 5 V
5 5 V
5 5 V
5 5 V
5 5 V
5 5 V
5 5 V
5 5 V
5 5 V
5 5 V
5 5 V
5 6 V
6 5 V
5 5 V
5 5 V
5 5 V
5 5 V
5 6 V
5 5 V
5 5 V
5 5 V
5 5 V
5 6 V
5 5 V
6 5 V
5 5 V
5 5 V
5 6 V
5 5 V
5 5 V
5 6 V
5 5 V
5 5 V
5 5 V
5 6 V
5 5 V
6 5 V
5 6 V
5 5 V
5 5 V
5 5 V
5 6 V
5 5 V
5 5 V
5 6 V
5 5 V
5 5 V
5 5 V
6 6 V
5 5 V
5 5 V
5 5 V
5 6 V
5 5 V
5 5 V
5 5 V
5 5 V
5 5 V
5 6 V
5 5 V
6 5 V
5 5 V
5 5 V
5 5 V
5 5 V
5 5 V
5 5 V
5 5 V
5 4 V
5 5 V
5 5 V
5 5 V
6 4 V
5 5 V
5 5 V
5 4 V
5 4 V
5 5 V
5 4 V
5 5 V
5 4 V
5 4 V
5 4 V
5 4 V
6 4 V
5 4 V
5 4 V
5 4 V
5 3 V
5 4 V
5 4 V
5 3 V
5 4 V
5 3 V
5 3 V
5 4 V
6 3 V
5 3 V
5 3 V
5 3 V
5 2 V
5 3 V
5 3 V
5 3 V
5 2 V
5 3 V
5 2 V
5 2 V
6 3 V
5 2 V
5 2 V
5 2 V
5 2 V
5 2 V
5 2 V
5 2 V
5 2 V
5 2 V
5 1 V
5 2 V
6 1 V
5 2 V
5 1 V
5 2 V
5 1 V
5 2 V
5 1 V
5 1 V
5 1 V
5 2 V
5 1 V
5 1 V
6 1 V
5 1 V
5 1 V
5 1 V
5 1 V
5 1 V
5 0 V
5 1 V
5 1 V
5 1 V
5 1 V
5 0 V
6 1 V
5 1 V
5 0 V
5 1 V
5 1 V
5 0 V
5 1 V
5 0 V
5 1 V
5 0 V
5 1 V
5 0 V
6 1 V
5 0 V
5 0 V
5 1 V
5 0 V
5 1 V
5 0 V
5 0 V
5 1 V
5 0 V
5 0 V
5 1 V
6 0 V
5 0 V
5 0 V
5 1 V
5 0 V
5 0 V
5 0 V
5 1 V
5 0 V
5 0 V
5 0 V
5 0 V
6 1 V
5 0 V
5 0 V
5 0 V
5 0 V
5 0 V
5 1 V
5 0 V
5 0 V
5 0 V
5 0 V
5 0 V
6 0 V
5 0 V
5 1 V
5 0 V
5 0 V
5 0 V
5 0 V
5 0 V
5 0 V
5 0 V
5 0 V
5 0 V
6 0 V
5 0 V
5 0 V
5 0 V
5 1 V
5 0 V
5 0 V
5 0 V
5 0 V
5 0 V
5 0 V
5 0 V
6 0 V
5 0 V
5 0 V
5 0 V
5 0 V
5 0 V
5 0 V
5 0 V
5 0 V
5 0 V
5 0 V
5 0 V
6 0 V
5 0 V
5 0 V
5 0 V
5 0 V
5 0 V
5 0 V
5 0 V
5 0 V
5 0 V
currentpoint stroke M
5 0 V
5 0 V
6 0 V
5 0 V
5 0 V
5 0 V
5 0 V
5 0 V
5 0 V
5 0 V
5 0 V
5 0 V
5 0 V
5 0 V
6 0 V
5 0 V
5 0 V
5 0 V
5 0 V
5 0 V
5 0 V
5 0 V
5 0 V
5 0 V
5 0 V
5 0 V
6 0 V
5 0 V
5 0 V
5 0 V
5 0 V
5 0 V
5 0 V
5 0 V
5 0 V
5 0 V
5 0 V
5 0 V
6 -1 V
5 0 V
5 0 V
5 0 V
5 0 V
5 0 V
5 0 V
5 0 V
5 0 V
5 0 V
5 0 V
5 0 V
6 0 V
5 0 V
5 0 V
5 0 V
5 0 V
5 0 V
5 0 V
5 0 V
5 0 V
5 0 V
5 0 V
5 0 V
6 0 V
5 0 V
5 0 V
5 0 V
5 0 V
5 0 V
5 0 V
5 0 V
5 0 V
5 0 V
5 0 V
5 0 V
6 0 V
5 0 V
5 0 V
5 0 V
5 0 V
5 -1 V
5 0 V
5 0 V
5 0 V
5 0 V
5 0 V
5 0 V
6 0 V
5 0 V
5 0 V
5 0 V
5 0 V
5 0 V
5 0 V
5 0 V
5 0 V
5 0 V
5 0 V
5 0 V
6 0 V
5 0 V
5 0 V
5 0 V
5 0 V
5 0 V
5 0 V
5 0 V
5 0 V
5 0 V
5 0 V
5 0 V
6 0 V
5 0 V
5 0 V
5 0 V
5 0 V
5 0 V
5 0 V
5 0 V
5 0 V
5 -1 V
5 0 V
5 0 V
6 0 V
5 0 V
5 0 V
5 0 V
5 0 V
5 0 V
5 0 V
5 0 V
5 0 V
5 0 V
5 0 V
5 0 V
6 0 V
5 0 V
5 0 V
5 0 V
5 0 V
5 0 V
5 0 V
5 0 V
5 0 V
5 0 V
5 0 V
5 0 V
6 0 V
5 0 V
5 0 V
5 0 V
5 0 V
5 0 V
5 0 V
5 0 V
5 0 V
5 0 V
5 0 V
5 0 V
6 0 V
5 0 V
5 0 V
5 0 V
5 0 V
5 0 V
5 0 V
5 0 V
5 0 V
5 -1 V
5 0 V
5 0 V
6 0 V
5 0 V
5 0 V
5 0 V
5 0 V
5 0 V
5 0 V
5 0 V
5 0 V
5 0 V
5 0 V
5 0 V
6 0 V
5 0 V
5 0 V
5 0 V
5 0 V
5 0 V
5 0 V
5 0 V
5 0 V
5 0 V
5 0 V
5 0 V
6 0 V
5 0 V
5 0 V
5 0 V
5 0 V
5 0 V
5 0 V
stroke
grestore
end
showpage
}}%
\put(1076,1660){\makebox(0,0)[l]{(v)}}%
\put(1351,1420){\makebox(0,0)[l]{(iv)}}%
\put(1595,1356){\makebox(0,0)[l]{(iii)}}%
\put(2077,1244){\makebox(0,0)[l]{(ii)}}%
\put(2942,804){\makebox(0,0)[l]{(i)}}%
\put(350,1557){\makebox(0,0)[r]{\large{$\frac{\pi}{2}$}}}%
\put(1925,50){\makebox(0,0){\Large{$vr$}}}%
\put(100,1180){%
\special{ps: gsave currentpoint currentpoint translate
270 rotate neg exch neg exch translate}%
\makebox(0,0)[b]{\shortstack{\Large{$F(r)$}}}%
\special{ps: currentpoint grestore moveto}%
}%
\put(3450,200){\makebox(0,0){6}}%
\put(2942,200){\makebox(0,0){5}}%
\put(2433,200){\makebox(0,0){4}}%
\put(1925,200){\makebox(0,0){3}}%
\put(1417,200){\makebox(0,0){2}}%
\put(1383,101){\makebox(0,0){$vr_{\rm H}$}}%
\put(908,200){\makebox(0,0){1}}%
\put(400,200){\makebox(0,0){0}}%
\put(350,1900){\makebox(0,0)[r]{2}}%
\put(350,1500){\makebox(0,0)[r]{}}%
\put(350,1100){\makebox(0,0)[r]{1}}%
\put(350,700){\makebox(0,0)[r]{0.5}}%
\put(350,300){\makebox(0,0)[r]{0}}%
\end{picture}%
\endgroup
 

%% file: srsfig2b.tex
\begingroup%
  \makeatletter%
  \newcommand{\GNUPLOTspecial}{%
    \@sanitize\catcode`\%=14\relax\special}%
  \setlength{\unitlength}{0.1bp}%
{\GNUPLOTspecial{!
/gnudict 256 dict def
gnudict begin
/Color false def
/Solid false def
/gnulinewidth 5.000 def
/userlinewidth gnulinewidth def
/vshift -33 def
/dl {10 mul} def
/hpt_ 31.5 def
/vpt_ 31.5 def
/hpt hpt_ def
/vpt vpt_ def
/M {moveto} bind def
/L {lineto} bind def
/R {rmoveto} bind def
/V {rlineto} bind def
/vpt2 vpt 2 mul def
/hpt2 hpt 2 mul def
/Lshow { currentpoint stroke M
  0 vshift R show } def
/Rshow { currentpoint stroke M
  dup stringwidth pop neg vshift R show } def
/Cshow { currentpoint stroke M
  dup stringwidth pop -2 div vshift R show } def
/UP { dup vpt_ mul /vpt exch def hpt_ mul /hpt exch def
  /hpt2 hpt 2 mul def /vpt2 vpt 2 mul def } def
/DL { Color {setrgbcolor Solid {pop []} if 0 setdash }
 {pop pop pop Solid {pop []} if 0 setdash} ifelse } def
/BL { stroke userlinewidth 2 mul setlinewidth } def
/AL { stroke userlinewidth 2 div setlinewidth } def
/UL { dup gnulinewidth mul /userlinewidth exch def
      10 mul /udl exch def } def
/PL { stroke userlinewidth setlinewidth } def
/LTb { BL [] 0 0 0 DL } def
/LTa { AL [1 udl mul 2 udl mul] 0 setdash 0 0 0 setrgbcolor } def
/LT0 { PL [] 1 0 0 DL } def
/LT1 { PL [4 dl 2 dl] 0 1 0 DL } def
/LT2 { PL [2 dl 3 dl] 0 0 1 DL } def
/LT3 { PL [1 dl 1.5 dl] 1 0 1 DL } def
/LT4 { PL [5 dl 2 dl 1 dl 2 dl] 0 1 1 DL } def
/LT5 { PL [4 dl 3 dl 1 dl 3 dl] 1 1 0 DL } def
/LT6 { PL [2 dl 2 dl 2 dl 4 dl] 0 0 0 DL } def
/LT7 { PL [2 dl 2 dl 2 dl 2 dl 2 dl 4 dl] 1 0.3 0 DL } def
/LT8 { PL [2 dl 2 dl 2 dl 2 dl 2 dl 2 dl 2 dl 4 dl] 0.5 0.5 0.5 DL } def
/Pnt { stroke [] 0 setdash
   gsave 1 setlinecap M 0 0 V stroke grestore } def
/Dia { stroke [] 0 setdash 2 copy vpt add M
  hpt neg vpt neg V hpt vpt neg V
  hpt vpt V hpt neg vpt V closepath stroke
  Pnt } def
/Pls { stroke [] 0 setdash vpt sub M 0 vpt2 V
  currentpoint stroke M
  hpt neg vpt neg R hpt2 0 V stroke
  } def
/Box { stroke [] 0 setdash 2 copy exch hpt sub exch vpt add M
  0 vpt2 neg V hpt2 0 V 0 vpt2 V
  hpt2 neg 0 V closepath stroke
  Pnt } def
/Crs { stroke [] 0 setdash exch hpt sub exch vpt add M
  hpt2 vpt2 neg V currentpoint stroke M
  hpt2 neg 0 R hpt2 vpt2 V stroke } def
/TriU { stroke [] 0 setdash 2 copy vpt 1.12 mul add M
  hpt neg vpt -1.62 mul V
  hpt 2 mul 0 V
  hpt neg vpt 1.62 mul V closepath stroke
  Pnt  } def
/Star { 2 copy Pls Crs } def
/BoxF { stroke [] 0 setdash exch hpt sub exch vpt add M
  0 vpt2 neg V  hpt2 0 V  0 vpt2 V
  hpt2 neg 0 V  closepath fill } def
/TriUF { stroke [] 0 setdash vpt 1.12 mul add M
  hpt neg vpt -1.62 mul V
  hpt 2 mul 0 V
  hpt neg vpt 1.62 mul V closepath fill } def
/TriD { stroke [] 0 setdash 2 copy vpt 1.12 mul sub M
  hpt neg vpt 1.62 mul V
  hpt 2 mul 0 V
  hpt neg vpt -1.62 mul V closepath stroke
  Pnt  } def
/TriDF { stroke [] 0 setdash vpt 1.12 mul sub M
  hpt neg vpt 1.62 mul V
  hpt 2 mul 0 V
  hpt neg vpt -1.62 mul V closepath fill} def
/DiaF { stroke [] 0 setdash vpt add M
  hpt neg vpt neg V hpt vpt neg V
  hpt vpt V hpt neg vpt V closepath fill } def
/Pent { stroke [] 0 setdash 2 copy gsave
  translate 0 hpt M 4 {72 rotate 0 hpt L} repeat
  closepath stroke grestore Pnt } def
/PentF { stroke [] 0 setdash gsave
  translate 0 hpt M 4 {72 rotate 0 hpt L} repeat
  closepath fill grestore } def
/Circle { stroke [] 0 setdash 2 copy
  hpt 0 360 arc stroke Pnt } def
/CircleF { stroke [] 0 setdash hpt 0 360 arc fill } def
/C0 { BL [] 0 setdash 2 copy moveto vpt 90 450  arc } bind def
/C1 { BL [] 0 setdash 2 copy        moveto
       2 copy  vpt 0 90 arc closepath fill
               vpt 0 360 arc closepath } bind def
/C2 { BL [] 0 setdash 2 copy moveto
       2 copy  vpt 90 180 arc closepath fill
               vpt 0 360 arc closepath } bind def
/C3 { BL [] 0 setdash 2 copy moveto
       2 copy  vpt 0 180 arc closepath fill
               vpt 0 360 arc closepath } bind def
/C4 { BL [] 0 setdash 2 copy moveto
       2 copy  vpt 180 270 arc closepath fill
               vpt 0 360 arc closepath } bind def
/C5 { BL [] 0 setdash 2 copy moveto
       2 copy  vpt 0 90 arc
       2 copy moveto
       2 copy  vpt 180 270 arc closepath fill
               vpt 0 360 arc } bind def
/C6 { BL [] 0 setdash 2 copy moveto
      2 copy  vpt 90 270 arc closepath fill
              vpt 0 360 arc closepath } bind def
/C7 { BL [] 0 setdash 2 copy moveto
      2 copy  vpt 0 270 arc closepath fill
              vpt 0 360 arc closepath } bind def
/C8 { BL [] 0 setdash 2 copy moveto
      2 copy vpt 270 360 arc closepath fill
              vpt 0 360 arc closepath } bind def
/C9 { BL [] 0 setdash 2 copy moveto
      2 copy  vpt 270 450 arc closepath fill
              vpt 0 360 arc closepath } bind def
/C10 { BL [] 0 setdash 2 copy 2 copy moveto vpt 270 360 arc closepath fill
       2 copy moveto
       2 copy vpt 90 180 arc closepath fill
               vpt 0 360 arc closepath } bind def
/C11 { BL [] 0 setdash 2 copy moveto
       2 copy  vpt 0 180 arc closepath fill
       2 copy moveto
       2 copy  vpt 270 360 arc closepath fill
               vpt 0 360 arc closepath } bind def
/C12 { BL [] 0 setdash 2 copy moveto
       2 copy  vpt 180 360 arc closepath fill
               vpt 0 360 arc closepath } bind def
/C13 { BL [] 0 setdash  2 copy moveto
       2 copy  vpt 0 90 arc closepath fill
       2 copy moveto
       2 copy  vpt 180 360 arc closepath fill
               vpt 0 360 arc closepath } bind def
/C14 { BL [] 0 setdash 2 copy moveto
       2 copy  vpt 90 360 arc closepath fill
               vpt 0 360 arc } bind def
/C15 { BL [] 0 setdash 2 copy vpt 0 360 arc closepath fill
               vpt 0 360 arc closepath } bind def
/Rec   { newpath 4 2 roll moveto 1 index 0 rlineto 0 exch rlineto
       neg 0 rlineto closepath } bind def
/Square { dup Rec } bind def
/Bsquare { vpt sub exch vpt sub exch vpt2 Square } bind def
/S0 { BL [] 0 setdash 2 copy moveto 0 vpt rlineto BL Bsquare } bind def
/S1 { BL [] 0 setdash 2 copy vpt Square fill Bsquare } bind def
/S2 { BL [] 0 setdash 2 copy exch vpt sub exch vpt Square fill Bsquare } bind def
/S3 { BL [] 0 setdash 2 copy exch vpt sub exch vpt2 vpt Rec fill Bsquare } bind def
/S4 { BL [] 0 setdash 2 copy exch vpt sub exch vpt sub vpt Square fill Bsquare } bind def
/S5 { BL [] 0 setdash 2 copy 2 copy vpt Square fill
       exch vpt sub exch vpt sub vpt Square fill Bsquare } bind def
/S6 { BL [] 0 setdash 2 copy exch vpt sub exch vpt sub vpt vpt2 Rec fill Bsquare } bind def
/S7 { BL [] 0 setdash 2 copy exch vpt sub exch vpt sub vpt vpt2 Rec fill
       2 copy vpt Square fill
       Bsquare } bind def
/S8 { BL [] 0 setdash 2 copy vpt sub vpt Square fill Bsquare } bind def
/S9 { BL [] 0 setdash 2 copy vpt sub vpt vpt2 Rec fill Bsquare } bind def
/S10 { BL [] 0 setdash 2 copy vpt sub vpt Square fill 2 copy exch vpt sub exch vpt Square fill
       Bsquare } bind def
/S11 { BL [] 0 setdash 2 copy vpt sub vpt Square fill 2 copy exch vpt sub exch vpt2 vpt Rec fill
       Bsquare } bind def
/S12 { BL [] 0 setdash 2 copy exch vpt sub exch vpt sub vpt2 vpt Rec fill Bsquare } bind def
/S13 { BL [] 0 setdash 2 copy exch vpt sub exch vpt sub vpt2 vpt Rec fill
       2 copy vpt Square fill Bsquare } bind def
/S14 { BL [] 0 setdash 2 copy exch vpt sub exch vpt sub vpt2 vpt Rec fill
       2 copy exch vpt sub exch vpt Square fill Bsquare } bind def
/S15 { BL [] 0 setdash 2 copy Bsquare fill Bsquare } bind def
/D0 { gsave translate 45 rotate 0 0 S0 stroke grestore } bind def
/D1 { gsave translate 45 rotate 0 0 S1 stroke grestore } bind def
/D2 { gsave translate 45 rotate 0 0 S2 stroke grestore } bind def
/D3 { gsave translate 45 rotate 0 0 S3 stroke grestore } bind def
/D4 { gsave translate 45 rotate 0 0 S4 stroke grestore } bind def
/D5 { gsave translate 45 rotate 0 0 S5 stroke grestore } bind def
/D6 { gsave translate 45 rotate 0 0 S6 stroke grestore } bind def
/D7 { gsave translate 45 rotate 0 0 S7 stroke grestore } bind def
/D8 { gsave translate 45 rotate 0 0 S8 stroke grestore } bind def
/D9 { gsave translate 45 rotate 0 0 S9 stroke grestore } bind def
/D10 { gsave translate 45 rotate 0 0 S10 stroke grestore } bind def
/D11 { gsave translate 45 rotate 0 0 S11 stroke grestore } bind def
/D12 { gsave translate 45 rotate 0 0 S12 stroke grestore } bind def
/D13 { gsave translate 45 rotate 0 0 S13 stroke grestore } bind def
/D14 { gsave translate 45 rotate 0 0 S14 stroke grestore } bind def
/D15 { gsave translate 45 rotate 0 0 S15 stroke grestore } bind def
/DiaE { stroke [] 0 setdash vpt add M
  hpt neg vpt neg V hpt vpt neg V
  hpt vpt V hpt neg vpt V closepath stroke } def
/BoxE { stroke [] 0 setdash exch hpt sub exch vpt add M
  0 vpt2 neg V hpt2 0 V 0 vpt2 V
  hpt2 neg 0 V closepath stroke } def
/TriUE { stroke [] 0 setdash vpt 1.12 mul add M
  hpt neg vpt -1.62 mul V
  hpt 2 mul 0 V
  hpt neg vpt 1.62 mul V closepath stroke } def
/TriDE { stroke [] 0 setdash vpt 1.12 mul sub M
  hpt neg vpt 1.62 mul V
  hpt 2 mul 0 V
  hpt neg vpt -1.62 mul V closepath stroke } def
/PentE { stroke [] 0 setdash gsave
  translate 0 hpt M 4 {72 rotate 0 hpt L} repeat
  closepath stroke grestore } def
/CircE { stroke [] 0 setdash 
  hpt 0 360 arc stroke } def
/Opaque { gsave closepath 1 setgray fill grestore 0 setgray closepath } def
/DiaW { stroke [] 0 setdash vpt add M
  hpt neg vpt neg V hpt vpt neg V
  hpt vpt V hpt neg vpt V Opaque stroke } def
/BoxW { stroke [] 0 setdash exch hpt sub exch vpt add M
  0 vpt2 neg V hpt2 0 V 0 vpt2 V
  hpt2 neg 0 V Opaque stroke } def
/TriUW { stroke [] 0 setdash vpt 1.12 mul add M
  hpt neg vpt -1.62 mul V
  hpt 2 mul 0 V
  hpt neg vpt 1.62 mul V Opaque stroke } def
/TriDW { stroke [] 0 setdash vpt 1.12 mul sub M
  hpt neg vpt 1.62 mul V
  hpt 2 mul 0 V
  hpt neg vpt -1.62 mul V Opaque stroke } def
/PentW { stroke [] 0 setdash gsave
  translate 0 hpt M 4 {72 rotate 0 hpt L} repeat
  Opaque stroke grestore } def
/CircW { stroke [] 0 setdash 
  hpt 0 360 arc Opaque stroke } def
/BoxFill { gsave Rec 1 setgray fill grestore } def
end
}}%
\begin{picture}(3600,2880)(0,0)%
{\GNUPLOTspecial{"
gnudict begin
gsave
0 0 translate
0.100 0.100 scale
0 setgray
newpath
1.000 UL
LTb
400 300 M
63 0 V
2987 0 R
-63 0 V
400 713 M
63 0 V
2987 0 R
-63 0 V
400 1127 M
63 0 V
2987 0 R
-63 0 V
400 1540 M
63 0 V
2987 0 R
-63 0 V
400 1953 M
63 0 V
2987 0 R
-63 0 V
400 2367 M
63 0 V
2987 0 R
-63 0 V
400 2780 M
63 0 V
2987 0 R
-63 0 V
400 300 M
0 63 V
0 2417 R
0 -63 V
908 300 M
0 63 V
0 2417 R
0 -63 V
1417 300 M
0 63 V
0 2417 R
0 -63 V
1925 300 M
0 63 V
0 2417 R
0 -63 V
2433 300 M
0 63 V
0 2417 R
0 -63 V
2942 300 M
0 63 V
0 2417 R
0 -63 V
3450 300 M
0 63 V
0 2417 R
0 -63 V
1.000 UL
LTb
400 300 M
3050 0 V
0 2480 V
-3050 0 V
400 300 L
1.000 UL
LT1
1385 300 M
0 2480 V
1.000 UL
LT2
400 2367 M
5 0 V
5 0 V
5 0 V
5 -1 V
5 0 V
6 0 V
5 0 V
5 0 V
5 -1 V
5 0 V
5 0 V
5 -1 V
5 0 V
5 -1 V
5 0 V
5 -1 V
5 0 V
6 -1 V
5 0 V
5 -1 V
5 -1 V
5 0 V
5 -1 V
5 -1 V
5 -1 V
5 0 V
5 -1 V
5 -1 V
5 -1 V
6 -1 V
5 -1 V
5 -1 V
5 -1 V
5 -1 V
5 -1 V
5 -1 V
5 -2 V
5 -1 V
5 -1 V
5 -1 V
5 -2 V
6 -1 V
5 -1 V
5 -2 V
5 -1 V
5 -2 V
5 -1 V
5 -2 V
5 -1 V
5 -2 V
5 -1 V
5 -2 V
5 -1 V
6 -2 V
5 -2 V
5 -1 V
5 -2 V
5 -2 V
5 -2 V
5 -2 V
5 -1 V
5 -2 V
5 -2 V
5 -2 V
5 -2 V
6 -2 V
5 -2 V
5 -2 V
5 -2 V
5 -2 V
5 -2 V
5 -2 V
5 -2 V
5 -3 V
5 -2 V
5 -2 V
5 -2 V
6 -2 V
5 -3 V
5 -2 V
5 -2 V
5 -3 V
5 -2 V
5 -2 V
5 -3 V
5 -2 V
5 -2 V
5 -3 V
5 -2 V
6 -3 V
5 -2 V
5 -3 V
5 -2 V
5 -3 V
5 -2 V
5 -3 V
5 -2 V
5 -3 V
5 -3 V
5 -2 V
5 -3 V
6 -3 V
5 -2 V
5 -3 V
5 -3 V
5 -2 V
5 -3 V
5 -3 V
5 -2 V
5 -3 V
5 -3 V
5 -3 V
5 -2 V
6 -3 V
5 -3 V
5 -3 V
5 -2 V
5 -3 V
5 -3 V
5 -3 V
5 -3 V
5 -2 V
5 -3 V
5 -3 V
5 -3 V
6 -3 V
5 -3 V
5 -2 V
5 -3 V
5 -3 V
5 -3 V
5 -3 V
5 -3 V
5 -2 V
5 -3 V
5 -3 V
5 -3 V
6 -3 V
5 -3 V
5 -3 V
5 -2 V
5 -3 V
5 -3 V
5 -3 V
5 -3 V
5 -3 V
5 -2 V
5 -3 V
5 -3 V
6 -3 V
5 -3 V
5 -2 V
5 -3 V
5 -3 V
5 -3 V
5 -3 V
5 -2 V
5 -3 V
5 -3 V
5 -2 V
5 -3 V
6 -3 V
5 -3 V
5 -2 V
5 -3 V
5 -3 V
5 -2 V
5 -3 V
5 -3 V
5 -2 V
5 -3 V
5 -2 V
5 -3 V
6 -2 V
5 -3 V
5 -3 V
5 -2 V
5 -3 V
5 -2 V
5 -2 V
5 -3 V
5 -2 V
5 -3 V
5 -2 V
5 -2 V
6 -3 V
5 -2 V
5 -2 V
5 -3 V
5 -2 V
5 -2 V
5 -2 V
5 -3 V
5 -2 V
5 -2 V
5 -2 V
5 -2 V
6 -2 V
5 -2 V
5 -2 V
5 -3 V
5 -2 V
5 -1 V
5 -2 V
5 -2 V
5 -2 V
5 -2 V
5 -2 V
5 -2 V
6 -2 V
5 -1 V
5 -2 V
5 -2 V
5 -2 V
5 -1 V
5 -2 V
5 -1 V
5 -2 V
5 -2 V
5 -1 V
5 -2 V
6 -1 V
5 -2 V
5 -1 V
5 -1 V
5 -2 V
5 -1 V
5 -1 V
5 -1 V
5 -2 V
5 -1 V
5 -1 V
5 -1 V
6 -1 V
5 -1 V
5 -1 V
5 -1 V
5 -1 V
5 -1 V
5 -1 V
5 -1 V
5 -1 V
5 -1 V
5 0 V
5 -1 V
6 -1 V
5 -1 V
5 0 V
5 -1 V
5 0 V
5 -1 V
5 0 V
5 -1 V
5 0 V
5 -1 V
5 0 V
5 0 V
6 -1 V
5 0 V
5 0 V
5 0 V
5 0 V
5 -1 V
5 0 V
5 0 V
5 0 V
5 0 V
5 0 V
5 1 V
6 0 V
5 0 V
5 0 V
5 0 V
5 1 V
5 0 V
5 0 V
5 1 V
5 0 V
5 1 V
5 0 V
5 1 V
6 0 V
5 1 V
5 1 V
5 0 V
5 1 V
5 1 V
5 1 V
5 0 V
5 1 V
5 1 V
5 1 V
5 1 V
6 1 V
5 1 V
5 1 V
5 1 V
5 2 V
5 1 V
5 1 V
5 1 V
5 2 V
5 1 V
5 2 V
5 1 V
6 1 V
5 2 V
5 2 V
5 1 V
5 2 V
5 1 V
5 2 V
5 2 V
5 2 V
5 1 V
5 2 V
5 2 V
6 2 V
5 2 V
5 2 V
5 2 V
5 2 V
5 2 V
5 2 V
5 3 V
5 2 V
5 2 V
5 2 V
5 3 V
6 2 V
5 3 V
5 2 V
5 3 V
5 2 V
5 3 V
5 2 V
5 3 V
5 2 V
5 3 V
5 3 V
5 3 V
6 2 V
5 3 V
5 3 V
5 3 V
5 3 V
5 3 V
5 3 V
5 3 V
5 3 V
5 3 V
5 4 V
5 3 V
6 3 V
5 3 V
5 4 V
5 3 V
5 3 V
5 4 V
5 3 V
5 4 V
5 3 V
5 4 V
5 3 V
5 4 V
6 3 V
5 4 V
5 4 V
5 4 V
5 3 V
5 4 V
5 4 V
5 4 V
5 4 V
5 4 V
5 4 V
5 4 V
6 4 V
5 4 V
5 4 V
5 4 V
5 5 V
5 4 V
5 4 V
5 4 V
5 5 V
5 4 V
5 4 V
5 5 V
6 4 V
5 5 V
5 4 V
5 5 V
5 4 V
5 5 V
5 5 V
5 4 V
5 5 V
5 5 V
currentpoint stroke M
5 5 V
5 4 V
6 5 V
5 5 V
5 5 V
5 5 V
5 5 V
5 5 V
5 5 V
5 5 V
5 5 V
5 5 V
5 5 V
5 6 V
6 5 V
5 5 V
5 5 V
5 6 V
5 5 V
5 5 V
5 6 V
5 5 V
5 6 V
5 5 V
5 6 V
5 5 V
6 6 V
5 5 V
5 6 V
5 6 V
5 5 V
5 6 V
5 6 V
5 6 V
5 6 V
5 5 V
5 6 V
5 6 V
6 6 V
5 6 V
5 6 V
5 6 V
5 6 V
5 6 V
5 6 V
5 6 V
5 7 V
5 6 V
5 6 V
5 6 V
6 7 V
5 6 V
5 6 V
5 7 V
5 6 V
5 6 V
5 7 V
5 6 V
5 7 V
5 6 V
5 7 V
5 6 V
6 7 V
5 7 V
5 6 V
5 7 V
5 7 V
5 7 V
5 6 V
5 7 V
5 7 V
5 7 V
5 7 V
5 7 V
6 7 V
5 6 V
5 7 V
5 7 V
5 8 V
5 7 V
5 7 V
5 7 V
5 7 V
5 7 V
5 7 V
5 8 V
6 7 V
5 7 V
5 7 V
5 8 V
5 7 V
5 8 V
5 7 V
1.000 UL
LT4
400 2367 M
5 0 V
5 0 V
5 -1 V
5 0 V
5 0 V
6 -1 V
5 0 V
5 -1 V
5 0 V
5 -1 V
5 -1 V
5 0 V
5 -1 V
5 -1 V
5 -1 V
5 -1 V
5 -2 V
6 -1 V
5 -1 V
5 -1 V
5 -2 V
5 -1 V
5 -2 V
5 -2 V
5 -1 V
5 -2 V
5 -2 V
5 -2 V
5 -2 V
6 -2 V
5 -2 V
5 -2 V
5 -3 V
5 -2 V
5 -3 V
5 -2 V
5 -3 V
5 -2 V
5 -3 V
5 -3 V
5 -2 V
6 -3 V
5 -3 V
5 -3 V
5 -3 V
5 -4 V
5 -3 V
5 -3 V
5 -3 V
5 -4 V
5 -3 V
5 -4 V
5 -3 V
6 -4 V
5 -4 V
5 -4 V
5 -4 V
5 -4 V
5 -4 V
5 -4 V
5 -4 V
5 -4 V
5 -4 V
5 -4 V
5 -5 V
6 -4 V
5 -5 V
5 -4 V
5 -5 V
5 -5 V
5 -4 V
5 -5 V
5 -5 V
5 -5 V
5 -5 V
5 -5 V
5 -5 V
6 -5 V
5 -5 V
5 -6 V
5 -5 V
5 -5 V
5 -6 V
5 -5 V
5 -6 V
5 -5 V
5 -6 V
5 -6 V
5 -5 V
6 -6 V
5 -6 V
5 -6 V
5 -6 V
5 -6 V
5 -6 V
5 -6 V
5 -6 V
5 -6 V
5 -6 V
5 -7 V
5 -6 V
6 -6 V
5 -7 V
5 -6 V
5 -7 V
5 -6 V
5 -7 V
5 -7 V
5 -6 V
5 -7 V
5 -7 V
5 -7 V
5 -7 V
6 -6 V
5 -7 V
5 -7 V
5 -7 V
5 -7 V
5 -7 V
5 -8 V
5 -7 V
5 -7 V
5 -7 V
5 -7 V
5 -8 V
6 -7 V
5 -7 V
5 -8 V
5 -7 V
5 -7 V
5 -8 V
5 -7 V
5 -8 V
5 -7 V
5 -8 V
5 -8 V
5 -7 V
6 -8 V
5 -7 V
5 -8 V
5 -8 V
5 -7 V
5 -8 V
5 -8 V
5 -8 V
5 -7 V
5 -8 V
5 -8 V
5 -8 V
6 -8 V
5 -7 V
5 -8 V
5 -8 V
5 -8 V
5 -8 V
5 -8 V
5 -8 V
5 -7 V
5 -8 V
5 -8 V
5 -8 V
6 -8 V
5 -8 V
5 -8 V
5 -8 V
5 -8 V
5 -7 V
5 -8 V
5 -8 V
5 -8 V
5 -8 V
5 -8 V
5 -7 V
6 -8 V
5 -8 V
5 -8 V
5 -8 V
5 -7 V
5 -8 V
5 -8 V
5 -8 V
5 -7 V
5 -8 V
5 -7 V
5 -8 V
6 -8 V
5 -7 V
5 -8 V
5 -7 V
5 -8 V
5 -7 V
5 -8 V
5 -7 V
5 -7 V
5 -8 V
5 -7 V
5 -7 V
6 -7 V
5 -7 V
5 -7 V
5 -8 V
5 -7 V
5 -7 V
5 -6 V
5 -7 V
5 -7 V
5 -7 V
5 -7 V
5 -6 V
6 -7 V
5 -7 V
5 -6 V
5 -7 V
5 -6 V
5 -6 V
5 -7 V
5 -6 V
5 -6 V
5 -6 V
5 -6 V
5 -6 V
6 -6 V
5 -6 V
5 -6 V
5 -6 V
5 -5 V
5 -6 V
5 -6 V
5 -5 V
5 -5 V
5 -6 V
5 -5 V
5 -5 V
6 -5 V
5 -5 V
5 -5 V
5 -5 V
5 -5 V
5 -5 V
5 -4 V
5 -5 V
5 -4 V
5 -5 V
5 -4 V
5 -4 V
6 -4 V
5 -5 V
5 -4 V
5 -3 V
5 -4 V
5 -4 V
5 -4 V
5 -3 V
5 -4 V
5 -3 V
5 -3 V
5 -4 V
6 -3 V
5 -3 V
5 -3 V
5 -3 V
5 -3 V
5 -2 V
5 -3 V
5 -2 V
5 -3 V
5 -2 V
5 -2 V
5 -3 V
6 -2 V
5 -2 V
5 -2 V
5 -2 V
5 -1 V
5 -2 V
5 -1 V
5 -2 V
5 -1 V
5 -2 V
5 -1 V
5 -1 V
6 -1 V
5 -1 V
5 -1 V
5 -1 V
5 0 V
5 -1 V
5 -1 V
5 0 V
5 0 V
5 -1 V
5 0 V
5 0 V
6 0 V
5 0 V
5 0 V
5 0 V
5 0 V
5 1 V
5 0 V
5 1 V
5 0 V
5 1 V
5 1 V
5 0 V
6 1 V
5 1 V
5 1 V
5 1 V
5 2 V
5 1 V
5 1 V
5 2 V
5 1 V
5 2 V
5 1 V
5 2 V
6 2 V
5 1 V
5 2 V
5 2 V
5 2 V
5 2 V
5 3 V
5 2 V
5 2 V
5 2 V
5 3 V
5 2 V
6 3 V
5 2 V
5 3 V
5 3 V
5 2 V
5 3 V
5 3 V
5 3 V
5 3 V
5 3 V
5 3 V
5 3 V
6 4 V
5 3 V
5 3 V
5 4 V
5 3 V
5 3 V
5 4 V
5 4 V
5 3 V
5 4 V
5 4 V
5 3 V
6 4 V
5 4 V
5 4 V
5 4 V
5 4 V
5 4 V
5 4 V
5 5 V
5 4 V
5 4 V
5 4 V
5 5 V
6 4 V
5 5 V
5 4 V
5 5 V
5 4 V
5 5 V
5 4 V
5 5 V
5 5 V
5 5 V
5 5 V
5 4 V
6 5 V
5 5 V
5 5 V
5 5 V
5 5 V
5 5 V
5 6 V
5 5 V
5 5 V
5 5 V
5 6 V
5 5 V
6 5 V
5 6 V
5 5 V
5 6 V
5 5 V
5 6 V
5 5 V
5 6 V
5 6 V
5 5 V
currentpoint stroke M
5 6 V
5 6 V
6 6 V
5 5 V
5 6 V
5 6 V
5 6 V
5 6 V
5 6 V
5 6 V
5 6 V
5 6 V
5 6 V
5 6 V
6 7 V
5 6 V
5 6 V
5 6 V
5 7 V
5 6 V
5 6 V
5 7 V
5 6 V
5 7 V
5 6 V
5 7 V
6 6 V
5 7 V
5 6 V
5 7 V
5 7 V
5 6 V
5 7 V
5 7 V
5 6 V
5 7 V
5 7 V
5 7 V
6 7 V
5 7 V
5 7 V
5 7 V
5 7 V
5 7 V
5 7 V
5 7 V
5 7 V
5 7 V
5 7 V
5 7 V
6 8 V
5 7 V
5 7 V
5 7 V
5 8 V
5 7 V
5 7 V
5 8 V
5 7 V
5 8 V
5 7 V
5 8 V
6 7 V
5 8 V
5 7 V
5 8 V
5 7 V
5 8 V
5 8 V
5 7 V
5 8 V
5 8 V
5 8 V
5 7 V
6 8 V
5 8 V
5 8 V
5 8 V
5 8 V
5 8 V
5 8 V
5 8 V
5 8 V
5 8 V
5 8 V
5 8 V
6 8 V
5 8 V
5 8 V
5 8 V
5 8 V
5 9 V
5 8 V
5 8 V
5 8 V
5 9 V
5 8 V
5 8 V
6 9 V
5 8 V
5 8 V
5 9 V
5 8 V
5 9 V
5 8 V
5 9 V
5 8 V
5 9 V
5 9 V
5 8 V
6 9 V
5 9 V
5 8 V
5 9 V
5 9 V
5 8 V
5 9 V
5 9 V
5 9 V
5 9 V
5 8 V
5 9 V
6 9 V
5 9 V
5 9 V
5 9 V
5 9 V
5 9 V
5 9 V
5 9 V
5 9 V
5 9 V
5 9 V
5 10 V
6 9 V
5 9 V
5 9 V
5 9 V
5 10 V
5 9 V
5 9 V
5 9 V
5 10 V
5 9 V
5 9 V
5 10 V
6 9 V
5 10 V
5 9 V
5 10 V
5 9 V
5 10 V
5 9 V
5 10 V
5 9 V
5 10 V
5 9 V
5 10 V
6 10 V
5 9 V
5 10 V
5 10 V
5 10 V
5 9 V
5 10 V
5 10 V
5 10 V
5 10 V
5 9 V
5 10 V
6 10 V
5 10 V
5 10 V
5 10 V
5 10 V
5 10 V
5 10 V
5 10 V
5 10 V
5 10 V
5 10 V
5 10 V
6 11 V
5 10 V
5 10 V
3 7 V
1.000 UL
LT5
400 2367 M
5 0 V
5 -1 V
5 0 V
5 0 V
5 -1 V
6 0 V
5 -1 V
5 0 V
5 -1 V
5 -1 V
5 -1 V
5 -1 V
5 -2 V
5 -1 V
5 -1 V
5 -2 V
5 -2 V
6 -1 V
5 -2 V
5 -2 V
5 -2 V
5 -2 V
5 -2 V
5 -3 V
5 -2 V
5 -3 V
5 -2 V
5 -3 V
5 -3 V
6 -2 V
5 -3 V
5 -4 V
5 -3 V
5 -3 V
5 -3 V
5 -4 V
5 -3 V
5 -4 V
5 -4 V
5 -4 V
5 -4 V
6 -4 V
5 -4 V
5 -4 V
5 -4 V
5 -5 V
5 -4 V
5 -5 V
5 -4 V
5 -5 V
5 -5 V
5 -5 V
5 -5 V
6 -5 V
5 -5 V
5 -6 V
5 -5 V
5 -6 V
5 -5 V
5 -6 V
5 -6 V
5 -6 V
5 -5 V
5 -7 V
5 -6 V
6 -6 V
5 -6 V
5 -7 V
5 -6 V
5 -7 V
5 -6 V
5 -7 V
5 -7 V
5 -7 V
5 -7 V
5 -7 V
5 -7 V
6 -7 V
5 -8 V
5 -7 V
5 -7 V
5 -8 V
5 -8 V
5 -7 V
5 -8 V
5 -8 V
5 -8 V
5 -8 V
5 -8 V
6 -8 V
5 -9 V
5 -8 V
5 -8 V
5 -9 V
5 -8 V
5 -9 V
5 -9 V
5 -9 V
5 -8 V
5 -9 V
5 -9 V
6 -9 V
5 -10 V
5 -9 V
5 -9 V
5 -9 V
5 -10 V
5 -9 V
5 -10 V
5 -9 V
5 -10 V
5 -10 V
5 -10 V
6 -10 V
5 -9 V
5 -10 V
5 -10 V
5 -10 V
5 -11 V
5 -10 V
5 -10 V
5 -10 V
5 -11 V
5 -10 V
5 -11 V
6 -10 V
5 -11 V
5 -10 V
5 -11 V
5 -11 V
5 -10 V
5 -11 V
5 -11 V
5 -11 V
5 -11 V
5 -10 V
5 -11 V
6 -11 V
5 -11 V
5 -12 V
5 -11 V
5 -11 V
5 -11 V
5 -11 V
5 -11 V
5 -11 V
5 -12 V
5 -11 V
5 -11 V
6 -12 V
5 -11 V
5 -11 V
5 -12 V
5 -11 V
5 -11 V
5 -12 V
5 -11 V
5 -11 V
5 -12 V
5 -11 V
5 -11 V
6 -12 V
5 -11 V
5 -12 V
5 -11 V
5 -11 V
5 -12 V
5 -11 V
5 -11 V
5 -11 V
5 -12 V
5 -11 V
5 -11 V
6 -11 V
5 -11 V
5 -11 V
5 -11 V
5 -11 V
5 -11 V
5 -11 V
5 -11 V
5 -11 V
5 -11 V
5 -11 V
5 -10 V
6 -11 V
5 -10 V
5 -11 V
5 -10 V
5 -11 V
5 -10 V
5 -10 V
5 -10 V
5 -10 V
5 -10 V
5 -10 V
5 -10 V
6 -9 V
5 -10 V
5 -9 V
5 -9 V
5 -10 V
5 -9 V
5 -9 V
5 -9 V
5 -8 V
5 -9 V
5 -9 V
5 -8 V
6 -8 V
5 -8 V
5 -8 V
5 -8 V
5 -8 V
5 -8 V
5 -7 V
5 -7 V
5 -7 V
5 -7 V
5 -7 V
5 -7 V
6 -6 V
5 -7 V
5 -6 V
5 -6 V
5 -6 V
5 -5 V
5 -6 V
5 -5 V
5 -5 V
5 -5 V
5 -5 V
5 -5 V
6 -4 V
5 -4 V
5 -4 V
5 -4 V
5 -4 V
5 -4 V
5 -3 V
5 -3 V
5 -3 V
5 -3 V
5 -3 V
5 -2 V
6 -3 V
5 -2 V
5 -2 V
5 -2 V
5 -1 V
5 -2 V
5 -1 V
5 -1 V
5 -1 V
5 -1 V
5 -1 V
5 -1 V
6 0 V
5 -1 V
5 0 V
5 0 V
5 0 V
5 1 V
5 0 V
5 0 V
5 1 V
5 1 V
5 1 V
5 1 V
6 1 V
5 1 V
5 1 V
5 2 V
5 1 V
5 2 V
5 1 V
5 2 V
5 2 V
5 2 V
5 2 V
5 2 V
6 3 V
5 2 V
5 2 V
5 3 V
5 2 V
5 3 V
5 3 V
5 3 V
5 2 V
5 3 V
5 3 V
5 3 V
6 4 V
5 3 V
5 3 V
5 3 V
5 4 V
5 3 V
5 4 V
5 3 V
5 4 V
5 3 V
5 4 V
5 4 V
6 4 V
5 4 V
5 3 V
5 4 V
5 4 V
5 4 V
5 5 V
5 4 V
5 4 V
5 4 V
5 4 V
5 5 V
6 4 V
5 4 V
5 5 V
5 4 V
5 5 V
5 4 V
5 5 V
5 5 V
5 4 V
5 5 V
5 5 V
5 4 V
6 5 V
5 5 V
5 5 V
5 5 V
5 5 V
5 5 V
5 5 V
5 5 V
5 5 V
5 5 V
5 5 V
5 5 V
6 5 V
5 5 V
5 6 V
5 5 V
5 5 V
5 6 V
5 5 V
5 5 V
5 6 V
5 5 V
5 6 V
5 5 V
6 6 V
5 5 V
5 6 V
5 5 V
5 6 V
5 6 V
5 5 V
5 6 V
5 6 V
5 6 V
5 6 V
5 5 V
6 6 V
5 6 V
5 6 V
5 6 V
5 6 V
5 6 V
5 6 V
5 6 V
5 6 V
5 6 V
5 6 V
5 6 V
6 6 V
5 7 V
5 6 V
5 6 V
5 6 V
5 6 V
5 7 V
5 6 V
5 6 V
5 7 V
5 6 V
5 7 V
6 6 V
5 6 V
5 7 V
5 6 V
5 7 V
5 7 V
5 6 V
5 7 V
5 6 V
5 7 V
currentpoint stroke M
5 7 V
5 6 V
6 7 V
5 7 V
5 7 V
5 6 V
5 7 V
5 7 V
5 7 V
5 7 V
5 7 V
5 6 V
5 7 V
5 7 V
6 7 V
5 7 V
5 7 V
5 7 V
5 7 V
5 7 V
5 8 V
5 7 V
5 7 V
5 7 V
5 7 V
5 7 V
6 8 V
5 7 V
5 7 V
5 7 V
5 8 V
5 7 V
5 7 V
5 8 V
5 7 V
5 8 V
5 7 V
5 8 V
6 7 V
5 8 V
5 7 V
5 8 V
5 7 V
5 8 V
5 7 V
5 8 V
5 8 V
5 7 V
5 8 V
5 8 V
6 7 V
5 8 V
5 8 V
5 8 V
5 8 V
5 7 V
5 8 V
5 8 V
5 8 V
5 8 V
5 8 V
5 8 V
6 8 V
5 8 V
5 8 V
5 8 V
5 8 V
5 8 V
5 8 V
5 8 V
5 8 V
5 8 V
5 9 V
5 8 V
6 8 V
5 8 V
5 9 V
5 8 V
5 8 V
5 8 V
5 9 V
5 8 V
5 8 V
5 9 V
5 8 V
5 9 V
6 8 V
5 9 V
5 8 V
5 9 V
5 8 V
5 9 V
5 8 V
5 9 V
5 8 V
5 9 V
5 9 V
5 8 V
6 9 V
5 9 V
5 9 V
5 8 V
5 9 V
5 9 V
5 9 V
5 8 V
5 9 V
5 9 V
5 9 V
5 9 V
6 9 V
5 9 V
5 9 V
5 9 V
5 9 V
5 9 V
5 9 V
5 9 V
5 9 V
5 9 V
5 9 V
5 9 V
6 9 V
5 10 V
5 9 V
5 9 V
5 9 V
5 9 V
5 10 V
5 9 V
5 9 V
5 10 V
5 9 V
5 9 V
6 10 V
5 9 V
5 10 V
5 9 V
5 9 V
5 10 V
5 9 V
5 10 V
5 9 V
5 10 V
5 10 V
5 9 V
6 10 V
5 9 V
5 10 V
5 10 V
5 9 V
5 10 V
5 10 V
5 10 V
5 9 V
5 10 V
5 10 V
5 10 V
6 10 V
5 9 V
5 10 V
5 10 V
5 10 V
5 10 V
5 10 V
5 10 V
5 10 V
5 10 V
5 10 V
5 10 V
6 10 V
5 10 V
5 10 V
5 10 V
5 11 V
5 10 V
5 10 V
5 10 V
5 10 V
5 11 V
5 10 V
5 10 V
6 10 V
5 11 V
5 10 V
5 10 V
5 11 V
5 10 V
5 11 V
5 10 V
2 3 V
1.000 UL
LT3
400 2367 M
5 0 V
5 -1 V
5 0 V
5 0 V
5 -1 V
6 -1 V
5 -1 V
5 -1 V
5 -1 V
5 -1 V
5 -1 V
5 -2 V
5 -1 V
5 -2 V
5 -2 V
5 -2 V
5 -2 V
6 -3 V
5 -2 V
5 -3 V
5 -2 V
5 -3 V
5 -3 V
5 -3 V
5 -3 V
5 -4 V
5 -3 V
5 -4 V
5 -4 V
6 -3 V
5 -4 V
5 -5 V
5 -4 V
5 -4 V
5 -5 V
5 -4 V
5 -5 V
5 -5 V
5 -5 V
5 -5 V
5 -5 V
6 -6 V
5 -5 V
5 -6 V
5 -6 V
5 -6 V
5 -6 V
5 -6 V
5 -6 V
5 -6 V
5 -7 V
5 -7 V
5 -6 V
6 -7 V
5 -7 V
5 -8 V
5 -7 V
5 -7 V
5 -8 V
5 -7 V
5 -8 V
5 -8 V
5 -8 V
5 -8 V
5 -8 V
6 -9 V
5 -8 V
5 -9 V
5 -9 V
5 -8 V
5 -9 V
5 -9 V
5 -10 V
5 -9 V
5 -9 V
5 -10 V
5 -9 V
6 -10 V
5 -10 V
5 -10 V
5 -10 V
5 -10 V
5 -11 V
5 -10 V
5 -11 V
5 -10 V
5 -11 V
5 -11 V
5 -11 V
6 -11 V
5 -11 V
5 -11 V
5 -12 V
5 -11 V
5 -12 V
5 -11 V
5 -12 V
5 -12 V
5 -12 V
5 -12 V
5 -12 V
6 -13 V
5 -12 V
5 -13 V
5 -12 V
5 -13 V
5 -13 V
5 -12 V
5 -13 V
5 -13 V
5 -13 V
5 -14 V
5 -13 V
6 -13 V
5 -14 V
5 -13 V
5 -14 V
5 -13 V
5 -14 V
5 -14 V
5 -14 V
5 -14 V
5 -14 V
5 -14 V
5 -14 V
6 -15 V
5 -14 V
5 -14 V
5 -15 V
5 -14 V
5 -15 V
5 -14 V
5 -15 V
5 -15 V
5 -14 V
5 -15 V
5 -15 V
6 -15 V
5 -15 V
5 -15 V
5 -15 V
5 -15 V
5 -15 V
5 -15 V
5 -15 V
5 -15 V
5 -15 V
5 -16 V
5 -15 V
6 -15 V
5 -15 V
5 -15 V
5 -15 V
5 -16 V
5 -15 V
5 -15 V
5 -15 V
5 -15 V
5 -15 V
5 -15 V
5 -15 V
6 -15 V
5 -15 V
5 -15 V
5 -14 V
5 -15 V
5 -15 V
5 -14 V
5 -15 V
5 -14 V
5 -14 V
5 -14 V
5 -14 V
6 -14 V
5 -14 V
5 -13 V
5 -14 V
5 -13 V
5 -13 V
5 -13 V
5 -13 V
5 -12 V
5 -12 V
5 -12 V
5 -12 V
6 -12 V
5 -11 V
5 -11 V
5 -10 V
5 -11 V
5 -10 V
5 -9 V
5 -10 V
5 -9 V
5 -8 V
5 -8 V
5 -8 V
6 -7 V
5 -7 V
5 -7 V
5 -6 V
5 -5 V
5 -5 V
5 -5 V
5 -4 V
5 -4 V
5 -3 V
5 -3 V
5 -3 V
6 -2 V
5 -2 V
5 -1 V
5 -1 V
5 -1 V
5 0 V
5 0 V
5 0 V
5 1 V
5 0 V
5 1 V
5 1 V
6 1 V
5 1 V
5 2 V
5 1 V
5 2 V
5 2 V
5 1 V
5 2 V
5 2 V
5 2 V
5 2 V
5 2 V
6 2 V
5 3 V
5 2 V
5 2 V
5 3 V
5 2 V
5 2 V
5 3 V
5 2 V
5 3 V
5 2 V
5 3 V
6 3 V
5 2 V
5 3 V
5 3 V
5 3 V
5 2 V
5 3 V
5 3 V
5 3 V
5 3 V
5 3 V
5 3 V
6 3 V
5 3 V
5 3 V
5 3 V
5 3 V
5 3 V
5 3 V
5 3 V
5 4 V
5 3 V
5 3 V
5 4 V
6 3 V
5 3 V
5 4 V
5 3 V
5 4 V
5 3 V
5 4 V
5 3 V
5 4 V
5 4 V
5 3 V
5 4 V
6 4 V
5 3 V
5 4 V
5 4 V
5 4 V
5 4 V
5 4 V
5 4 V
5 4 V
5 4 V
5 4 V
5 4 V
6 4 V
5 4 V
5 4 V
5 4 V
5 4 V
5 5 V
5 4 V
5 4 V
5 5 V
5 4 V
5 4 V
5 5 V
6 4 V
5 5 V
5 4 V
5 5 V
5 4 V
5 5 V
5 4 V
5 5 V
5 4 V
5 5 V
5 5 V
5 5 V
6 4 V
5 5 V
5 5 V
5 5 V
5 5 V
5 5 V
5 4 V
5 5 V
5 5 V
5 5 V
5 5 V
5 5 V
6 5 V
5 6 V
5 5 V
5 5 V
5 5 V
5 5 V
5 5 V
5 6 V
5 5 V
5 5 V
5 6 V
5 5 V
6 5 V
5 6 V
5 5 V
5 6 V
5 5 V
5 6 V
5 5 V
5 6 V
5 5 V
5 6 V
5 5 V
5 6 V
6 6 V
5 5 V
5 6 V
5 6 V
5 6 V
5 5 V
5 6 V
5 6 V
5 6 V
5 6 V
5 6 V
5 6 V
6 6 V
5 6 V
5 6 V
5 6 V
5 6 V
5 6 V
5 6 V
5 6 V
5 6 V
5 6 V
5 6 V
5 6 V
6 7 V
5 6 V
5 6 V
5 6 V
5 7 V
5 6 V
5 6 V
5 7 V
5 6 V
5 7 V
5 6 V
5 7 V
6 6 V
5 7 V
5 6 V
5 7 V
5 6 V
5 7 V
5 6 V
5 7 V
5 7 V
5 6 V
currentpoint stroke M
5 7 V
5 7 V
6 7 V
5 6 V
5 7 V
5 7 V
5 7 V
5 7 V
5 7 V
5 6 V
5 7 V
5 7 V
5 7 V
5 7 V
6 7 V
5 7 V
5 7 V
5 7 V
5 7 V
5 8 V
5 7 V
5 7 V
5 7 V
5 7 V
5 7 V
5 8 V
6 7 V
5 7 V
5 7 V
5 8 V
5 7 V
5 7 V
5 8 V
5 7 V
5 8 V
5 7 V
5 8 V
5 7 V
6 7 V
5 8 V
5 8 V
5 7 V
5 8 V
5 7 V
5 8 V
5 8 V
5 7 V
5 8 V
5 8 V
5 7 V
6 8 V
5 8 V
5 8 V
5 7 V
5 8 V
5 8 V
5 8 V
5 8 V
5 8 V
5 8 V
5 8 V
5 8 V
6 8 V
5 8 V
5 8 V
5 8 V
5 8 V
5 8 V
5 8 V
5 8 V
5 8 V
5 8 V
5 8 V
5 9 V
6 8 V
5 8 V
5 8 V
5 9 V
5 8 V
5 8 V
5 9 V
5 8 V
5 8 V
5 9 V
5 8 V
5 9 V
6 8 V
5 9 V
5 8 V
5 9 V
5 8 V
5 9 V
5 8 V
5 9 V
5 8 V
5 9 V
5 9 V
5 8 V
6 9 V
5 9 V
5 9 V
5 8 V
5 9 V
5 9 V
5 9 V
5 8 V
5 9 V
5 9 V
5 9 V
5 9 V
6 9 V
5 9 V
5 9 V
5 9 V
5 9 V
5 9 V
5 9 V
5 9 V
5 9 V
5 9 V
5 9 V
5 9 V
6 9 V
5 10 V
5 9 V
5 9 V
5 9 V
5 10 V
5 9 V
5 9 V
5 9 V
5 10 V
5 9 V
5 9 V
6 10 V
5 9 V
5 10 V
5 9 V
5 10 V
5 9 V
5 10 V
5 9 V
5 10 V
5 9 V
5 10 V
5 9 V
6 10 V
5 10 V
5 9 V
5 10 V
5 10 V
5 9 V
5 10 V
5 10 V
5 10 V
5 9 V
5 10 V
5 10 V
6 10 V
5 10 V
5 9 V
5 10 V
5 10 V
5 10 V
5 10 V
5 10 V
5 10 V
5 10 V
5 10 V
5 10 V
6 10 V
5 10 V
5 10 V
5 11 V
5 10 V
5 10 V
5 10 V
5 10 V
5 11 V
5 10 V
5 10 V
5 10 V
6 11 V
5 10 V
5 10 V
5 11 V
5 10 V
5 10 V
5 11 V
5 10 V
1 2 V
1.000 UL
LT0
400 2367 M
5 0 V
5 -1 V
5 0 V
5 -1 V
5 0 V
6 -1 V
5 -2 V
5 -1 V
5 -1 V
5 -2 V
5 -2 V
5 -2 V
5 -2 V
5 -2 V
5 -3 V
5 -2 V
5 -3 V
6 -3 V
5 -3 V
5 -3 V
5 -4 V
5 -4 V
5 -3 V
5 -4 V
5 -5 V
5 -4 V
5 -4 V
5 -5 V
5 -5 V
6 -5 V
5 -5 V
5 -5 V
5 -6 V
5 -6 V
5 -5 V
5 -6 V
5 -7 V
5 -6 V
5 -6 V
5 -7 V
5 -7 V
6 -7 V
5 -7 V
5 -7 V
5 -8 V
5 -8 V
5 -7 V
5 -8 V
5 -9 V
5 -8 V
5 -8 V
5 -9 V
5 -9 V
6 -9 V
5 -9 V
5 -9 V
5 -10 V
5 -9 V
5 -10 V
5 -10 V
5 -10 V
5 -10 V
5 -11 V
5 -10 V
5 -11 V
6 -11 V
5 -11 V
5 -11 V
5 -11 V
5 -12 V
5 -11 V
5 -12 V
5 -12 V
5 -12 V
5 -12 V
5 -13 V
5 -12 V
6 -13 V
5 -13 V
5 -13 V
5 -13 V
5 -13 V
5 -13 V
5 -14 V
5 -14 V
5 -13 V
5 -14 V
5 -14 V
5 -15 V
6 -14 V
5 -15 V
5 -14 V
5 -15 V
5 -15 V
5 -15 V
5 -15 V
5 -16 V
5 -15 V
5 -16 V
5 -15 V
5 -16 V
6 -16 V
5 -16 V
5 -16 V
5 -17 V
5 -16 V
5 -17 V
5 -16 V
5 -17 V
5 -17 V
5 -17 V
5 -17 V
5 -17 V
6 -18 V
5 -17 V
5 -18 V
5 -17 V
5 -18 V
5 -18 V
5 -18 V
5 -18 V
5 -18 V
5 -18 V
5 -18 V
5 -19 V
6 -18 V
5 -19 V
5 -18 V
5 -19 V
5 -18 V
5 -19 V
5 -19 V
5 -19 V
5 -19 V
5 -19 V
5 -19 V
5 -19 V
6 -19 V
5 -19 V
5 -19 V
5 -19 V
5 -19 V
5 -19 V
5 -19 V
5 -19 V
5 -19 V
5 -18 V
5 -19 V
5 -19 V
6 -19 V
5 -18 V
5 -19 V
5 -18 V
5 -18 V
5 -18 V
5 -18 V
5 -17 V
5 -18 V
5 -17 V
5 -16 V
5 -17 V
6 -15 V
5 -16 V
5 -15 V
5 -14 V
5 -13 V
5 -13 V
5 -12 V
5 -10 V
5 -10 V
5 -8 V
5 -7 V
5 -5 V
6 -3 V
5 -2 V
5 0 V
5 0 V
5 1 V
5 1 V
5 1 V
5 0 V
5 1 V
5 0 V
5 0 V
5 -1 V
6 -1 V
5 -1 V
5 -1 V
5 -1 V
5 -1 V
5 -2 V
5 -1 V
5 -1 V
5 0 V
5 0 V
5 0 V
5 0 V
6 1 V
5 0 V
5 0 V
5 1 V
5 0 V
5 1 V
5 1 V
5 1 V
5 0 V
5 1 V
5 1 V
5 1 V
6 1 V
5 1 V
5 1 V
5 2 V
5 1 V
5 1 V
5 2 V
5 1 V
5 1 V
5 2 V
5 2 V
5 1 V
6 2 V
5 2 V
5 1 V
5 2 V
5 2 V
5 2 V
5 2 V
5 2 V
5 2 V
5 2 V
5 2 V
5 3 V
6 2 V
5 2 V
5 2 V
5 3 V
5 2 V
5 3 V
5 2 V
5 3 V
5 2 V
5 3 V
5 2 V
5 3 V
6 3 V
5 3 V
5 2 V
5 3 V
5 3 V
5 3 V
5 3 V
5 3 V
5 3 V
5 3 V
5 3 V
5 3 V
6 3 V
5 3 V
5 4 V
5 3 V
5 3 V
5 4 V
5 3 V
5 3 V
5 4 V
5 3 V
5 4 V
5 3 V
6 4 V
5 3 V
5 4 V
5 4 V
5 3 V
5 4 V
5 4 V
5 3 V
5 4 V
5 4 V
5 4 V
5 4 V
6 4 V
5 4 V
5 4 V
5 4 V
5 4 V
5 4 V
5 4 V
5 4 V
5 4 V
5 4 V
5 4 V
5 5 V
6 4 V
5 4 V
5 5 V
5 4 V
5 4 V
5 5 V
5 4 V
5 5 V
5 4 V
5 5 V
5 4 V
5 5 V
6 4 V
5 5 V
5 4 V
5 5 V
5 5 V
5 5 V
5 4 V
5 5 V
5 5 V
5 5 V
5 5 V
5 4 V
6 5 V
5 5 V
5 5 V
5 5 V
5 5 V
5 5 V
5 5 V
5 5 V
5 5 V
5 5 V
5 6 V
5 5 V
6 5 V
5 5 V
5 5 V
5 6 V
5 5 V
5 5 V
5 6 V
5 5 V
5 5 V
5 6 V
5 5 V
5 6 V
6 5 V
5 6 V
5 5 V
5 6 V
5 5 V
5 6 V
5 6 V
5 5 V
5 6 V
5 6 V
5 5 V
5 6 V
6 6 V
5 5 V
5 6 V
5 6 V
5 6 V
5 6 V
5 6 V
5 6 V
5 6 V
5 6 V
5 5 V
5 6 V
6 7 V
5 6 V
5 6 V
5 6 V
5 6 V
5 6 V
5 6 V
5 6 V
5 6 V
5 7 V
5 6 V
5 6 V
6 7 V
5 6 V
5 6 V
5 7 V
5 6 V
5 6 V
5 7 V
5 6 V
5 7 V
5 6 V
5 7 V
5 6 V
6 7 V
5 6 V
5 7 V
5 6 V
5 7 V
5 7 V
5 6 V
5 7 V
5 7 V
5 6 V
currentpoint stroke M
5 7 V
5 7 V
6 7 V
5 7 V
5 6 V
5 7 V
5 7 V
5 7 V
5 7 V
5 7 V
5 7 V
5 7 V
5 7 V
5 7 V
6 7 V
5 7 V
5 7 V
5 7 V
5 7 V
5 8 V
5 7 V
5 7 V
5 7 V
5 7 V
5 8 V
5 7 V
6 7 V
5 7 V
5 8 V
5 7 V
5 8 V
5 7 V
5 7 V
5 8 V
5 7 V
5 8 V
5 7 V
5 8 V
6 7 V
5 8 V
5 7 V
5 8 V
5 8 V
5 7 V
5 8 V
5 8 V
5 7 V
5 8 V
5 8 V
5 7 V
6 8 V
5 8 V
5 8 V
5 8 V
5 7 V
5 8 V
5 8 V
5 8 V
5 8 V
5 8 V
5 8 V
5 8 V
6 8 V
5 8 V
5 8 V
5 8 V
5 8 V
5 8 V
5 8 V
5 9 V
5 8 V
5 8 V
5 8 V
5 8 V
6 9 V
5 8 V
5 8 V
5 9 V
5 8 V
5 8 V
5 9 V
5 8 V
5 8 V
5 9 V
5 8 V
5 9 V
6 8 V
5 9 V
5 8 V
5 9 V
5 8 V
5 9 V
5 8 V
5 9 V
5 9 V
5 8 V
5 9 V
5 9 V
6 8 V
5 9 V
5 9 V
5 9 V
5 8 V
5 9 V
5 9 V
5 9 V
5 9 V
5 9 V
5 9 V
5 8 V
6 9 V
5 9 V
5 9 V
5 9 V
5 9 V
5 9 V
5 9 V
5 10 V
5 9 V
5 9 V
5 9 V
5 9 V
6 9 V
5 9 V
5 10 V
5 9 V
5 9 V
5 9 V
5 10 V
5 9 V
5 9 V
5 10 V
5 9 V
5 9 V
6 10 V
5 9 V
5 10 V
5 9 V
5 10 V
5 9 V
5 10 V
5 9 V
5 10 V
5 9 V
5 10 V
5 9 V
6 10 V
5 10 V
5 9 V
5 10 V
5 10 V
5 9 V
5 10 V
5 10 V
5 10 V
5 9 V
5 10 V
5 10 V
6 10 V
5 10 V
5 10 V
5 10 V
5 10 V
5 10 V
5 9 V
5 10 V
5 10 V
5 10 V
5 11 V
5 10 V
6 10 V
5 10 V
5 10 V
5 10 V
5 10 V
5 10 V
5 11 V
5 10 V
5 10 V
5 10 V
5 10 V
5 11 V
6 10 V
5 10 V
5 11 V
5 10 V
5 10 V
5 11 V
5 10 V
5 11 V
3 6 V
stroke
grestore
end
showpage
}}%
\put(1010,507){\makebox(0,0)[l]{(v)}}%
\put(1330,610){\makebox(0,0)[l]{(iv)}}%
\put(1595,651){\makebox(0,0)[l]{(iii)}}%
\put(1823,1023){\makebox(0,0)[l]{(ii)}}%
\put(1823,2015){\makebox(0,0)[l]{(i)}}%
\put(1925,50){\makebox(0,0){\Large$vr$}}%
\put(100,1540){%
\special{ps: gsave currentpoint currentpoint translate
270 rotate neg exch neg exch translate}%
\makebox(0,0)[b]{\shortstack{\Large$B(r)$}}%
\special{ps: currentpoint grestore moveto}%
}%
\put(3450,200){\makebox(0,0){6}}%
\put(2942,200){\makebox(0,0){5}}%
\put(2433,200){\makebox(0,0){4}}%
\put(1925,200){\makebox(0,0){3}}%
\put(1417,200){\makebox(0,0){2}}%
\put(1435,101){\makebox(0,0){$vr_{\rm H}$}}%
\put(908,200){\makebox(0,0){1}}%
\put(400,200){\makebox(0,0){0}}%
\put(350,2780){\makebox(0,0)[r]{1.2}}%
\put(350,2367){\makebox(0,0)[r]{1}}%
\put(350,1953){\makebox(0,0)[r]{0.8}}%
\put(350,1540){\makebox(0,0)[r]{0.6}}%
\put(350,1127){\makebox(0,0)[r]{0.4}}%
\put(350,713){\makebox(0,0)[r]{0.2}}%
\put(350,300){\makebox(0,0)[r]{0}}%
\end{picture}%
\endgroup
 